\documentclass[twocolumn, resetfootnote]{aastex701}

\usepackage{calrsfs}
\usepackage{makecell}

\usepackage{hyperref}

\newcommand{\degrees}{$^\circ$}
\newcommand{\msun}{M$_{\sun}$}

\newcommand{\mstar}{M$_{\star}$}

\newcommand{\gaia}{$d_{\rm Gaia}$}

\newcommand{\msunyr}{\msun\,yr$^{-1}$}

\newcommand{\tshock}{$\rm T_{shock}$}
\newcommand{\tmax}{$\rm T_{max}$}

\newcommand{\cosi}{$\cos(i_{\rm mag})$}
\newcommand{\imag}{$i_{\rm mag}$}
\newcommand{\idisk}{$i_{\rm disk,gas}$}
\newcommand{\idust}{$i_{\rm disk,dust}$}

\newcommand{\halpha}{H$\alpha$}

\newcommand{\chisq}{$\chi^2$}
\newcommand{\mdot}{$\dot{M}$}
\newcommand{\ri}{$R_{\rm i}$}
\newcommand{\rw}{$W_{\rm r}$}

\newcommand{\rco}{$R_{\rm co}$}

\newcommand{\prot}{$P_{\rm rot}$}

\newcommand{\av}{$A_V$}

\newcommand\be {\begin{equation}}
\newcommand\en{\end{equation}}
\def\micron{$\mu$m}

\def\fp1{$f_{p1}$}

\usepackage[utf8]{inputenc}
\usepackage{tikz}
\usepackage[percent]{overpic}
\usepackage{graphicx}
\usepackage{amsmath,amssymb}
\usepackage{breqn}
\usepackage{booktabs}
\usepackage{longtable}
\usepackage{lipsum}
\usepackage{amsmath}
\usepackage{subcaption}

\shorttitle{ODYSSEUS Star-Cloud Connection}
\shortauthors{Pittman et al.}

\graphicspath{{./}{figures/}}

\begin{document}

\title{The ODYSSEUS Survey. Spatial correlation of magnetospheric inclinations points to parsec-scale star-cloud connection}

\correspondingauthor{Caeley V. Pittman}
\email{cpittman@bu.edu}

\author[0000-0001-9301-6252]{Caeley V. Pittman}\altaffiliation{NSF Graduate Research Fellow}
\affiliation{Department of Astronomy, Boston University, 725 Commonwealth Avenue, Boston, MA 02215, USA}
\affiliation{Institute for Astrophysical Research, Boston University, 725 Commonwealth Avenue, Boston, MA 02215, USA}
\email{cpittman@bu.edu}

\author[0000-0001-9227-5949]{Catherine C. Espaillat}
\affiliation{Department of Astronomy, Boston University, 725 Commonwealth Avenue, Boston, MA 02215, USA}
\affiliation{Institute for Astrophysical Research, Boston University, 725 Commonwealth Avenue, Boston, MA 02215, USA}
\email{cce@bu.edu}

\author[0000-0003-4507-1710]{Thanawuth Thanathibodee}
\affiliation{Department of Physics, Faculty of Science, Chulalongkorn University, 254 Phayathai Road, Pathumwan, Bangkok 10330, Thailand}
\email{Thanawuth.T@chula.ac.th}

\author[0000-0002-3950-5386]{Nuria Calvet}
\affiliation{Department of Astronomy, University of Michigan, 311 West Hall, 1085 S. University Avenue, Ann Arbor, MI 48109, USA}
\email{ncalvet@umich.edu}

\author[0000-0003-1430-8519]{Lee W. Hartmann}
\affiliation{Department of Astronomy, University of Michigan, 311 West Hall, 1085 S. University Avenue, Ann Arbor, MI 48109, USA}
\email{lhartm@umich.edu}

\author[0000-0002-1593-3693]{Sylvie Cabrit}
\affiliation{Observatoire de Paris - PSL University, Sorbonne Université, LERMA, CNRS, Paris, France}
\affiliation{Univ. Grenoble Alpes, CNRS, IPAG, Grenoble, France}
\email{sylvie.cabrit@obspm.fr}

\submitjournal{ApJL}
\received{25 September 2025}
\revised{27 October 2025}
\accepted{29 October 2025}

\begin{abstract}
The properties of stars and planets are shaped by the initial conditions of their natal clouds.
However, the spatial scales over which the initial conditions can exert a significant influence are not well constrained.
We report the first evidence for parsec-scale spatial correlations of stellar magnetospheric inclinations (\imag), observed in the Lupus low-mass star forming region.
Applying consensus clustering with a hierarchical density-based clustering algorithm, we demonstrate that the detected spatial dependencies are stable against perturbations by measurement uncertainties.
The \imag\ correlation scales are on the order of $\sim$3~pc, which aligns with the typical scales of the Lupus molecular cloud filaments. Our results reveal a connection between large-scale forces|in the form of expanding shells from the Upper Scorpius and Upper-Centaurus-Lupus regions|and sub-au scale system configurations. 
We find that Lupus has a non-uniform \imag\ distribution and suggest that this results from the preferential elongation of protostellar cores along filamentary axes.
Non-uniformity would have significant implications for exoplanet occurrence rate calculations, so future work should explore the longevity of these biases driven by the star-cloud connection.

\end{abstract}


\section{Introduction} \label{sec:intro}

The development of pre-main-sequence stars and their surrounding protoplanetary disks occurs in the context of their larger-scale natal molecular clouds. The mass, velocity, angular momentum, and magnetic field structures of the progenitor interstellar medium (ISM) may have observable effects on the same features of T Tauri stars (TTSs) and disks. 
It is well known that TTSs are influenced by their surrounding environments, such as through external irradiation from massive stars \citep[e.g.,][]{WinterHaworth2022,Mauco2025,Allen2025}, Bondi-Hoyle-Lyttleton infall from the surrounding interstellar medium \citep[ISM; e.g.,][]{Winter2024Lupus}, and interactions with neighboring bodies such as binary companions \citep[e.g.,][]{Kurtovic2022} and dynamical encounters from flybys \citep[e.g.,][]{Cuello2023,Winter2024Encounters}. Additionally, star formation trends have been found to be connected across parsec to kiloparsec scales \citep{Mendigutia2018}.
The importance of the parsec-scale natal cloud conditions to the observed characteristics in the Class II stage (when the magnetospherically-accreting systems are called classical TTSs; CTTSs) has not been well established. Additionally, the physical scales over which initial conditions can create coherent structures are unclear.

The Lupus star-forming region (SFR) provides a strong sample within which to search for spatial dependencies, as it is relatively close \citep[$\sim$160~pc,][]{Luhman2020} and young \citep[1--3~Myr][]{Galli2020}, and it contains many well-characterized CTTSs in relatively low extinction regions \citep{manara23PPVII}.
Lupus contains four subregions with CTTSs (Lupus I--IV), % spread across approximately 50~pc \cite{Tothill2009} [all of Lupus, not just I-IV], 
and its cloud morphologies have been shaped by expanding shells from high-mass stellar activity in the nearby Upper Scorpius (USco) and Upper-Centaurus-Lupus (UCL) regions \citep{deGeus1992,Tachihara1996,Tachihara2001,MoreiraYun2002,Preibisch2002,Comeron2008,Gaczkowski2015,Gaczkowski2017}. Such large-scale flows are well-established ``molecular cloud factories'' \citep{Dawson2011,Dawson2015,Dawson2013,Krause2013,Pineda2023} and may be able to produce associated large-scale coherent kinematic structures. 
% Molecular cloud studies have found that $\sim70$\% of pre- and proto-stellar sources are located on filaments \citep[][]{Polychroni2013,Schisano2014}

Lupus I contains a primary filament that is 5--7~pc in length and 1.6 parsec in width. It is parallel to, and co-moving with, the edge of the USco HI shell \citep{Tothill2009,Benedettini2015,Gaczkowski2015,Gaczkowski2017}. 
Its CO maps suggest the presence of coherent rotation around the axis of the southeast portion of the filament, spanning over 1~pc \citep{Tothill2009}, as has been found in other ISM filaments \citep[e.g.][]{Heyer1987,Dhabal2018,Pineda2023}. Its large-scale magnetic field in the plane of the sky is oriented perpendicular to the filament axis \citep{MyersGoodman1991,Rizzo1998,FrancoAlves2015,Soler2016}, as has also been commonly found in other molecular cloud filaments \citep{Heyer1987,PereyraMagalhaes2004,Alves2008}. It appears to be undergoing active star formation \citep{VilasBoas2000,Benedettini2012,Rygl2013}.

The main Lupus II filament spans 40\arcmin\ \citep{MoreiraYun2002}, which corresponds to 1.8~pc in projection when assuming a distance of 157~pc.
Its local magnetic field is nearly parallel to the filamentary axis \citep{MyersGoodman1991}, which is more common in diffuse matter \citep[][]{FrancoAlves2015,Planck2016Diffuse}. This is consistent with Lupus II's lower density compared to the other three subregions \citep[][]{Cambresy1999,Tachihara1996}. Velocity gradients in its CO observations may indicate large-scale streaming motions along the filament resulting from interaction with the USco and/or UCL shells \citep{MoreiraYun2002}.

The Lupus III region is the most densely populated in terms of CTTSs, and it is likely the most evolved region \citep{Benedettini2012,Rygl2013,Benedettini2018}. Its local magnetic field orientation is approximately perpendicular to the main filament \citep{MyersGoodman1991,Planck2016a,Aizawa2020}, which spans $\sim$4~pc with a width of 1~pc \citep[][]{Tothill2009}. Finally, Lupus IV has a fairly complex filamentary structure, but its longest filaments span $\sim$3~pc \citep{Benedettini2015}. Its local magnetic field is particularly well-ordered and oriented perpendicular to the eastern filamentary axis and nearly parallel to the western axis \citep{Rizzo1998}. It also shows velocity gradients that may originate from interaction with the UCL shell shock front \citep{MoreiraYun2002}.

In this work, we examine spatial correlations of CTTS magnetospheric inclinations in Lupus. This connects scales below 0.1~au in individual systems to the parsec scales of the host region.
In Section~\ref{sec:SampleAnalysis}, we describe our sample and analysis methods. In Section~\ref{sec:results}, we present our results, and we discuss the implications in Section~\ref{sec:discussion}. Finally, we present a summary in Section~\ref{sec:summary}.

\section{Sample \& Analysis} \label{sec:SampleAnalysis}

This work continues the series of \cite{Pittman2025,Pittman2025b}, examining accretion in the well-characterized HST ULLYSES sample.
In \cite{Pittman2025}, we presented a self-consistent analysis of magnetospheric accretion flow and shock signatures. By applying the flow model of \cite{hartmann94, muzerolle98b, muzerolle01} to velocity-resolved \halpha\ profiles, we obtained the accretion rate (\mdot), the inner magnetospheric truncation radius (\ri), the width of the truncation region at the midplane (\rw), the maximum temperature of the flow (\tmax), and the viewing inclination of the magnetosphere (\imag). Using the accretion shock model of \cite{cg98} as updated by \cite{re19}, we determined the $V$-band extinction (\av), the surface coverage and energy flux densities of multi-component shocks, and approximate hot spot temperature (\tshock). In \cite{Pittman2025b}, we measured the rotation periods (\prot), and associated corotation radii (\rco), of the sample and explored the implications for angular momentum regulation in the inner disk.

Following these works, we now expand the sample of consistently-characterized Lupus CTTSs and incorporate 3D spatial information to test for correlations with the larger-scale environment. 
The \cite{Pittman2025} sample contains 25 CTTSs in Lupus with reliable Gaia DR3 distances, \gaia\ (RUWE$\lesssim$1.4).
We have selected an additional 36 Lupus CTTSs with archival VLT spectra, stellar parameters derived in the same manner as our sample by \cite{alcala14,alcala17,manara23PPVII}, and reliable \gaia.\footnote{RUWE$\lesssim$1.4 applies to all CTTS except V1094~Sco. This target has RUWE=2.1, but its distance and proper motions place it in Lupus III as expected, and its parallax uncertainty is small (0.03~mas).} 

\startlongtable
\begin{deluxetable*}{llllllll}
\tablecaption{Sample and System Characteristics \label{tab:sample}}
\tablehead{
\colhead{Object} & \colhead{RA} & \colhead{Dec} & \colhead{\gaia} & \colhead{\imag} & \colhead{\idisk} & \colhead{PA} & \colhead{\texttt{HDBSCAN} Group}\\
\colhead{} & \colhead{(hh:mm:ss.ss)} & \colhead{(dd:mm:ss.ss)} & \colhead{(pc)} & \colhead{(\degrees)} & \colhead{(\degrees)} & \colhead{(\degrees)} & \colhead{}
}
\startdata
\hline \textbf{Lupus I} \\ \hline
2MASSJ15445789-3423392 & 15:44:57.89 & -34:23:39.35 & $153.9\pm1.9$ & $80\pm7$ & \dots & \dots & Group I NW \\
2MASSJ15450887-3417333\tablenotemark{a} & 15:45:08.87 & -34:17:33.45 & $154.8\pm3.5$ & $80\pm5$ & $60^{+5}_{-15}$ & $170^{+10}_{-40}$ & Group I NW \\
SSTc2dJ154518.5-342125 & 15:45:18.52 & -34:21:24.55 & $153.0\pm2.2$ & $85\pm5$ & \dots & \dots & Group I NW \\
Sz 65\tablenotemark{b} & 15:39:27.77 & -34:46:17.21 & $153.5\pm0.6$ & $80\pm9$ & $64.6$ & $288.2$ & Group I NW \\
Sz 66\tablenotemark{b} & 15:39:28.28 & -34:46:18.08 & $155.9\pm0.7$ & $79\pm5$ & $67.8$ & $260.0$ & Group I NW \\
Sz 68\tablenotemark{a} & 15:45:12.87 & -34:17:30.64 & $152.7\pm4.3$ & $69\pm5$ & $<$45 & $110^{+10}_{-5}$ & Group I NW \\
Sz 71\tablenotemark{a} & 15:46:44.73 & -34:30:35.68 & $155.2\pm0.4$ & $50\pm5$ & $<$30 & $35^{+10}_{-5}$ & Group I SE \\
Sz 72\tablenotemark{b} & 15:47:50.63 & -35:28:35.40 & $156.7\pm0.5$ & $51\pm5$ & $36.9$ & $199.9$ & Group I SE \\
Sz 73\tablenotemark{a} & 15:47:56.94 & -35:14:34.80 & $157.8\pm0.7$ & $48\pm5$ & $<$50 & $255^{+5}_{-10}$ & Group I SE \\
Sz 75 & 15:49:12.11 & -35:39:05.06 & $154.1\pm0.7$ & $52\pm5$ & \dots & \dots & Group I SE \\
Sz 69\tablenotemark{a} & 15:45:17.41 & -34:18:28.29 & $152.6\pm1.6$ & $22\pm15$ & $<$40 & $315^{+10}_{-10}$ & Not grouped \\
\hline \textbf{Lupus II} \\ \hline
Sz 81 A & 15:55:50.28 & -38:01:33.70 & $158.2\pm0.6$ & $65\pm10$ & \dots & \dots & Group II \\
Sz 82 & 15:56:09.21 & -37:56:06.13 & $155.8\pm0.5$ & $62\pm5$ & \dots & \dots & Group II \\
Sz 84\tablenotemark{a} & 15:58:02.52 & -37:36:02.73 & $155.6\pm1.1$ & $70\pm5$ & $50^{+5}_{-15}$ & $355^{+5}_{-5}$ & Group II \\
\hline \textbf{Lupus III} \\ \hline
2MASSJ16073773-3921388 & 16:07:37.73 & -39:21:38.74 & $162.4\pm3.0$ & $85\pm5$ & \dots & \dots & Group III \\
2MASSJ16080017-3902595 & 16:08:00.17 & -39:02:59.50 & $161.1\pm1.6$ & $69\pm11$ & \dots & \dots & Group III \\
2MASSJ16084940-3905393 & 16:08:49.40 & -39:05:39.45 & $160.2\pm2.5$ & $72\pm10$ & \dots & \dots & Group III \\
2MASSJ16085324-3914401 & 16:08:53.24 & -39:14:40.16 & $163.0\pm1.3$ & $77\pm10$ & \dots & \dots & Group III \\
2MASSJ16085529-3848481 & 16:08:55.29 & -38:48:48.17 & $155.6\pm2.2$ & $82\pm5$ & \dots & \dots & Group III \\
2MASSJ16090141-3925119\tablenotemark{a} & 16:09:01.41 & -39:25:11.92 & $159.2\pm1.2$ & $76\pm5$ & $60^{+5}_{-5}$ & $355^{+5}_{-5}$ & Group III \\
2MASSJ16092697-3836269\tablenotemark{a} & 16:09:26.98 & -38:36:26.93 & $159.2\pm1.7$ & $63\pm14$ & $65^{+5}_{-5}$ & $130^{+20}_{-10}$ & Group III \\
2MASSJ16095628-3859518 & 16:09:56.30 & -38:59:51.58 & $157.1\pm2.0$ & $80\pm7$ & \dots & \dots & Group III \\
2MASSJ16101857-3836125 & 16:10:18.57 & -38:36:12.60 & $160.6\pm2.6$ & $65\pm19$ & \dots & \dots & Group III \\
2MASSJ16101984-3836065\tablenotemark{a} & 16:10:19.83 & -38:36:06.59 & $158.8\pm3.0$ & $77\pm10$ & $55^{+5}_{-10}$ & $335^{+10}_{-5}$ & Group III \\
2MASSJ16102955-3922144\tablenotemark{a} & 16:10:29.55 & -39:22:14.45 & $160.4\pm1.1$ & $73\pm7$ & $65^{+5}_{-10}$ & $120^{+10}_{-5}$ & Group III \\
SSTc2dJ160830.7-382827\tablenotemark{a} & 16:08:30.70 & -38:28:26.85 & $153.4\pm0.7$ & $56\pm5$ & $55^{+5}_{-5}$ & $110^{+5}_{-5}$ & Group III \\
Sz 90\tablenotemark{a} & 16:07:10.07 & -39:11:03.26 & $160.4\pm0.5$ & $78\pm12$ & $50^{+5}_{-15}$ & $130^{+5}_{-5}$ & Group III \\
Sz 91 & 16:07:11.59 & -39:03:47.49 & $159.4\pm0.7$ & $80\pm5$ & \dots & \dots & Group III \\
Sz 95\tablenotemark{b} & 16:07:52.31 & -38:58:06.09 & $160.5\pm0.7$ & $80\pm10$ & $56.8$ & $65.0$ & Group III \\
Sz 96\tablenotemark{a} & 16:08:12.63 & -39:08:33.47 & $156.0\pm0.5$ & $77\pm5$ & $<$50 & $25^{+5}_{-5}$ & Group III \\
Sz 97 & 16:08:21.80 & -39:04:21.49 & $157.3\pm0.6$ & $83\pm5$ & \dots & \dots & Group III \\
Sz 98\tablenotemark{a} & 16:08:22.49 & -39:04:46.43 & $156.3\pm0.6$ & $58\pm8$ & $50^{+5}_{-15}$ & $35^{+10}_{-5}$ & Group III \\
Sz 99 & 16:08:24.04 & -39:05:49.43 & $158.3\pm1.1$ & $82\pm5$ & \dots & \dots & Group III \\
Sz 103\tablenotemark{a} & 16:08:30.27 & -39:06:11.18 & $157.2\pm1.0$ & $82\pm5$ & $50^{+5}_{-20}$ & $50^{+5}_{-10}$ & Group III \\
Sz 104 & 16:08:30.81 & -39:05:48.84 & $159.8\pm1.1$ & $88\pm5$ & \dots & \dots & Group III \\
Sz 106 & 16:08:39.76 & -39:06:25.32 & $158.7\pm0.8$ & $67\pm22$ & \dots & \dots & Group III \\
Sz 110 & 16:08:51.57 & -39:03:17.71 & $157.5\pm0.6$ & $77\pm5$ & \dots & \dots & Group III \\
Sz 111\tablenotemark{a} & 16:08:54.68 & -39:37:43.15 & $158.4\pm0.5$ & $65\pm5$ & $<$35 & $40^{+5}_{-5}$ & Group III \\
Sz 112 & 16:08:55.53 & -39:02:33.94 & $159.3\pm0.8$ & $71\pm5$ & \dots & \dots & Group III \\
Sz 113 & 16:08:57.80 & -39:02:22.85 & $160.5\pm1.4$ & $84\pm5$ & \dots & \dots & Group III \\
Sz 114\tablenotemark{a} & 16:09:01.85 & -39:05:12.41 & $156.8\pm0.6$ & $51\pm5$ & $15^{+5}_{-5}$ & $170^{+5}_{-10}$ & Group III \\
Sz 115 & 16:09:06.21 & -39:08:51.87 & $155.8\pm0.9$ & $82\pm5$ & \dots & \dots & Group III \\
Sz 117 & 16:09:44.36 & -39:13:30.17 & $156.9\pm0.5$ & $82\pm5$ & \dots & \dots & Group III \\
Sz 118\tablenotemark{a} & 16:09:48.66 & -39:11:16.85 & $161.5\pm0.7$ & $61\pm5$ & $55^{+5}_{-15}$ & $155^{+10}_{-5}$ & Group III \\
V1094 Sco\tablenotemark{b} & 16:08:36.18 & -39:23:02.46 & $154.8\pm0.8$ & $60\pm12$ & $50.0$ & $110.0$ & Group III \\
\hline \textbf{Lupus IV} \\ \hline
2MASSJ16000236-4222145\tablenotemark{a} & 16:00:02.36 & -42:22:14.60 & $160.4\pm1.1$ & $68\pm10$ & $30^{+5}_{-10}$ & $340^{+5}_{-5}$ & Group IV \\
2MASSJ16002612-4153553 & 16:00:26.12 & -41:53:55.37 & $163.2\pm1.3$ & $73\pm9$ & \dots & \dots & Group IV \\
SSTc2dJ160000.6-422158 & 16:00:00.60 & -42:21:56.82 & $159.4\pm0.8$ & $72\pm5$ & \dots & \dots & Group IV \\
Sz 129\tablenotemark{a} & 15:59:16.47 & -41:57:10.30 & $160.1\pm0.4$ & $62\pm5$ & $70^{+5}_{-10}$ & $170^{+40}_{-10}$ & Group IV \\
Sz 130\tablenotemark{a} & 16:00:31.04 & -41:43:36.99 & $159.2\pm0.5$ & $75\pm5$ & $55^{+10}_{-15}$ & $325^{+10}_{-20}$ & Group IV \\
Sz 131\tablenotemark{b} & 16:00:49.43 & -41:30:03.92 & $160.6\pm0.7$ & $74\pm10$ & $59.0$ & $335.8$ & Group IV \\
MY Lup\tablenotemark{a} & 16:00:44.52 & -41:55:30.93 & $157.2\pm0.9$ & $89\pm5$ & $55^{+5}_{-5}$ & $200^{+10}_{-5}$ & Not grouped \\
\hline \textbf{Off-Cloud} \\ \hline
Sz 77\tablenotemark{c} & 15:51:46.96 & -35:56:44.11 & $155.3\pm0.4$ & $50\pm9$ & $26.0\pm0.8$ & \dots & Group I SE \\
RXJ 1556.1-3655 & 15:56:02.10 & -36:55:28.27 & $157.9\pm0.6$ & $64\pm6$ & \dots & \dots & Group II \\
SSTc2dJ161243.8-381503\tablenotemark{b} & 16:12:43.75 & -38:15:03.08 & $159.9\pm0.5$ & $73\pm5$ & $55.2$ & $198.4$ & Group III \\
2MASSJ15414081-3345188\tablenotemark{d} & 15:41:40.81 & -33:45:18.85 & $151.9\pm6.5$ & $70\pm8$ & \dots & \dots & Not grouped \\
2MASSJ16081497-3857145\tablenotemark{a} & 16:08:14.98 & -38:57:14.41 & $150.8\pm16.6$ & $84\pm5$ & $75^{+5}_{-5}$ & $35^{+10}_{-5}$ & Not grouped \\
2MASSJ16085373-3914367\tablenotemark{a} & 16:08:53.74 & -39:14:36.80 & $148.7\pm66.4$ & $79\pm9$ & $65^{+5}_{-5}$ & $305^{+20}_{-5}$ & Not grouped \\
2MASSJ16100133-3906449\tablenotemark{a} & 16:10:01.33 & -39:06:44.82 & $184.1\pm5.8$ & $80\pm7$ & $<$35 & $190^{+10}_{-5}$ & Not grouped \\
EX Lup & 16:03:05.49 & -40:18:25.43 & $154.7\pm0.4$ & $63\pm5$ & \dots & \dots & Not grouped \\
SSTc2dJ161344.1-373646 & 16:13:44.10 & -37:36:46.26 & $158.6\pm1.3$ & $58\pm5$ & \dots & \dots & Not grouped \\
 \enddata
 \tablenotetext{}{Sample and system characteristics divided by Lupus subregion. The coordinates are in the J2000 reference frame. Gaia distances (\gaia) and associated uncertainties come from Gaia Data Release 3 \citep{GaiaDR3}. The magnetospheric inclination (\imag) comes from \cite{Pittman2025} and this work; all accretion flow model results can be found in Table~\ref{tab:flowresults} in Appendix~\ref{Appsec:flowresults}. Gas disk inclinations (\idisk) and position angles (PA) come from \cite{Yen2018} and \cite{Trapman2025} as indicated. The \texttt{HDBSCAN} group assignment comes from the 4D clustering procedure, taking both the 3D spatial location and \imag\ values into account, as shown in Figure~\ref{fig:LupusIncl} (right).
 $a$: PA and \idisk\ from \cite{Yen2018}, $b$: PA and \idisk\ from \cite{Trapman2025}, $c$: \idisk\ from \cite{Vioque2025} due to stronger constraint on \idisk\ than in \cite{Trapman2025}, $d$: Membership status not included in \cite{Luhman2020}, but marked here as off-cloud due to its potentially older age \citep[8.3~Myr,][]{alcala17b} and smaller \gaia\ compared to confirmed Lupus I members.
 }
\end{deluxetable*}
\clearpage

We confirm that all CTTSs show detectable accretion emission above the chromosphere in \halpha, and then perform accretion flow model fits to obtain \imag\ following \cite{Pittman2025} and using the stellar parameters of \cite{alcala14,alcala17,manara23PPVII}, adjusted to \gaia.
Because HST spectra are not available for these non-ULLYSES targets, we use a uniform \tshock\ value of 5260~K, which is the median \tshock\ found for the Lupus targets in \cite{Pittman2025}. As before, we mask out any absorption features that cannot be attributed to a magnetospheric origin.
We calculate these new model grids with \imag\ ranging from 0\degrees--85\degrees\ in steps of 5\degrees\ for all CTTSs, and we calculate the mean and standard deviation of \imag\ from the top 1000 best-fit models weighted by their associated exp(-\chisq/2). When multiple epochs are available, the final fit values are taken to be the weighted means of the individual epochs. 
Though our statistical uncertainties can be very small, we enforce a minimum \imag\ uncertainty of 5\degrees\ as a more conservative estimate that reflects the model grid resolution.
We perform tests in Appendix~\ref{Appsec:flowresults} to validate our \imag\ results.
The final sample of 61 CTTSs is listed in Table~\ref{tab:sample}. Eleven are in Lupus I, three are in Lupus II, 31 are in Lupus III, seven are in Lupus IV, and nine are off-cloud \citep[][]{Luhman2020}. 

\subsection{Consensus clustering analysis with \texttt{HDBSCAN}} \label{sec:HDBSCAN}

To test for spatial correlations of \imag, we apply the density-based hierarchical clustering algorithm \texttt{HDBSCAN} \citep[][]{HDBSCANCampello2013} as implemented in Python by \cite{HDBSCANMcInnes2017}. \texttt{HDBSCAN} has been used for open cluster identification \citep[e.g.,][]{KounkelCovey2019,HuntReffert2021,Vioque2023} and is particularly useful for detecting clusters of different densities \citep[][]{HDBSCANCampello2013}. For our case, we use \texttt{HDBSCAN} to assign each CTTS to a group based on its 4D characteristics (the three spatial coordinates plus \imag). To produce equal parameter weights, we apply robust scaling to the 4D inputs as implemented in the \texttt{RobustScaler} of \texttt{scikit-learn}, adjusting them to each have a median of zero and an interquartile range of one. This also ensures that the absolute range of observed \imag\ values does not influence the clustering, as even a sample biased towards high or low \imag\ will be normalized such that its values are spread out according to the relative scatter around the median.

We incorporate measurement uncertainties by creating 5000 Monte Carlo (MC) samples, produced by perturbing the nominal \gaia\ and \imag\ values by normal distributions\footnote{We use a truncated normal distribution to prevent nonphysical values (negative distances, \imag\ below 0\degrees, and \imag\ above 90\degrees).} centered on zero with a standard deviation given by the uncertainties. Each of these MC samples is a possible true distribution of \gaia\ and \imag\ for the 61 CTTSs studied here. We classify each MC sample independently with \texttt{HDBSCAN}, similar to \cite{Ou2023}, to ensure that the group classification is stable against perturbations.

For each MC realization, \texttt{HDBSCAN} assigns each star a label that either places it in a group or marks it as noise. Group labels are not connected across different \texttt{HDBSCAN} instantiations, so we determine the final groups using consensus clustering \citep{Monti2003}, which is a probabilistic analysis of the pairwise relationships between individual CTTSs. For each pair, we determine the co-membership probability ($P_{ij}$) by calculating the percentage of the 5000 \texttt{HDBSCAN} instantiations in which the pair is classified into the same group (ignoring cases in which both are marked as noise). If $P_{ij}$ is above an empirically-defined probability threshold ($P_{\rm min}$), then the pair is marked as connected. 

We construct an adjacency matrix $A$ in which each element $A_{ij}$ is 1 if $P_{ij}>P_{\rm min}$ and 0 if $P_{ij}<P_{\rm min}$. We then apply the \texttt{connected_components} algorithm from \texttt{scipy.sparse.csgraph} to $A$ to create the stable groups of connected CTTSs. We define the high-probability group members to be the CTTSs whose mean pairwise connection probability with all of the other group members ($P_{ij,{\rm mean}}$) is above $P_{\rm min}$. To remove very low-probability group members, we calculate the percentage of the 5000 \texttt{HDBSCAN} instantiations in which each CTTS is marked as noise ($P_{\rm noise}$) and remove any with $P_{\rm noise}$ larger than an empirically-defined noise threshold ($P_{\rm noise,max}$). Appendix~\ref{Appsec:clusteringDesign} describes the determination of $P_{\rm min}$ and $P_{\rm noise,max}$ in detail.

\texttt{HDBSCAN} requires three main user inputs: \texttt{min_cluster_size}, which is the minimum number of members required to be classified as a group; \texttt{min_samples}, which is by default equivalent to \texttt{min_cluster_size} and sets how conservative the clustering will be, with larger \texttt{min_samples} values causing more points to be marked as noise rather than group members; and \texttt{cluster_selection_method}, which can be set to the default of \textit{excess of mass} or changed to \textit{leaf}. The former method is more likely to group based on larger-scale, stable structure, whereas the latter is more likely to produce smaller, more homogeneous groups that may be less stable (where stability describes the persistence of the group over a range of density thresholds).\footnote{\url{https://hdbscan.readthedocs.io/en/latest/index.html}} We choose the \textit{excess of mass} method to prioritize the most stable groups.

To choose the appropriate hyperparameters for \texttt{HDBSCAN}, we first optimize the model to group the CTTSs into their appropriate Lupus subregions based only on their 3D coordinates, which we call the \textit{3D groups}. We create 5000 MC samples by perturbing \gaia\ based on the associated uncertainties. We then choose physically-motivated \texttt{min_cluster_size} and \texttt{min_samples} values that create four stable groups (Lupus I, II, III, and IV), optimizing the $P_{\rm min}$ and $P_{\rm noise,max}$ thresholds during consensus clustering so that all on-cloud CTTSs are successfully grouped, and any CTTS marked as noise is an off-cloud source. Details can be found in Appendix~\ref{Appsec:clusteringDesign}.

Once we determine the best hyperparameter values and consensus clustering thresholds, we incorporate \imag\ as the fourth dimension and repeat the above process, the result of which we call the \textit{4D groups}. If the optimized algorithm finds stable groups that correspond to the nominal spatial correlations of \imag\ in the plane of the sky, this provides evidence that the correlations are physical and robust against perturbation within the uncertainties.
We ensure that the 5000 MC samples produce converged results by testing runs with 10,000, 20,000, and 50,000 MC samples in size and confirming that the reported group assignments remain unchanged for both the 3D groups and the 4D groups.

To test for any systematic effects from introducing \imag\ as a fourth dimension, we repeat the 4D clustering process by assigning new randomly-generated \imag\ values that are consistent with the observed \imag\ distribution during each MC realization. Then, to test for any spurious correlations that could be introduced by perturbing \imag\ by the associated uncertainties ($\sigma_{i_{\rm mag}}$), we repeat the 4D clustering process a final time by creating an initial sample with randomly-assigned \imag, and then perturb it by our $\sigma_{i_{\rm mag}}$ values. We repeat this with 100 different initial \imag\ samples. These tests will also ensure that the 4D groups do not result purely from the 3D distribution of the sources, and the results can be found in Appendix~\ref{Appsec:clusteringDesign}.

Here is a summary of the procedure:

\begin{enumerate}
    \item Optimize the \texttt{HDBSCAN} user inputs by determining the values that produce a successful classification of the known spatial structure in the region (the Lupus I-IV subregions) when considering only their 3D spatial locations perturbed by their \gaia\ uncertainties.
    \item Apply this optimized procedure with the addition of \imag\ as the fourth dimension, perturbed by the associated \imag\ uncertainties, to confirm the spatial correlations of \imag\ seen in the plane of the sky.
    \item Validate the addition of \imag\ as the fourth dimension by applying the same procedure, except with a new random set of \imag\ assigned during each MC realization.
    \item Validate the \imag\ measurement uncertainties by applying the same procedure, except by perturbing a single random set of \imag\ by the \imag\ measurement uncertainties during each MC realization (and repeat with different initial random sets).
\end{enumerate}

\begin{figure*}
\begin{tikzpicture}
    \centering
    \node[anchor=south west, inner sep=0] (image) at (0,0) {\includegraphics[width=0.535\linewidth]{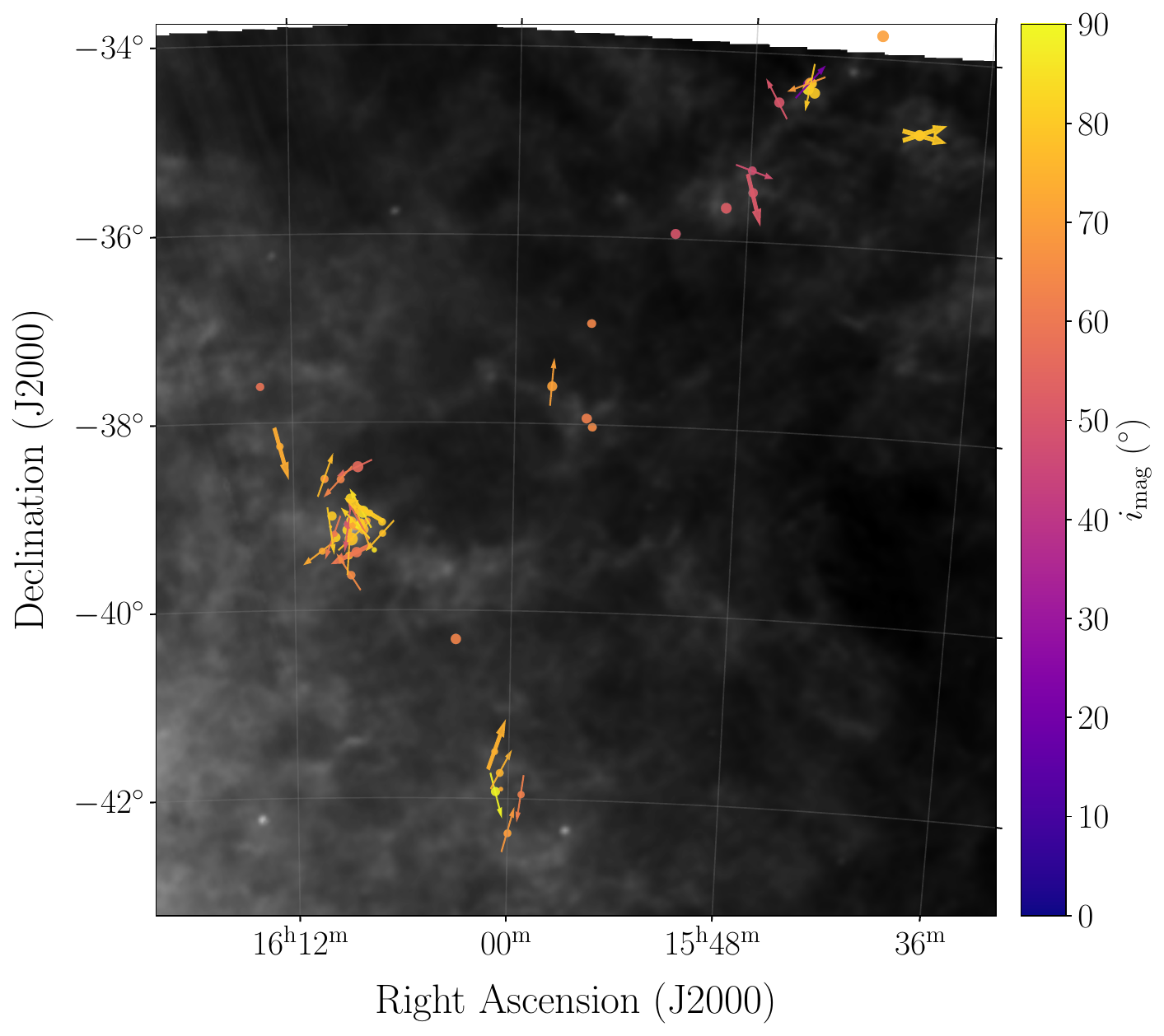}};
    \begin{scope}[x={(image.south east)},y={(image.north west)}]
            % Annotations relative to the image coordinates (0,0) to (1,1)
            \node[white] at (0.57, 0.89) {Lupus I};
            \node[white] at (0.55, 0.64) {Lupus II};
            \node[white] at (0.42, 0.49) {Lupus III};
            \node[white] at (0.54, 0.24) {Lupus IV};
        \end{scope}
    \end{tikzpicture}
    \includegraphics[width=0.465\linewidth]{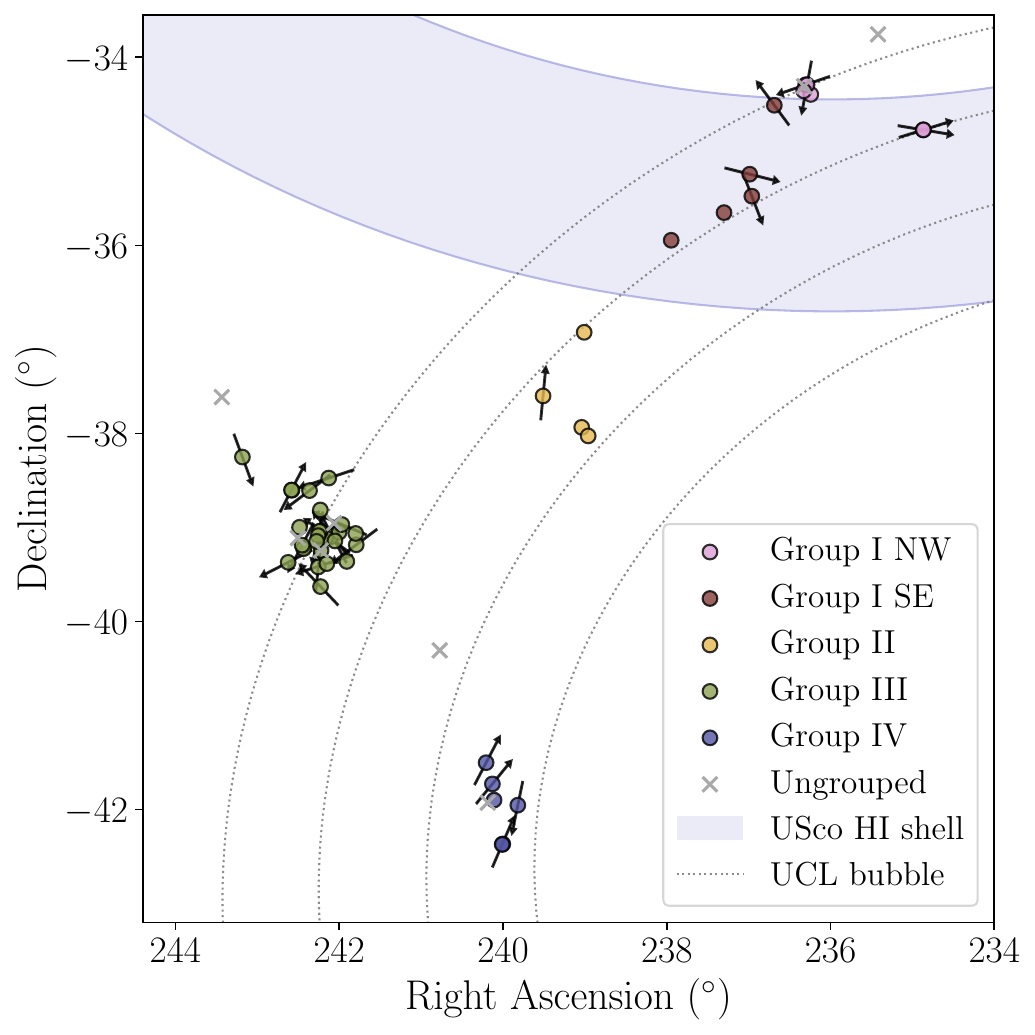}\\
    \begin{interactive}{animation}{PaperIII_Figure1_26Oct.mp4}\end{interactive}
    \includegraphics[trim=10pt 20pt 25pt 70pt, clip, width=0.5\linewidth]{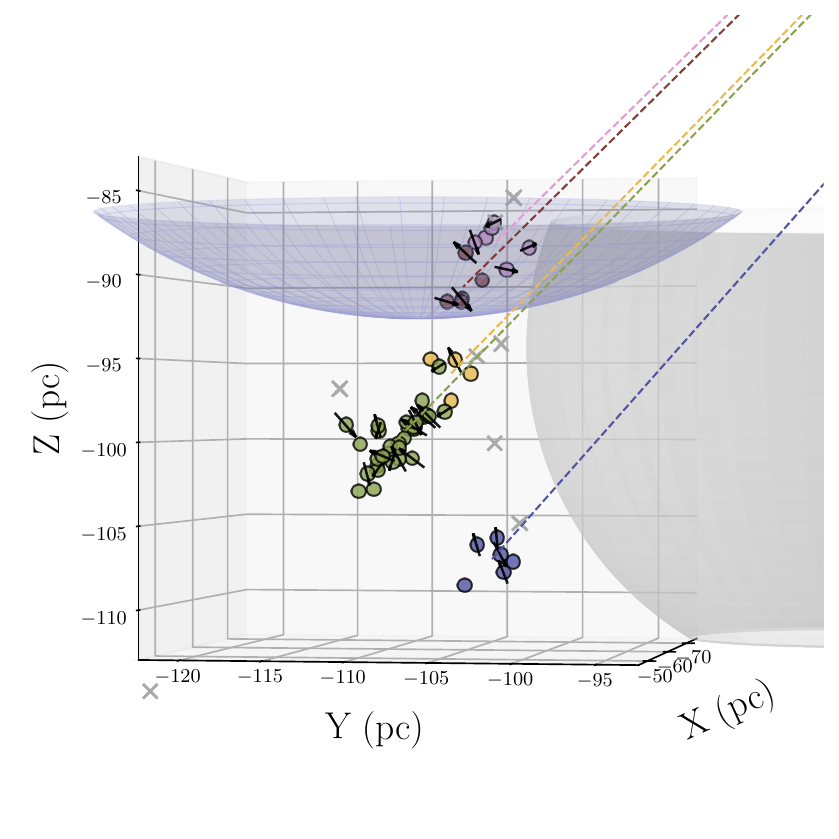} \hspace{1cm}
    \includegraphics[trim=50pt 20pt 15pt 70pt, clip, width=0.465\linewidth]{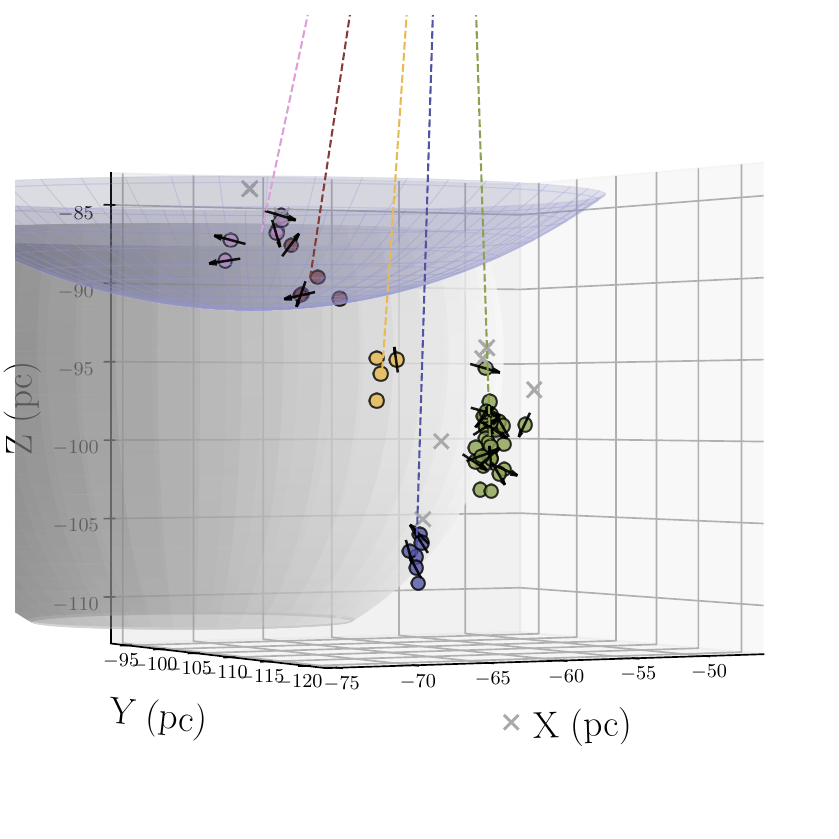}\\ % left, bottom, right, top
    \caption{Top left panel: Map of the Lupus targets analyzed in this work with the IRIS 100 \micron\ map \citep{IRIS2005} shown in grayscale. 
    The color bar shows the nominal magnetospheric inclination \imag\ found from the accretion flow model (see Tables~\ref{tab:sample} and \ref{tab:flowresults}). Point sizes indicate \gaia\ such that more distant sources appear smaller. 
    A labeled version can be found in Figure~\ref{fig:LupusApp} in Appendix~\ref{sec:Appendix}, along with zoomed images of individual subregions.
    Position angles (PAs) are shown as colored arrows, indicating the projected major axes of the gas disks. They come from \citet[][thin arrows]{Yen2018} and \citet[][thick arrows]{Trapman2025} and are measured from $0^\circ\leq {\rm PA} < 360^\circ$, defined counterclockwise from north (see compass in lower right). In the remaining subfigures, the PAs are indicated by black arrows.
    Top right panel: \texttt{HDBSCAN} group assignments (colors) of the Lupus CTTSs, taking 3D spatial location and \imag\ into account with equal weight. Non-grouped CTTSs are indicated by gray x markers. The shaded blue region indicates the inner and outer boundaries of the Upper Scorpius HI shell projected onto the plane of the sky, and the dotted lines show the projected orientation of the Upper-Centaurus-Lupus wind shell \citep{deGeus1992,Gaczkowski2017}.
    Bottom panels: Three-dimensional versions of the top right panel plotted in ICRS cartesian coordinates using the nominal \gaia\ values, with each axis spanning 30~pc. The blue curve indicates the outer edge of the USco shell, at 36~pc from its center. The white curve indicates the UCL bubble, which does not have a fixed extent but is plotted at a radius of 21~pc from its center for ease of comparison. The dashed lines show lines of sight between the cluster centroids and Earth. An animated version of this figure is available in the online journal, which rotates the figure 360\degrees\ around the $z$ axis to show the 3D structure. The video shows three full rotations over a duration of 1 minute 12 seconds.
    }
    \label{fig:LupusIncl}
\end{figure*}

\section{Results} \label{sec:results}

Figure~\ref{fig:LupusIncl} (top left) shows the nominal \imag\ spatial distribution in the context of IRIS 100~\micron\ dust map \citep{IRIS2005}.
We can immediately see spatial correlations in the plane of the sky. For example, the components of the wide binary in the western region of Lupus I, Sz~65/Sz~66, have nearly equivalent \imag\ values (yellow points in Figure~\ref{fig:LupusIncl}; also see the labeled version in Figure~\ref{fig:LupusApp} in Appendix~\ref{sec:Appendix}).
Then, the five most southeastern CTTSs around Lupus I all have a nominal \imag\ between $\sim$48\degrees--52\degrees\ (pink points).
The four CTTSs around Lupus II have nominal \imag\ values in the slightly higher range between $\sim$62\degrees--70\degrees\ (orange points). In Lupus III, we find a high prevalence of high-inclination magnetospheres, with a median nominal \imag\ of 77\degrees\ and a median absolute deviation (MAD)\footnote{${\rm MAD=median}(|x_i-\bar{x}|)$, where $x_i$ is each measurement and $\bar{x}$ is the sample's median.} of 6\degrees. Finally, we find that five of seven Lupus IV CTTSs have a nominal \imag\ between $\sim$68\degrees--75\degrees\ (orange points; there are two CTTSs co-located in the plane of the sky at the southernmost point).

\begin{deluxetable}{llllll}[t]
\tablewidth{\columnwidth}
\tabletypesize{\footnotesize}
\tablecaption{\texttt{HDBSCAN} Group Characteristics \label{tab:clusters}}
\tablehead{
\colhead{Group} & \colhead{$N$} & \colhead{$d_{\rm med}$ (pc)} & \colhead{$V$ (pc$^3$)} & \colhead{$s_{\rm med}$ (pc)} & \colhead{\imag\ (\degrees)}
}
\startdata
I NW & 6 & $154.05\pm0.64$ & $1.06\pm0.32$ & $3.78\pm0.32$ & $79\pm7$ \\
I SE & 5 & $155.47\pm0.22$ & $3.14\pm0.55$ & $3.24\pm0.18$ & $50\pm6$ \\
II & 4 & $156.95\pm0.27$ & $1.19\pm0.29$ & $3.03\pm0.26$ & $65\pm7$ \\
III & 32 & $158.68\pm0.25$ & $27.44\pm2.42$ & $3.01\pm0.15$ & $75\pm12$ \\
IV & 6 & $160.16\pm0.25$ & $1.21\pm0.25$ & $2.20\pm0.26$ & $71\pm8$ \\
 \enddata
 \tablenotetext{}{Characteristics of the 4D groups shown in Figure~\ref{fig:LupusIncl} (right). $N$ gives the number of group members; $d_{\rm med}$ gives the distance to the members; $V$ gives the volume of a convex hull that surrounds all members; and $s_{\rm med}$ gives the pairwise separation between members.
 Quantities are the medians calculated from the 5000 MC samples, with uncertainties given by the MAD for all except \imag, for which we instead report the root-mean-squared dispersion (RMS) to indicate the magnitude of the spread.
 }
\end{deluxetable}

As described in Appendix~\ref{Appsec:clusteringDesign}, the optimal \texttt{HDBSCAN} hyperparameters for Lupus are \texttt{min_cluster_size}=3 and \texttt{min_samples}=3, and the best consensus clustering thresholds are $P_{\rm min}$=0.5, and $P_{\rm noise,max}$=0.3. We add \imag\ as the fourth dimension and apply this optimized algorithm to determine the 4D groups. This results in five groups, shown in Figure~\ref{fig:LupusIncl} (right). These groups confirm that the correlation of the nominal \imag\ values in the plane of the sky are also present in 3D space, even when taking the uncertainties of \gaia\ and \imag\ into account. The validation tests in Appendix~\ref{Appsec:clusteringDesign} further confirm that these 4D groups are significant, being inconsistent with random inclinations and not attributable to the measurement uncertainties or the 3D CTTS distribution alone.

The 4D group characteristics are summarized in Table~\ref{tab:clusters}.
We find two distinct sub-groups within Lupus I, which we call Group I Northwest (NW) and Group I Southeast (SE). The NW group has a higher inclination, with a median and root-mean-squared (RMS) \imag\ of $79^\circ\pm7^\circ$ calculated from the 5000 MC samples. The SE group has a lower median \imag\ of $50^\circ\pm6^\circ$. One geometric Lupus I member, Sz~69, is marked as ungrouped here due to its comparatively low \imag\ of $22^\circ\pm15^\circ$. Group II contains four CTTSs, three of which are on-cloud Lupus II members. This group, as well as Group III near Lupus III, contains the same members as in the 3D group results. 
Finally, Group IV contains all Lupus IV sources except MY~Lup, which has a high \imag\ of $89^\circ$ and small \gaia\ compared to the other Lupus IV members.

It is important to note that the Group II and Group III members do not remain unchanged when a random set of \imag\ is assigned to the CTTSs, as shown in Figure~\ref{fig:clusterTests} in Appendix~\ref{Appsec:clusteringDesign}. Depending on the random seed used to generate the \imag\ values, we observe mergers of Groups I--IV, Groups I--III, Groups I and II, and Groups III and IV. It is only in rare cases that more than three unique groups are identified.
Therefore, these unchanged groups should not be interpreted to mean that the \imag\ correlations are irrelevant, but rather that the \imag\ coherence in the region is strong.
To characterize the physical scales of these coherent groups, we calculate the volume $V$ of a convex hull that surrounds all members and report the median/MAD from the 5000 MC realizations. We additionally report the median pairwise separations between members, $s_{\rm med}$.

\section{Discussion} \label{sec:discussion}

In Section~\ref{subsec:OtherWork}, we discuss complementary evidence for spatial dependencies in Lupus. In Section~\ref{subsec:context}, we discuss the large-scale context of the Lupus region and their influence on our observed \imag\ distributions.

\subsection{Complementary Evidence for Spatial Correlations} \label{subsec:OtherWork}

\cite{Winter2024Lupus} found a spatial dependency of stellar accretion rates (\mdot) in a homogeneously-characterized sample of 57 CTTSs from Lupus \citep{manara23PPVII}, of which 32 are in Lupus III and 25 are distributed throughout Lupus I, II, and IV. 
Using angular separations, these authors find higher \mdot\ in Lupus III stars that are closer to the center of the cluster, as well as a higher similarity of \mdot s between closer pairs of stars in the more distributed sample.
This is evidence of ongoing infall from the ISM, indicative of Bondi-Hoyle-Lyttleton accretion \citep[][]{HoyleLyttleton1941,Bondi1952}.

The \cite{Winter2024Lupus} sample is nearly equivalent to ours, except for their removal of off-cloud sources and inclusion of RU~Lup (which has a high RUWE and was therefore excluded here). 
We test whether our accretion results are compatible by comparing the \mdot s of each CTTS from \citep{manara23PPVII} with those from our accretion flow modeling (from which we obtained \imag; see Table~\ref{tab:flowresults}). We find strong agreement, with a log-space Pearson correlation coefficient of $r$$=$0.9 and a best-fit slope of 1.07. Our accretion modeling results thus reinforce the \mdot-based spatial correlation in \cite{Winter2024Lupus} and reveal that the same CTTSs with \imag\ spatial dependencies also show evidence of ongoing infall from the ISM.

\cite{Aizawa2020} found a tentative clustering of the gas disk inclination measurements (\idisk) around $\cos i_{\rm disk}=0.6$ for the Lupus III \idisk\ measurements in \cite{Yen2018} and \cite{Ansdell2016}.
There are 33 CTTS in our sample whose gas disk inclinations (\idisk) were estimated by \cite{Yen2018}, \cite{Vioque2025}, and \cite{Trapman2025} from ALMA $^{13}$CO \citep{Ansdell2016} and $^{12}$CO \citep[][AGE-PRO]{Zhang2025} observations. Using gas emission can provide strong constraints on disk geometry for small disks, due to its typically larger extent compared to dust emission \citep{Vioque2025}. It is also better suited for comparison with \imag\ than dust emission, as \imag\ is measured from the magnetospheric disruption of the inner gas disk.
While the existing sample of Lupus \idisk\ measurements is insufficient for applying our full clustering analysis, it can provide additional evidence that our \imag\ values are tracing physical system orientations.

\begin{figure}
    \centering
    \includegraphics[width=0.95\linewidth]{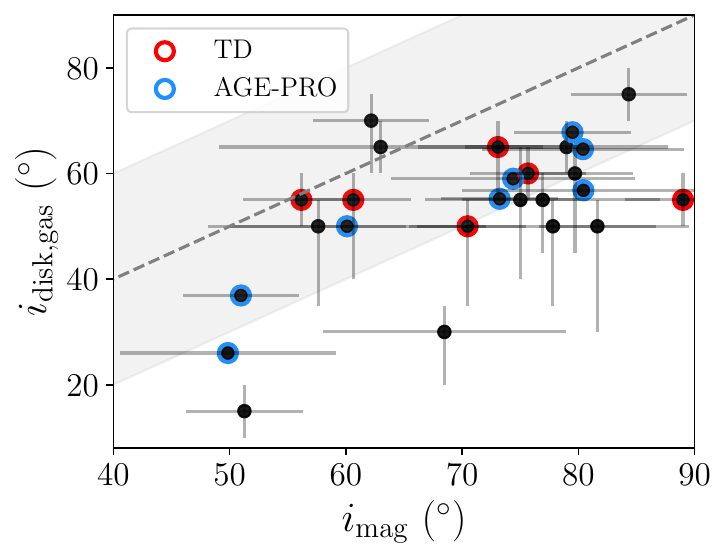}
    \caption{Comparison between \imag\ and \idisk. The dashed line indicates equality, and the shaded region marks $\pm$20\degrees. Red outlines indicate transition disks, and blue outlines indicate measurements from AGE-PRO \citep{Zhang2025,Vioque2025,Trapman2025}.}
    \label{fig:incl_comp}
\end{figure}

When considering only the 26 \idisk\ measurements that are not upper limits, the nominal \imag\ and \idisk\ values are correlated with a Pearson correlation coefficient of $r$$=$0.6 (see Figure~\ref{fig:incl_comp}).
Twenty-two of those 26 \idisk\ measurements are within 20\degrees\ of our \imag\ measurements, within uncertainties, and we find a median absolute difference of 16\degrees\ between \imag\ and \idisk.
Similarly, comparing our \imag\ values with the dust disk inclinations (\idust) measured from ALMA observations at 890~\micron\ \citep{Tazzari2017} and 1.3~mm \citep{GuerraAlvarado2025}, we find that 25 out of 35 are in agreement within 20\degrees, within uncertainties.

Our \imag\ measurements are biased towards being higher than the \idisk\ measurements, and this can in large part be explained by observational and survey biases. $^{12}$CO is optically thick, and therefore strongly dependent on the viewing inclination \citep{Bruderer2013}. For this reason, \citep{Zhang2025} explicitly excluded disks with inclinations above 70\degrees. While $^{13}$CO is optically thinner \citep{Bruderer2013}, the observed flux still depends on the amount of dust in the line of sight, so more inclined disks are less likely to be detected in $^{13}$CO. The \cite{Ansdell2016} sample has 96\% completeness in Lupus, so it does not introduce a strong selection bias. However, they detect $^{13}$CO in only 40\% of sources, and the reprocessing of the same $^{13}$CO data by \cite{Yen2018} results in a detection rate of only 47\%. However, when considering only transition disks (TDs), which are depleted in dust in the inner disk, the \cite{Ansdell2016} $^{13}$CO detection rate is 100\%. 
Therefore, it is likely that the $^{13}$CO detections 
come from disks that are less-inclined, and therefore experience less dust obscuration. This produces a bias that disks that are more inclined than our \imag\ values are less likely to be detected than those that are less inclined than our \imag\ values.

In addition to observational biases, intrinsic deviations between \imag, \idisk, and \idust\ on the order of tens of degrees are expected given observed obliquities between the magnetic axis $\mu_\star$ and the stellar rotation axis $\Omega_\star$ \citep[often by 10--20\degrees\ and even up to $\sim80^\circ$, see][]{Hussain2009,Donati2010,Donati2011a,Donati2011c,McGinnis2020,Nelissen2023a,Nelissen2023b}, between $\Omega_\star$ and the disk \citep{Biddle2025}, and between the inner and outer disk \citep[][]{Pinilla2018,Ansdell2020,Bohn2022,Nelissen2023a,Espaillat2024}. Misalignment between $\Omega_\star$ and exoplanet orbital planes is also common \citep{Bowler2023,Biddle2025}. 
The general agreement between \imag, \idisk, and \idust\ suggests that spatial correlations of one property may correspond to associated spatial correlations of the other properties. 

\cite{Yen2018} and \cite{Trapman2025} also determined disk position angles (PAs) for 32 CTTS in our sample, and these are indicated by the arrows in Figure~\ref{fig:LupusIncl}.
These PAs were spectroscopically-determined, which breaks rotational degeneracy and allows measurement in the 0\degrees--360\degrees\ range.
\cite{Aizawa2020} found that PAs of on-cloud Lupus III CTTSs, shown in greater detail in Figure~\ref{fig:LupusApp}, are non-uniformly distributed at the 2$\sigma$ level.
Using the method of \cite{Aizawa2020}, we find that the \cite{Yen2018} and \cite{Trapman2025} PAs have a mean value of $80^\circ\pm68^\circ$.
This is approximately parallel to the main filament in the plane of the sky in Lupus III (see Figure~\ref{fig:LupusApp}), and nearly perpendicular to the region's magnetic field \citep[which is within $\sim$10\degrees\ of 0\degrees;][]{Benedettini2015,Planck2016a,Aizawa2020}.

These signatures of infall-driven accretion and non-uniformity in \idisk\ and PA provide complementary evidence that the star-cloud connection is of central importance in Lupus, and our analysis confirms that correlations of these larger-scale CTTS characteristics extend down to the innermost sub-au regions from which \imag\ is measured.

\subsection{Clusters in Context} \label{subsec:context}
Figure~\ref{fig:LupusIncl} (left) shows our sample in the context of dust overdensities in the Lupus clouds, and Figure~\ref{fig:LupusIncl} (right and bottom) provide the context of the two primary external influences on the region: the expanding Upper Scorpius (USco) HI shell from the north, which likely originated from a supernova \citep{Blaauw1961,Preibisch2002}, and the Upper-Centaurus-Lupus (UCL) wind bubble from the southwest, which is actively driven by the UCL B stars \citep{Gaczkowski2015,Gaczkowski2017}. These shells are plotted using the geometric information in \cite{Gaczkowski2017}, where the inner and outer radii of the USco shell are 30~pc and 36~pc, respectively. Radii between 15--22~pc are plotted for the UCL bubble for ease of comparison with the Lupus CTTSs. Its total extent reaches a radius of $\sim110$~pc \citep{deGeus1992}, with the initial expansion coming from an HI shell. 
The current UCL bubble is perpetually driven by the stellar emission, so it does not have a fixed radial extent. 

The observed disk PAs may tend towards being parallel to one of the shells. This can be seen with respect to the UCL bubble in the PAs of Sz~65/Sz~66, all Lupus IV members, and the Lupus III systems that are more distant from the main filament (see Figure~\ref{fig:LupusApp}, bottom left). The wide binary of Sz~65 and Sz~66 is particularly interesting, showing PA alignment within 28.2\degrees, \idisk\ alignment within 3.2\degrees, and nearly equivalent \imag\ values. The USco and UCL shell surfaces are nearly perpendicular near Sz~65/Sz~66, and this may have produced a coherent force that resulted in their similar configurations.

In Table~\ref{tab:clusters}, we see that $s_{\rm med}$ decreases monotonically from Group I NW through Group IV. This also corresponds to increasing distance from the USco shell. While the sample is small, this may indicate that proximity to the USco shell can produce larger correlation scales. The USco shell also marks a natural division between Group I NW, Group I SE, and Group II, suggesting a physical origin for these observed \imag\ groups.
In the following sections, we will discuss potential physical origins beginning with the largest subregion, Lupus III.

\subsubsection{Lupus III} \label{sec:LupusIII}

Lupus III is significantly more populated than the other regions, with 32 members included in the associated \texttt{HDBSCAN} group. The Group III CTTSs show a strong tendency towards high \imag\ values, with a median/RMS of $75^\circ\pm12^\circ$. In the plane of the sky, the cause of these high inclinations is not apparent. However, when viewed in three dimensions, it becomes clear that we are looking almost directly down the axis of a filamentary configuration of CTTSs (see Figure~\ref{fig:LupusIncl}, bottom), with the axis within $\sim$3\degrees\ of our line of sight. 

Given that we find \imag\ is typically within 20\degrees\ of \idisk, high inclination magnetospheres should also correspond to moderate-to-high disk inclinations. Indeed, we find that out of the twelve non-upper-limit \idisk\ measurements for the Lupus III sample, eleven have nominal values in the range between 50\degrees--65\degrees\ \citep{Yen2018,Trapman2025}. Similarly, fourteen out of sixteen Lupus III \idust\ measurements have nominal values between 47\degrees--81\degrees\ \citep{Tazzari2017,GuerraAlvarado2025}.
The viewing geometry, combined with the high inclinations, implies that the disk planes are biased against being perpendicular to the major axis of the filamentary structure.
This is consistent with expectations for cloud collapse and angular momentum conservation in filaments, as protostellar cores tend to be elongated along the major axis \citep{Myers1991,LeeMyers1999,Jones2001,Hartmann2002} with rotation axes perpendicular to the major axis \citep{Misugi2023,Misugi2024}. Additionally, protostellar outflows, which are approximately perpendicular to the disk, tend to be orthogonal to their host filaments \citep{Kong2019} and have been found to show alignment on $\sim$1~pc scales \citep{Green2024}.
Our observation of high-\imag\ sources when looking down the filamentary axis suggests that the elongation observed in protostellar cores may propagate through to the Class II stage.

\subsubsection{Lupus I} \label{sec:LupusI}

The Lupus I subregion has likely been compressed between the USco shell and UCL bubble, causing its filamentary structure to be parallel to the USco shell \cite{Gaczkowski2015,Gaczkowski2017} as expected from models of molecular cloud and star formation \citep{Hartmann2001,Heitsch2008}. 
The northern portion has been found to have lower densities compared to the southern region \citep{Gaczkowski2015}, suggesting that material has been swept up and moved southward by the USco shell.
Our analysis of the 3D distribution indicates that the expanding shell has surpassed the northernmost off-cloud source, 2MASSJ15414081-3345188, which is now within the shell's inner cavity. The six Group I NW group members tend to be near the inner wall, whereas the five Group I SE group members tend to be near the outer front. The southernmost source, Sz~77, is considered off-cloud and may be in, or just outside, the shell front.

We see that the six Group~I~NW members lie in two lines along our line of sight. One line consists of the wide binary Sz~65 and Sz~66, and the other consists of the remaining four members. Analysis of the 3D structure shows that they are physically aligned in 3D space as well, in two approximately parallel lines. These systems are also observed to be highly inclined, with a median/RMS \imag\ of $79^\circ\pm7^\circ$. Considering the above discussion of Lupus III, it is reasonable to surmise that these systems also formed along a filament whose major axis lies along our line of sight, causing the initial core elongation to occur along the filament axis.

Conversely, the five Group~I~SE members are much more distributed in 3D space, with the strongest linear alignment being in the plane of the sky. Along with this, this group has the lowest \imag\ in Lupus, with a median/RMS \imag\ of $50^\circ\pm6^\circ$.

\subsubsection{Lupus II} \label{sec:LupusII}

Lupus II is the smallest group, with 4 members, consistent with it being the lowest-mass Lupus subregion \citep[][]{Tachihara1996}. In the plane of the sky, the three on-cloud members fall on a line parallel to the main filament and the local magnetic field (see Figure~\ref{fig:LupusApp}, middle right). Lupus II contains one additional CTTS (RU~Lup) that falls in line with the other three on-cloud members, but its large RUWE value prohibits its inclusion in the 3D analysis. This region's CO velocity gradient is oriented within 5\degrees\ of the UCL expansion direction \citep{MoreiraYun2002}. This suggests a long-lasting effect on gas motions resulting from the initial interaction $\sim$2~Myr ago, and/or the continued influence of the UCL stellar-wind-driven bubble. The stars show only moderate alignment along our line of sight, consistent with their moderate median/RMS \imag\ of $65^\circ\pm7^\circ$, and most likely did not form from a strong filamentary structure.

\subsubsection{Lupus IV} \label{sec:LupusIV}

Star formation in the Lupus IV subregion was likely triggered by the initial UCL HI shell, as its C$^{18}$O velocity gradients are consistent with a propagating shock front oriented within 5\degrees\ of the direction of the bubble's spherical expansion \citep{MoreiraYun2002}.
The four PAs in Lupus IV are all approximately parallel to the UCL expansion front, providing further validation that the initial conditions of their formation might have produced parsec-scale spatial correlations. 
These appear in a line along our line of sight, but their overall 3D distribution is more coplanar than filamentary.
Their relatively high median/RMS \imag\ of $71^\circ\pm8^\circ$ may result from the initial collision between the UCL HI shell and the local ISM, rather than the purely filamentary collapse that seems more likely in Group I NW and Group III.

\subsubsection{Conclusions} \label{sec:conclusions}

The spatial correlations that we have found in Lupus I--IV suggest that the effects of initial conditions may persist on Myr timescales. Future work should search for similar spatial correlations in regions of different ages to determine whether these trends are eventually disrupted by turbulence. Additionally, diffuse regions with minimal influence from massive star feedback should be studied to test whether these spatial correlations can be produced by weaker forces. As discussed in Appendix~\ref{Appsec:otherregions}, the Orion OB1b and Taurus CTTSs in our sample show hints of \imag\ and PA spatial correlations, so our results may apply to regions with very different characteristics than Lupus. 

Our analysis requires large samples of CTTSs with high spectral resolution \halpha\ observations, homogeneously-determined stellar parameters, and reliable \gaia\ measurements. Such data are currently available for $\sim$50 CTTSs in the Chamaeleon I region from \citet[][and references therein]{manara23PPVII}, as well as 37 CTTSs in Taurus from the GHOsT project \citep{Alcala2021,Gangi2022}. We will pursue this analysis in future work. Orion OB1b contains 65 CTTSs \citep{briceno19}, but very few have high-resolution observations and homogeneous stellar parameter determinations. Therefore, additional observations and analysis are required to further study spatial correlations of \imag\ in Orion OB1b.

Our results have the important implication that while \imag\ (and other system orientations) may be uniform when considering all parts of the sky together (see Appendix~\ref{Appsec:InclReliability}), surveys that focus on individual regions may not be able to assume uniformity.
If these inclination trends propagate to the ultimate star-planet orbital planes, there would be significant implications for exoplanet survey completeness corrections, which typically assume uniformly distributed viewing inclinations.

\section{Summary} \label{sec:summary}

For the first time, we have drawn a direct observational connection between parsec-scale molecular cloud characteristics and sub-au scale CTTS properties. Future work should examine whether these trends occur in the absence of high-mass stellar feedback processes, as well as whether they persist over longer timescales (such as in older regions and in exoplanet orbital inclinations). 

To summarize:

\begin{itemize}
    \item We have measured magnetospheric inclinations, \imag, for 61 CTTSs in Lupus, combining our previous results from the HST ULLYSES sample in \cite{Pittman2025} with 36 new \imag\ measurements from this work.
    \item We applied consensus clustering with the hierarchical density-based clustering algorithm \texttt{HDBSCAN} to confirm a spatial correlation of \imag\ on parsec scales.
    \item The observed correlation scales, cluster morphologies, and disk position angles point to a connection between the expanding shells that shape the Lupus clouds, the filamentary structure of the subregions, and the CTTS properties observed on sub-au scales.
\end{itemize}

\begin{acknowledgments}
We wish to recognize the work of Will Fischer, who passed away in 2024. His dedication at STScI ensured the successful implementation of ULLYSES, and his kindness made him a friend to many. 

We are grateful to the anonymous reviewer for an insightful and constructive report that improved the manuscript.
We thank Uma Gorti, Lynne Hillenbrand, Carlo Manara, and Karina Mauc{\'o} for valuable discussion about this work.
This work was supported by HST AR-16129 and benefited from discussions with the ODYSSEUS team \citep[\url{https://sites.bu.edu/odysseus/},][]{espaillat22}. C.P. acknowledges funding from the NSF Graduate Research Fellowship Program under grant No. DGE-1840990.
Based on data obtained from the ESO Science Archive Facility with DOIs: \url{https://doi.org/10.18727/archive/88} (PENELLOPE), 
\url{https://doi.org/10.18727/archive/21} (ESPRESSO), \url{https://doi.org/10.18727/archive/50} (UVES), \url{https://doi.org/10.18727/archive/71} (X-Shooter).

\end{acknowledgments}

\facilities{VLT:Antu, VLT:Kueyen, VLT:Melipal, VLT:Yepun (ESPRESSO, UVES, X-Shooter), CTIO:1.5m}

\software{astropy \citep{astropy:2013,astropy:2018,astropy:2022},  
          hdbscan \citep{HDBSCANMcInnes2017},
          scikit-learn \citep{scikit-learn},
          SciPy \citep{SciPy2020}
          }

\bibliography{bib}{}

\begin{thebibliography}{}
\expandafter\ifx\csname natexlab\endcsname\relax\def\natexlab#1{#1}\fi
\providecommand{\url}[1]{\href{#1}{#1}}
\providecommand{\dodoi}[1]{doi:~\href{http://doi.org/#1}{\nolinkurl{#1}}}
\providecommand{\doeprint}[1]{\href{http://ascl.net/#1}{\nolinkurl{http://ascl.net/#1}}}
\providecommand{\doarXiv}[1]{\href{https://arxiv.org/abs/#1}{\nolinkurl{https://arxiv.org/abs/#1}}}

\bibitem[{{Aizawa} {et~al.}(2020){Aizawa}, {Suto}, {Oya}, {Ikeda}, \& {Nakazato}}]{Aizawa2020}
{Aizawa}, M., {Suto}, Y., {Oya}, Y., {Ikeda}, S., \& {Nakazato}, T. 2020, \apj, 899, 55, \dodoi{10.3847/1538-4357/aba43d}

\bibitem[{{Alcal{\'a}} {et~al.}(2014){Alcal{\'a}}, {Natta}, {Manara}, {Spezzi}, {Stelzer}, {Frasca}, {Biazzo}, {Covino}, {Randich}, {Rigliaco}, {Testi}, {Comer{\'o}n}, {Cupani}, \& {D'Elia}}]{alcala14}
{Alcal{\'a}}, J.~M., {Natta}, A., {Manara}, C.~F., {et~al.} 2014, \aap, 561, A2, \dodoi{10.1051/0004-6361/201322254}

\bibitem[{{Alcal{\'a}} {et~al.}(2017){Alcal{\'a}}, {Manara}, {Natta}, {Frasca}, {Testi}, {Nisini}, {Stelzer}, {Williams}, {Antoniucci}, {Biazzo}, {Covino}, {Esposito}, {Getman}, \& {Rigliaco}}]{alcala17}
{Alcal{\'a}}, J.~M., {Manara}, C.~F., {Natta}, A., {et~al.} 2017, \aap, 600, A20, \dodoi{10.1051/0004-6361/201629929}

\bibitem[{{Alcal{\'a}} {et~al.}(2021){Alcal{\'a}}, {Gangi}, {Biazzo}, {Antoniucci}, {Frasca}, {Giannini}, {Munari}, {Nisini}, {Harutyunyan}, {Manara}, \& {Vitali}}]{Alcala2021}
{Alcal{\'a}}, J.~M., {Gangi}, M., {Biazzo}, K., {et~al.} 2021, \aap, 652, A72, \dodoi{10.1051/0004-6361/202140918}

\bibitem[{{Allen} {et~al.}(2025){Allen}, {Anania}, {Andersen}, {Aru}, {Ballabio}, {Ballering}, {Beccari}, {Bern{\'e}}, {Bik}, {Boyden}, {Coleman}, {D{\'\i}az-Berrios}, {Eatson}, {Frediani}, {Forbrich}, {Gkimisi}, {Goicoechea}, {Gupta}, {Guarcello}, {Haworth}, {Henney}, {Isella}, {Itrich}, {Keyte}, {Kim}, {Kuhn}, {Le Petit}, {Luo}, {Manara}, {Mauco}, {Meshaka}, {Millstone}, {Owen}, {Paine}, {Parker}, {Peake}, {Peatt}, {Pinilla}, {Qiao}, {Ram{\'\i}rez-Tannus}, {Ramsay}, {Reiter}, {Rogers}, {Rosotti}, {Schroetter}, {Sellek}, {Testi}, {van Terwisga}, {Vicente}, {Walsh}, {Winter}, {Wright}, \& {Zeidler}}]{Allen2025}
{Allen}, M., {Anania}, R., {Andersen}, M., {et~al.} 2025, The Open Journal of Astrophysics, 8, 54, \dodoi{10.33232/001c.137538}

\bibitem[{{Alves} {et~al.}(2008){Alves}, {Franco}, \& {Girart}}]{Alves2008}
{Alves}, F.~O., {Franco}, G.~A.~P., \& {Girart}, J.~M. 2008, \aap, 486, L13, \dodoi{10.1051/0004-6361:200810091}

\bibitem[{{Andr{\'e}} {et~al.}(2010){Andr{\'e}}, {Men'shchikov}, {Bontemps}, {K{\"o}nyves}, {Motte}, {Schneider}, {Didelon}, {Minier}, {Saraceno}, {Ward-Thompson}, {di Francesco}, {White}, {Molinari}, {Testi}, {Abergel}, {Griffin}, {Henning}, {Royer}, {Mer{\'\i}n}, {Vavrek}, {Attard}, {Arzoumanian}, {Wilson}, {Ade}, {Aussel}, {Baluteau}, {Benedettini}, {Bernard}, {Blommaert}, {Cambr{\'e}sy}, {Cox}, {di Giorgio}, {Hargrave}, {Hennemann}, {Huang}, {Kirk}, {Krause}, {Launhardt}, {Leeks}, {Le Pennec}, {Li}, {Martin}, {Maury}, {Olofsson}, {Omont}, {Peretto}, {Pezzuto}, {Prusti}, {Roussel}, {Russeil}, {Sauvage}, {Sibthorpe}, {Sicilia-Aguilar}, {Spinoglio}, {Waelkens}, {Woodcraft}, \& {Zavagno}}]{Andre2010}
{Andr{\'e}}, P., {Men'shchikov}, A., {Bontemps}, S., {et~al.} 2010, \aap, 518, L102, \dodoi{10.1051/0004-6361/201014666}

\bibitem[{{Ansdell} {et~al.}(2016){Ansdell}, {Williams}, {van der Marel}, {Carpenter}, {Guidi}, {Hogerheijde}, {Mathews}, {Manara}, {Miotello}, {Natta}, {Oliveira}, {Tazzari}, {Testi}, {van Dishoeck}, \& {van Terwisga}}]{Ansdell2016}
{Ansdell}, M., {Williams}, J.~P., {van der Marel}, N., {et~al.} 2016, \apj, 828, 46, \dodoi{10.3847/0004-637X/828/1/46}

\bibitem[{{Ansdell} {et~al.}(2020){Ansdell}, {Gaidos}, {Hedges}, {Tazzari}, {Kraus}, {Wyatt}, {Kennedy}, {Williams}, {Mann}, {Angelo}, {D{\^u}chene}, {Mamajek}, {Carpenter}, {Esplin}, \& {Rizzuto}}]{Ansdell2020}
{Ansdell}, M., {Gaidos}, E., {Hedges}, C., {et~al.} 2020, \mnras, 492, 572, \dodoi{10.1093/mnras/stz3361}

\bibitem[{{Astropy Collaboration} {et~al.}(2013){Astropy Collaboration}, {Robitaille}, {Tollerud}, {Greenfield}, {Droettboom}, {Bray}, {Aldcroft}, {Davis}, {Ginsburg}, {Price-Whelan}, {Kerzendorf}, {Conley}, {Crighton}, {Barbary}, {Muna}, {Ferguson}, {Grollier}, {Parikh}, {Nair}, {Unther}, {Deil}, {Woillez}, {Conseil}, {Kramer}, {Turner}, {Singer}, {Fox}, {Weaver}, {Zabalza}, {Edwards}, {Azalee Bostroem}, {Burke}, {Casey}, {Crawford}, {Dencheva}, {Ely}, {Jenness}, {Labrie}, {Lim}, {Pierfederici}, {Pontzen}, {Ptak}, {Refsdal}, {Servillat}, \& {Streicher}}]{astropy:2013}
{Astropy Collaboration}, {Robitaille}, T.~P., {Tollerud}, E.~J., {et~al.} 2013, \aap, 558, A33, \dodoi{10.1051/0004-6361/201322068}

\bibitem[{{Astropy Collaboration} {et~al.}(2018){Astropy Collaboration}, {Price-Whelan}, {Sip{\H{o}}cz}, {G{\"u}nther}, {Lim}, {Crawford}, {Conseil}, {Shupe}, {Craig}, {Dencheva}, {Ginsburg}, {Vand erPlas}, {Bradley}, {P{\'e}rez-Su{\'a}rez}, {de Val-Borro}, {Aldcroft}, {Cruz}, {Robitaille}, {Tollerud}, {Ardelean}, {Babej}, {Bach}, {Bachetti}, {Bakanov}, {Bamford}, {Barentsen}, {Barmby}, {Baumbach}, {Berry}, {Biscani}, {Boquien}, {Bostroem}, {Bouma}, {Brammer}, {Bray}, {Breytenbach}, {Buddelmeijer}, {Burke}, {Calderone}, {Cano Rodr{\'\i}guez}, {Cara}, {Cardoso}, {Cheedella}, {Copin}, {Corrales}, {Crichton}, {D'Avella}, {Deil}, {Depagne}, {Dietrich}, {Donath}, {Droettboom}, {Earl}, {Erben}, {Fabbro}, {Ferreira}, {Finethy}, {Fox}, {Garrison}, {Gibbons}, {Goldstein}, {Gommers}, {Greco}, {Greenfield}, {Groener}, {Grollier}, {Hagen}, {Hirst}, {Homeier}, {Horton}, {Hosseinzadeh}, {Hu}, {Hunkeler}, {Ivezi{\'c}}, {Jain}, {Jenness}, {Kanarek}, {Kendrew}, {Kern}, {Kerzendorf}, {Khvalko}, {King}, {Kirkby}, {Kulkarni},
  {Kumar}, {Lee}, {Lenz}, {Littlefair}, {Ma}, {Macleod}, {Mastropietro}, {McCully}, {Montagnac}, {Morris}, {Mueller}, {Mumford}, {Muna}, {Murphy}, {Nelson}, {Nguyen}, {Ninan}, {N{\"o}the}, {Ogaz}, {Oh}, {Parejko}, {Parley}, {Pascual}, {Patil}, {Patil}, {Plunkett}, {Prochaska}, {Rastogi}, {Reddy Janga}, {Sabater}, {Sakurikar}, {Seifert}, {Sherbert}, {Sherwood-Taylor}, {Shih}, {Sick}, {Silbiger}, {Singanamalla}, {Singer}, {Sladen}, {Sooley}, {Sornarajah}, {Streicher}, {Teuben}, {Thomas}, {Tremblay}, {Turner}, {Terr{\'o}n}, {van Kerkwijk}, {de la Vega}, {Watkins}, {Weaver}, {Whitmore}, {Woillez}, {Zabalza}, \& {Astropy Contributors}}]{astropy:2018}
{Astropy Collaboration}, {Price-Whelan}, A.~M., {Sip{\H{o}}cz}, B.~M., {et~al.} 2018, \aj, 156, 123, \dodoi{10.3847/1538-3881/aabc4f}

\bibitem[{{Astropy Collaboration} {et~al.}(2022){Astropy Collaboration}, {Price-Whelan}, {Lim}, {Earl}, {Starkman}, {Bradley}, {Shupe}, {Patil}, {Corrales}, {Brasseur}, {N{"o}the}, {Donath}, {Tollerud}, {Morris}, {Ginsburg}, {Vaher}, {Weaver}, {Tocknell}, {Jamieson}, {van Kerkwijk}, {Robitaille}, {Merry}, {Bachetti}, {G{"u}nther}, {Aldcroft}, {Alvarado-Montes}, {Archibald}, {B{'o}di}, {Bapat}, {Barentsen}, {Baz{'a}n}, {Biswas}, {Boquien}, {Burke}, {Cara}, {Cara}, {Conroy}, {Conseil}, {Craig}, {Cross}, {Cruz}, {D'Eugenio}, {Dencheva}, {Devillepoix}, {Dietrich}, {Eigenbrot}, {Erben}, {Ferreira}, {Foreman-Mackey}, {Fox}, {Freij}, {Garg}, {Geda}, {Glattly}, {Gondhalekar}, {Gordon}, {Grant}, {Greenfield}, {Groener}, {Guest}, {Gurovich}, {Handberg}, {Hart}, {Hatfield-Dodds}, {Homeier}, {Hosseinzadeh}, {Jenness}, {Jones}, {Joseph}, {Kalmbach}, {Karamehmetoglu}, {Ka{l}uszy{'n}ski}, {Kelley}, {Kern}, {Kerzendorf}, {Koch}, {Kulumani}, {Lee}, {Ly}, {Ma}, {MacBride}, {Maljaars}, {Muna}, {Murphy}, {Norman}, {O'Steen},
  {Oman}, {Pacifici}, {Pascual}, {Pascual-Granado}, {Patil}, {Perren}, {Pickering}, {Rastogi}, {Roulston}, {Ryan}, {Rykoff}, {Sabater}, {Sakurikar}, {Salgado}, {Sanghi}, {Saunders}, {Savchenko}, {Schwardt}, {Seifert-Eckert}, {Shih}, {Jain}, {Shukla}, {Sick}, {Simpson}, {Singanamalla}, {Singer}, {Singhal}, {Sinha}, {Sip{H{o}}cz}, {Spitler}, {Stansby}, {Streicher}, {{\v{S}}umak}, {Swinbank}, {Taranu}, {Tewary}, {Tremblay}, {Val-Borro}, {Van Kooten}, {Vasovi{'c}}, {Verma}, {de Miranda Cardoso}, {Williams}, {Wilson}, {Winkel}, {Wood-Vasey}, {Xue}, {Yoachim}, {Zhang}, {Zonca}, \& {Astropy Project Contributors}}]{astropy:2022}
{Astropy Collaboration}, {Price-Whelan}, A.~M., {Lim}, P.~L., {et~al.} 2022, \apj, 935, 167, \dodoi{10.3847/1538-4357/ac7c74}

\bibitem[{{Benedettini} {et~al.}(2012){Benedettini}, {Pezzuto}, {Burton}, {Viti}, {Molinari}, {Caselli}, \& {Testi}}]{Benedettini2012}
{Benedettini}, M., {Pezzuto}, S., {Burton}, M.~G., {et~al.} 2012, \mnras, 419, 238, \dodoi{10.1111/j.1365-2966.2011.19687.x}

\bibitem[{{Benedettini} {et~al.}(2015){Benedettini}, {Schisano}, {Pezzuto}, {Elia}, {Andr{\'e}}, {K{\"o}nyves}, {Schneider}, {Tremblin}, {Arzoumanian}, {di Giorgio}, {Di Francesco}, {Hill}, {Molinari}, {Motte}, {Nguyen-Luong}, {Palmeirim}, {Rivera-Ingraham}, {Roy}, {Rygl}, {Spinoglio}, {Ward-Thompson}, \& {White}}]{Benedettini2015}
{Benedettini}, M., {Schisano}, E., {Pezzuto}, S., {et~al.} 2015, \mnras, 453, 2036, \dodoi{10.1093/mnras/stv1750}

\bibitem[{{Benedettini} {et~al.}(2018){Benedettini}, {Pezzuto}, {Schisano}, {Andr{\'e}}, {K{\"o}nyves}, {Men'shchikov}, {Ladjelate}, {Di Francesco}, {Elia}, {Arzoumanian}, {Louvet}, {Palmeirim}, {Rygl}, {Schneider}, {Spinoglio}, \& {Ward-Thompson}}]{Benedettini2018}
{Benedettini}, M., {Pezzuto}, S., {Schisano}, E., {et~al.} 2018, \aap, 619, A52, \dodoi{10.1051/0004-6361/201833364}

\bibitem[{{Biddle} {et~al.}(2025){Biddle}, {Bowler}, {Morgan}, {Tran}, \& {Wu}}]{Biddle2025}
{Biddle}, L.~I., {Bowler}, B.~P., {Morgan}, M., {Tran}, Q.~H., \& {Wu}, Y.-L. 2025, \nat, 644, 356, \dodoi{10.1038/s41586-025-09324-0}

\bibitem[{{Blaauw}(1961)}]{Blaauw1961}
{Blaauw}, A. 1961, \bain, 15, 265

\bibitem[{{Bohn} {et~al.}(2022){Bohn}, {Benisty}, {Perraut}, {van der Marel}, {W{\"o}lfer}, {van Dishoeck}, {Facchini}, {Manara}, {Teague}, {Francis}, {Berger}, {Garcia-Lopez}, {Ginski}, {Henning}, {Kenworthy}, {Kraus}, {M{\'e}nard}, {M{\'e}rand}, \& {P{\'e}rez}}]{Bohn2022}
{Bohn}, A.~J., {Benisty}, M., {Perraut}, K., {et~al.} 2022, \aap, 658, A183, \dodoi{10.1051/0004-6361/202142070}

\bibitem[{{Bondi}(1952)}]{Bondi1952}
{Bondi}, H. 1952, \mnras, 112, 195, \dodoi{10.1093/mnras/112.2.195}

\bibitem[{Bowler {et~al.}(2023)Bowler, Tran, Zhang, Morgan, Ashok, Blunt, Bryan, Evans, Franson, Huber, Nagpal, Wu, \& Zhou}]{Bowler2023}
Bowler, B.~P., Tran, Q.~H., Zhang, Z., {et~al.} 2023, The Astronomical Journal, 165, 164, \dodoi{10.3847/1538-3881/acbd34}

\bibitem[{{Brice{\~n}o} {et~al.}(2019){Brice{\~n}o}, {Calvet}, {Hern{\'a}ndez}, {Vivas}, {Mateu}, {Downes}, {Loerincs}, {P{\'e}rez-Blanco}, {Berlind}, {Espaillat}, {Allen}, {Hartmann}, {Mateo}, \& {Bailey}}]{briceno19}
{Brice{\~n}o}, C., {Calvet}, N., {Hern{\'a}ndez}, J., {et~al.} 2019, \aj, 157, 85, \dodoi{10.3847/1538-3881/aaf79b}

\bibitem[{{Bruderer}(2013)}]{Bruderer2013}
{Bruderer}, S. 2013, \aap, 559, A46, \dodoi{10.1051/0004-6361/201321171}

\bibitem[{{Calvet} \& {Gullbring}(1998)}]{cg98}
{Calvet}, N., \& {Gullbring}, E. 1998, \apj, 509, 802, \dodoi{10.1086/306527}

\bibitem[{{Cambr{\'e}sy}(1999)}]{Cambresy1999}
{Cambr{\'e}sy}, L. 1999, \aap, 345, 965, \dodoi{10.48550/arXiv.astro-ph/9903149}

\bibitem[{Campello {et~al.}(2013)Campello, Moulavi, \& Sander}]{HDBSCANCampello2013}
Campello, R. J. G.~B., Moulavi, D., \& Sander, J. 2013, in Advances in Knowledge Discovery and Data Mining, ed. J.~Pei, V.~S. Tseng, L.~Cao, H.~Motoda, \& G.~Xu (Berlin, Heidelberg: Springer Berlin Heidelberg), 160--172

\bibitem[{{Comer{\'o}n}(2008)}]{Comeron2008}
{Comer{\'o}n}, F. 2008, in Handbook of Star Forming Regions, Volume II, ed. B.~{Reipurth}, Vol.~5, 295

\bibitem[{{Cuello} {et~al.}(2023){Cuello}, {M{\'e}nard}, \& {Price}}]{Cuello2023}
{Cuello}, N., {M{\'e}nard}, F., \& {Price}, D.~J. 2023, European Physical Journal Plus, 138, 11, \dodoi{10.1140/epjp/s13360-022-03602-w}

\bibitem[{{Dawson}(2013)}]{Dawson2013}
{Dawson}, J.~R. 2013, \pasa, 30, e025, \dodoi{10.1017/pas.2013.002}

\bibitem[{{Dawson} {et~al.}(2011){Dawson}, {McClure-Griffiths}, {Kawamura}, {Mizuno}, {Onishi}, {Mizuno}, \& {Fukui}}]{Dawson2011}
{Dawson}, J.~R., {McClure-Griffiths}, N.~M., {Kawamura}, A., {et~al.} 2011, \apj, 728, 127, \dodoi{10.1088/0004-637X/728/2/127}

\bibitem[{{Dawson} {et~al.}(2015){Dawson}, {Ntormousi}, {Fukui}, {Hayakawa}, \& {Fierlinger}}]{Dawson2015}
{Dawson}, J.~R., {Ntormousi}, E., {Fukui}, Y., {Hayakawa}, T., \& {Fierlinger}, K. 2015, \apj, 799, 64, \dodoi{10.1088/0004-637X/799/1/64}

\bibitem[{{de Geus}(1992)}]{deGeus1992}
{de Geus}, E.~J. 1992, \aap, 262, 258

\bibitem[{{Dhabal} {et~al.}(2018){Dhabal}, {Mundy}, {Rizzo}, {Storm}, \& {Teuben}}]{Dhabal2018}
{Dhabal}, A., {Mundy}, L.~G., {Rizzo}, M.~J., {Storm}, S., \& {Teuben}, P. 2018, \apj, 853, 169, \dodoi{10.3847/1538-4357/aaa76b}

\bibitem[{{Donati} {et~al.}(2010){Donati}, {Skelly}, {Bouvier}, {Gregory}, {Grankin}, {Jardine}, {Hussain}, {M{\'e}nard}, {Dougados}, {Unruh}, {Mohanty}, {Auri{\`e}re}, {Morin}, {Far{\`e}s}, \& {MAPP Collaboration}}]{Donati2010}
{Donati}, J.~F., {Skelly}, M.~B., {Bouvier}, J., {et~al.} 2010, \mnras, 409, 1347, \dodoi{10.1111/j.1365-2966.2010.17409.x}

\bibitem[{{Donati} {et~al.}(2011{\natexlab{a}}){Donati}, {Bouvier}, {Walter}, {Gregory}, {Skelly}, {Hussain}, {Flaccomio}, {Argiroffi}, {Grankin}, {Jardine}, {M{\'e}nard}, {Dougados}, \& {Romanova}}]{Donati2011a}
{Donati}, J.~F., {Bouvier}, J., {Walter}, F.~M., {et~al.} 2011{\natexlab{a}}, \mnras, 412, 2454, \dodoi{10.1111/j.1365-2966.2010.18069.x}

\bibitem[{{Donati} {et~al.}(2011{\natexlab{b}}){Donati}, {Gregory}, {Montmerle}, {Maggio}, {Argiroffi}, {Sacco}, {Hussain}, {Kastner}, {Alencar}, {Audard}, {Bouvier}, {Damiani}, {G{\"u}del}, {Huenemoerder}, \& {Wade}}]{Donati2011c}
{Donati}, J.~F., {Gregory}, S.~G., {Montmerle}, T., {et~al.} 2011{\natexlab{b}}, \mnras, 417, 1747, \dodoi{10.1111/j.1365-2966.2011.19366.x}

\bibitem[{Espaillat {et~al.}(2022)Espaillat, Herczeg, Thanathibodee, Pittman, Calvet, Arulanantham, France, Serna, Hernandez, Kospal, Walter, Frasca, Fischer, Johns-Krull, Schneider, Robinson, Edwards, Abraham, Fang, Erkal, Manara, Alcala, Alecian, Alexander, Alonso-Santiago, Antoniucci, Ardila, Banzatti, Benisty, Bergin, Biazzo, Briceno, Campbell-White, Cleeves, Coffey, Eisloffel, Facchini, Fedele, Fiorellino, Froebrich, Gangi, Giannini, Grankin, Gunther, Guo, Hartmann, Hillenbrand, Hinton, Kastner, Koen, Mauco, Mendigutia, Nisini, Panwar, Principe, Robberto, Sicilia-Aguilar, Valenti, Wendeborn, Williams, Xu, \& Yadav}]{espaillat22}
Espaillat, C.~C., Herczeg, G.~J., Thanathibodee, T., {et~al.} 2022, The ODYSSEUS Survey. Motivation and First Results: Accretion, Ejection, and Disk Irradiation of CVSO 109.
\newblock \doarXiv{2201.06502}

\bibitem[{{Espaillat} {et~al.}(2024){Espaillat}, {Thanathibodee}, {Zhu}, {Rabago}, {Wendeborn}, {Calvet}, {Zamudio-Ruvalcaba}, {Volz}, {Pittman}, {McClure}, {Babb}, {Franco-Hern{\'a}ndez}, {Mac{\'\i}as}, {Reynolds}, \& {Yan}}]{Espaillat2024}
{Espaillat}, C.~C., {Thanathibodee}, T., {Zhu}, Z., {et~al.} 2024, \apjl, 973, L16, \dodoi{10.3847/2041-8213/ad76a5}

\bibitem[{{Franco} \& {Alves}(2015)}]{FrancoAlves2015}
{Franco}, G.~A.~P., \& {Alves}, F.~O. 2015, \apj, 807, 5, \dodoi{10.1088/0004-637X/807/1/5}

\bibitem[{{Frasca} {et~al.}(2017){Frasca}, {Biazzo}, {Alcal{\'a}}, {Manara}, {Stelzer}, {Covino}, \& {Antoniucci}}]{alcala17b}
{Frasca}, A., {Biazzo}, K., {Alcal{\'a}}, J.~M., {et~al.} 2017, \aap, 602, A33, \dodoi{10.1051/0004-6361/201630108}

\bibitem[{{Gaczkowski} {et~al.}(2015){Gaczkowski}, {Preibisch}, {Stanke}, {Krause}, {Burkert}, {Diehl}, {Fierlinger}, {Kroell}, {Ngoumou}, \& {Roccatagliata}}]{Gaczkowski2015}
{Gaczkowski}, B., {Preibisch}, T., {Stanke}, T., {et~al.} 2015, \aap, 584, A36, \dodoi{10.1051/0004-6361/201526527}

\bibitem[{{Gaczkowski} {et~al.}(2017){Gaczkowski}, {Roccatagliata}, {Flaischlen}, {Kr{\"o}ll}, {Krause}, {Burkert}, {Diehl}, {Fierlinger}, {Ngoumou}, \& {Preibisch}}]{Gaczkowski2017}
{Gaczkowski}, B., {Roccatagliata}, V., {Flaischlen}, S., {et~al.} 2017, \aap, 608, A102, \dodoi{10.1051/0004-6361/201628508}

\bibitem[{{Gaia Collaboration} {et~al.}(2023){Gaia Collaboration}, {Vallenari}, {Brown}, {Prusti}, {de Bruijne}, {Arenou}, {Babusiaux}, {Biermann}, {Creevey}, {Ducourant}, {Evans}, {Eyer}, {Guerra}, {Hutton}, {Jordi}, {Klioner}, {Lammers}, {Lindegren}, {Luri}, {Mignard}, {Panem}, {Pourbaix}, {Randich}, {Sartoretti}, {Soubiran}, {Tanga}, {Walton}, {Bailer-Jones}, {Bastian}, {Drimmel}, {Jansen}, {Katz}, {Lattanzi}, {van Leeuwen}, {Bakker}, {Cacciari}, {Casta{\~n}eda}, {De Angeli}, {Fabricius}, {Fouesneau}, {Fr{\'e}mat}, {Galluccio}, {Guerrier}, {Heiter}, {Masana}, {Messineo}, {Mowlavi}, {Nicolas}, {Nienartowicz}, {Pailler}, {Panuzzo}, {Riclet}, {Roux}, {Seabroke}, {Sordo}, {Th{\'e}venin}, {Gracia-Abril}, {Portell}, {Teyssier}, {Altmann}, {Andrae}, {Audard}, {Bellas-Velidis}, {Benson}, {Berthier}, {Blomme}, {Burgess}, {Busonero}, {Busso}, {C{\'a}novas}, {Carry}, {Cellino}, {Cheek}, {Clementini}, {Damerdji}, {Davidson}, {de Teodoro}, {Nu{\~n}ez Campos}, {Delchambre}, {Dell'Oro}, {Esquej},
  {Fern{\'a}ndez-Hern{\'a}ndez}, {Fraile}, {Garabato}, {Garc{\'\i}a-Lario}, {Gosset}, {Haigron}, {Halbwachs}, {Hambly}, {Harrison}, {Hern{\'a}ndez}, {Hestroffer}, {Hodgkin}, {Holl}, {Jan{\ss}en}, {Jevardat de Fombelle}, {Jordan}, {Krone-Martins}, {Lanzafame}, {L{\"o}ffler}, {Marchal}, {Marrese}, {Moitinho}, {Muinonen}, {Osborne}, {Pancino}, {Pauwels}, {Recio-Blanco}, {Reyl{\'e}}, {Riello}, {Rimoldini}, {Roegiers}, {Rybizki}, {Sarro}, {Siopis}, {Smith}, {Sozzetti}, {Utrilla}, {van Leeuwen}, {Abbas}, {{\'A}brah{\'a}m}, {Abreu Aramburu}, {Aerts}, {Aguado}, {Ajaj}, {Aldea-Montero}, {Altavilla}, {{\'A}lvarez}, {Alves}, {Anders}, {Anderson}, {Anglada Varela}, {Antoja}, {Baines}, {Baker}, {Balaguer-N{\'u}{\~n}ez}, {Balbinot}, {Balog}, {Barache}, {Barbato}, {Barros}, {Barstow}, {Bartolom{\'e}}, {Bassilana}, {Bauchet}, {Becciani}, {Bellazzini}, {Berihuete}, {Bernet}, {Bertone}, {Bianchi}, {Binnenfeld}, {Blanco-Cuaresma}, {Blazere}, {Boch}, {Bombrun}, {Bossini}, {Bouquillon}, {Bragaglia}, {Bramante}, {Breedt},
  {Bressan}, {Brouillet}, {Brugaletta}, {Bucciarelli}, {Burlacu}, {Butkevich}, {Buzzi}, {Caffau}, {Cancelliere}, {Cantat-Gaudin}, {Carballo}, {Carlucci}, {Carnerero}, {Carrasco}, {Casamiquela}, {Castellani}, {Castro-Ginard}, {Chaoul}, {Charlot}, {Chemin}, {Chiaramida}, {Chiavassa}, {Chornay}, {Comoretto}, {Contursi}, {Cooper}, {Cornez}, {Cowell}, {Crifo}, {Cropper}, {Crosta}, {Crowley}, {Dafonte}, {Dapergolas}, {David}, {David}, {de Laverny}, {De Luise}, {De March}, {De Ridder}, {de Souza}, {de Torres}, {del Peloso}, {del Pozo}, {Delbo}, {Delgado}, {Delisle}, {Demouchy}, {Dharmawardena}, {Di Matteo}, {Diakite}, {Diener}, {Distefano}, {Dolding}, {Edvardsson}, {Enke}, {Fabre}, {Fabrizio}, {Faigler}, {Fedorets}, {Fernique}, {Fienga}, {Figueras}, {Fournier}, {Fouron}, {Fragkoudi}, {Gai}, {Garcia-Gutierrez}, {Garcia-Reinaldos}, {Garc{\'\i}a-Torres}, {Garofalo}, {Gavel}, {Gavras}, {Gerlach}, {Geyer}, {Giacobbe}, {Gilmore}, {Girona}, {Giuffrida}, {Gomel}, {Gomez}, {Gonz{\'a}lez-N{\'u}{\~n}ez},
  {Gonz{\'a}lez-Santamar{\'\i}a}, {Gonz{\'a}lez-Vidal}, {Granvik}, {Guillout}, {Guiraud}, {Guti{\'e}rrez-S{\'a}nchez}, {Guy}, {Hatzidimitriou}, {Hauser}, {Haywood}, {Helmer}, {Helmi}, {Sarmiento}, {Hidalgo}, {Hilger}, {H{\l}adczuk}, {Hobbs}, {Holland}, {Huckle}, {Jardine}, {Jasniewicz}, {Jean-Antoine Piccolo}, {Jim{\'e}nez-Arranz}, {Jorissen}, {Juaristi Campillo}, {Julbe}, {Karbevska}, {Kervella}, {Khanna}, {Kontizas}, {Kordopatis}, {Korn}, {K{\'o}sp{\'a}l}, {Kostrzewa-Rutkowska}, {Kruszy{\'n}ska}, {Kun}, {Laizeau}, {Lambert}, {Lanza}, {Lasne}, {Le Campion}, {Lebreton}, {Lebzelter}, {Leccia}, {Leclerc}, {Lecoeur-Taibi}, {Liao}, {Licata}, {Lindstr{\o}m}, {Lister}, {Livanou}, {Lobel}, {Lorca}, {Loup}, {Madrero Pardo}, {Magdaleno Romeo}, {Managau}, {Mann}, {Manteiga}, {Marchant}, {Marconi}, {Marcos}, {Marcos Santos}, {Mar{\'\i}n Pina}, {Marinoni}, {Marocco}, {Marshall}, {Martin Polo}, {Mart{\'\i}n-Fleitas}, {Marton}, {Mary}, {Masip}, {Massari}, {Mastrobuono-Battisti}, {Mazeh}, {McMillan}, {Messina}, {Michalik},
  {Millar}, {Mints}, {Molina}, {Molinaro}, {Moln{\'a}r}, {Monari}, {Mongui{\'o}}, {Montegriffo}, {Montero}, {Mor}, {Mora}, {Morbidelli}, {Morel}, {Morris}, {Muraveva}, {Murphy}, {Musella}, {Nagy}, {Noval}, {Oca{\~n}a}, {Ogden}, {Ordenovic}, {Osinde}, {Pagani}, {Pagano}, {Palaversa}, {Palicio}, {Pallas-Quintela}, {Panahi}, {Payne-Wardenaar}, {Pe{\~n}alosa Esteller}, {Penttil{\"a}}, {Pichon}, {Piersimoni}, {Pineau}, {Plachy}, {Plum}, {Poggio}, {Pr{\v{s}}a}, {Pulone}, {Racero}, {Ragaini}, {Rainer}, {Raiteri}, {Rambaux}, {Ramos}, {Ramos-Lerate}, {Re Fiorentin}, {Regibo}, {Richards}, {Rios Diaz}, {Ripepi}, {Riva}, {Rix}, {Rixon}, {Robichon}, {Robin}, {Robin}, {Roelens}, {Rogues}, {Rohrbasser}, {Romero-G{\'o}mez}, {Rowell}, {Royer}, {Ruz Mieres}, {Rybicki}, {Sadowski}, {S{\'a}ez N{\'u}{\~n}ez}, {Sagrist{\`a} Sell{\'e}s}, {Sahlmann}, {Salguero}, {Samaras}, {Sanchez Gimenez}, {Sanna}, {Santove{\~n}a}, {Sarasso}, {Schultheis}, {Sciacca}, {Segol}, {Segovia}, {S{\'e}gransan}, {Semeux}, {Shahaf}, {Siddiqui}, {Siebert},
  {Siltala}, {Silvelo}, {Slezak}, {Slezak}, {Smart}, {Snaith}, {Solano}, {Solitro}, {Souami}, {Souchay}, {Spagna}, {Spina}, {Spoto}, {Steele}, {Steidelm{\"u}ller}, {Stephenson}, {S{\"u}veges}, {Surdej}, {Szabados}, {Szegedi-Elek}, {Taris}, {Taylor}, {Teixeira}, {Tolomei}, {Tonello}, {Torra}, {Torra}, {Torralba Elipe}, {Trabucchi}, {Tsounis}, {Turon}, {Ulla}, {Unger}, {Vaillant}, {van Dillen}, {van Reeven}, {Vanel}, {Vecchiato}, {Viala}, {Vicente}, {Voutsinas}, {Weiler}, {Wevers}, {Wyrzykowski}, {Yoldas}, {Yvard}, {Zhao}, {Zorec}, {Zucker}, \& {Zwitter}}]{GaiaDR3}
{Gaia Collaboration}, {Vallenari}, A., {Brown}, A.~G.~A., {et~al.} 2023, \aap, 674, A1, \dodoi{10.1051/0004-6361/202243940}

\bibitem[{{Galli} {et~al.}(2020){Galli}, {Bouy}, {Olivares}, {Miret-Roig}, {Vieira}, {Sarro}, {Barrado}, {Berihuete}, {Bertout}, {Bertin}, \& {Cuillandre}}]{Galli2020}
{Galli}, P.~A.~B., {Bouy}, H., {Olivares}, J., {et~al.} 2020, \aap, 643, A148, \dodoi{10.1051/0004-6361/202038717}

\bibitem[{{Gangi} {et~al.}(2022){Gangi}, {Antoniucci}, {Biazzo}, {Frasca}, {Nisini}, {Alcal{\'a}}, {Giannini}, {Manara}, {Giunta}, {Harutyunyan}, {Munari}, \& {Vitali}}]{Gangi2022}
{Gangi}, M., {Antoniucci}, S., {Biazzo}, K., {et~al.} 2022, \aap, 667, A124, \dodoi{10.1051/0004-6361/202244042}

\bibitem[{{Green} {et~al.}(2024){Green}, {Pontoppidan}, {Reiter}, {Watson}, {Shenoy}, {Manoj}, \& {Narang}}]{Green2024}
{Green}, J.~D., {Pontoppidan}, K.~M., {Reiter}, M., {et~al.} 2024, \apj, 972, 5, \dodoi{10.3847/1538-4357/ad5a02}

\bibitem[{{Guerra-Alvarado} {et~al.}(2025){Guerra-Alvarado}, {van der Marel}, {Williams}, {Pinilla}, {Mulders}, {Lambrechts}, \& {Sanchez}}]{GuerraAlvarado2025}
{Guerra-Alvarado}, O.~M., {van der Marel}, N., {Williams}, J.~P., {et~al.} 2025, \aap, 696, A232, \dodoi{10.1051/0004-6361/202453338}

\bibitem[{{Hartmann}(2002)}]{Hartmann2002}
{Hartmann}, L. 2002, \apj, 578, 914, \dodoi{10.1086/342657}

\bibitem[{{Hartmann} {et~al.}(2001){Hartmann}, {Ballesteros-Paredes}, \& {Bergin}}]{Hartmann2001}
{Hartmann}, L., {Ballesteros-Paredes}, J., \& {Bergin}, E.~A. 2001, \apj, 562, 852, \dodoi{10.1086/323863}

\bibitem[{{Hartmann} {et~al.}(1994){Hartmann}, {Hewett}, \& {Calvet}}]{hartmann94}
{Hartmann}, L., {Hewett}, R., \& {Calvet}, N. 1994, \apj, 426, 669, \dodoi{10.1086/174104}

\bibitem[{{Heitsch} \& {Hartmann}(2008)}]{Heitsch2008}
{Heitsch}, F., \& {Hartmann}, L. 2008, \apj, 689, 290, \dodoi{10.1086/592491}

\bibitem[{{Heyer} {et~al.}(1987){Heyer}, {Vrba}, {Snell}, {Schloerb}, {Strom}, {Goldsmith}, \& {Strom}}]{Heyer1987}
{Heyer}, M.~H., {Vrba}, F.~J., {Snell}, R.~L., {et~al.} 1987, \apj, 321, 855, \dodoi{10.1086/165678}

\bibitem[{{Hoyle} \& {Lyttleton}(1941)}]{HoyleLyttleton1941}
{Hoyle}, F., \& {Lyttleton}, R.~A. 1941, \mnras, 101, 227, \dodoi{10.1093/mnras/101.4.227}

\bibitem[{{Hunt} \& {Reffert}(2021)}]{HuntReffert2021}
{Hunt}, E.~L., \& {Reffert}, S. 2021, \aap, 646, A104, \dodoi{10.1051/0004-6361/202039341}

\bibitem[{Hussain {et~al.}(2009)Hussain, Collier~Cameron, Jardine, Dunstone, Velez, Stempels, Donati, Semel, Aulanier, Harries, Bouvier, Dougados, Ferreira, Carter, \& Lawson}]{Hussain2009}
Hussain, G. A.~J., Collier~Cameron, A., Jardine, M.~M., {et~al.} 2009, Monthly Notices of the Royal Astronomical Society, 398, 189, \dodoi{10.1111/j.1365-2966.2009.14881.x}

\bibitem[{{Jones} {et~al.}(2001){Jones}, {Basu}, \& {Dubinski}}]{Jones2001}
{Jones}, C.~E., {Basu}, S., \& {Dubinski}, J. 2001, \apj, 551, 387, \dodoi{10.1086/320093}

\bibitem[{{Kong} {et~al.}(2019){Kong}, {Arce}, {Maureira}, {Caselli}, {Tan}, \& {Fontani}}]{Kong2019}
{Kong}, S., {Arce}, H.~G., {Maureira}, M.~J., {et~al.} 2019, \apj, 874, 104, \dodoi{10.3847/1538-4357/ab07b9}

\bibitem[{{Kounkel} \& {Covey}(2019)}]{KounkelCovey2019}
{Kounkel}, M., \& {Covey}, K. 2019, \aj, 158, 122, \dodoi{10.3847/1538-3881/ab339a}

\bibitem[{{Krause} {et~al.}(2013){Krause}, {Fierlinger}, {Diehl}, {Burkert}, {Voss}, \& {Ziegler}}]{Krause2013}
{Krause}, M., {Fierlinger}, K., {Diehl}, R., {et~al.} 2013, \aap, 550, A49, \dodoi{10.1051/0004-6361/201220060}

\bibitem[{{Kurtovic} {et~al.}(2022){Kurtovic}, {Pinilla}, {Penzlin}, {Benisty}, {P{\'e}rez}, {Ginski}, {Isella}, {Kley}, {Menard}, {P{\'e}rez}, \& {Bayo}}]{Kurtovic2022}
{Kurtovic}, N.~T., {Pinilla}, P., {Penzlin}, A. B.~T., {et~al.} 2022, \aap, 664, A151, \dodoi{10.1051/0004-6361/202243505}

\bibitem[{{Lee} \& {Myers}(1999)}]{LeeMyers1999}
{Lee}, C.~W., \& {Myers}, P.~C. 1999, \apjs, 123, 233, \dodoi{10.1086/313234}

\bibitem[{{Luhman}(2020)}]{Luhman2020}
{Luhman}, K.~L. 2020, \aj, 160, 186, \dodoi{10.3847/1538-3881/abb12f}

\bibitem[{{Manara} {et~al.}(2023){Manara}, {Ansdell}, {Rosotti}, {Hughes}, {Armitage}, {Lodato}, \& {Williams}}]{manara23PPVII}
{Manara}, C.~F., {Ansdell}, M., {Rosotti}, G.~P., {et~al.} 2023, in Astronomical Society of the Pacific Conference Series, Vol. 534, Protostars and Planets VII, ed. S.~{Inutsuka}, Y.~{Aikawa}, T.~{Muto}, K.~{Tomida}, \& M.~{Tamura}, 539, \dodoi{10.48550/arXiv.2203.09930}

\bibitem[{{Mauc{\'o}} {et~al.}(2025){Mauc{\'o}}, {Manara}, {Bayo}, {Hern{\'a}ndez}, {Campbell-White}, {Calvet}, {Ballabio}, {Aru}, {Alcal{\'a}}, {Ansdell}, {Brice{\~n}o}, {Facchini}, {Haworth}, {McClure}, \& {Williams}}]{Mauco2025}
{Mauc{\'o}}, K., {Manara}, C.~F., {Bayo}, A., {et~al.} 2025, \aap, 693, A87, \dodoi{10.1051/0004-6361/202452386}

\bibitem[{McGinnis {et~al.}(2020)McGinnis, Bouvier, \& Gallet}]{McGinnis2020}
McGinnis, P., Bouvier, J., \& Gallet, F. 2020, Monthly Notices of the Royal Astronomical Society, 497, 2142, \dodoi{10.1093/mnras/staa2041}

\bibitem[{{McInnes} {et~al.}(2017){McInnes}, {Healy}, \& {Astels}}]{HDBSCANMcInnes2017}
{McInnes}, L., {Healy}, J., \& {Astels}, S. 2017, The Journal of Open Source Software, 2, 205, \dodoi{10.21105/joss.00205}

\bibitem[{{Mendigut{\'\i}a} {et~al.}(2018){Mendigut{\'\i}a}, {Lada}, \& {Oudmaijer}}]{Mendigutia2018}
{Mendigut{\'\i}a}, I., {Lada}, C.~J., \& {Oudmaijer}, R.~D. 2018, \aap, 618, A119, \dodoi{10.1051/0004-6361/201833166}

\bibitem[{{Misugi} {et~al.}(2023){Misugi}, {Inutsuka}, \& {Arzoumanian}}]{Misugi2023}
{Misugi}, Y., {Inutsuka}, S.-i., \& {Arzoumanian}, D. 2023, \apj, 943, 76, \dodoi{10.3847/1538-4357/aca88d}

\bibitem[{{Misugi} {et~al.}(2024){Misugi}, {Inutsuka}, {Arzoumanian}, \& {Tsukamoto}}]{Misugi2024}
{Misugi}, Y., {Inutsuka}, S.-i., {Arzoumanian}, D., \& {Tsukamoto}, Y. 2024, \apj, 963, 106, \dodoi{10.3847/1538-4357/ad1990}

\bibitem[{{Miville-Desch{\^e}nes} \& {Lagache}(2005)}]{IRIS2005}
{Miville-Desch{\^e}nes}, M.-A., \& {Lagache}, G. 2005, \apjs, 157, 302, \dodoi{10.1086/427938}

\bibitem[{Monti {et~al.}(2003)Monti, Tamayo, Mesirov, \& Golub}]{Monti2003}
Monti, S., Tamayo, P., Mesirov, J., \& Golub, T. 2003, Mach. Learn., 52, 91

\bibitem[{{Moreira} \& {Yun}(2002)}]{MoreiraYun2002}
{Moreira}, M.~C., \& {Yun}, J.~L. 2002, \aap, 381, 628, \dodoi{10.1051/0004-6361:20011558}

\bibitem[{{Muzerolle} {et~al.}(2001){Muzerolle}, {Calvet}, \& {Hartmann}}]{muzerolle01}
{Muzerolle}, J., {Calvet}, N., \& {Hartmann}, L. 2001, \apj, 550, 944, \dodoi{10.1086/319779}

\bibitem[{{Muzerolle} {et~al.}(1998){Muzerolle}, {Hartmann}, \& {Calvet}}]{muzerolle98b}
{Muzerolle}, J., {Hartmann}, L., \& {Calvet}, N. 1998, \aj, 116, 455, \dodoi{10.1086/300428}

\bibitem[{{Myers} {et~al.}(1991){Myers}, {Fuller}, {Goodman}, \& {Benson}}]{Myers1991}
{Myers}, P.~C., {Fuller}, G.~A., {Goodman}, A.~A., \& {Benson}, P.~J. 1991, \apj, 376, 561, \dodoi{10.1086/170305}

\bibitem[{{Myers} \& {Goodman}(1991)}]{MyersGoodman1991}
{Myers}, P.~C., \& {Goodman}, A.~A. 1991, \apj, 373, 509, \dodoi{10.1086/170070}

\bibitem[{{Nelissen} {et~al.}(2023{\natexlab{a}}){Nelissen}, {McGinnis}, {Folsom}, {Ray}, {Vidotto}, {Alecian}, {Bouvier}, {Morin}, {Donati}, \& {Devaraj}}]{Nelissen2023a}
{Nelissen}, M., {McGinnis}, P., {Folsom}, C.~P., {et~al.} 2023{\natexlab{a}}, \aap, 670, A165, \dodoi{10.1051/0004-6361/202245194}

\bibitem[{{Nelissen} {et~al.}(2023{\natexlab{b}}){Nelissen}, {Natta}, {McGinnis}, {Pittman}, {Delvaux}, \& {Ray}}]{Nelissen2023b}
{Nelissen}, M., {Natta}, A., {McGinnis}, P., {et~al.} 2023{\natexlab{b}}, \aap, 677, A64, \dodoi{10.1051/0004-6361/202347231}

\bibitem[{{Ou} {et~al.}(2023){Ou}, {Necib}, \& {Frebel}}]{Ou2023}
{Ou}, X., {Necib}, L., \& {Frebel}, A. 2023, \mnras, 521, 2623, \dodoi{10.1093/mnras/stad706}

\bibitem[{Pedregosa {et~al.}(2011)Pedregosa, Varoquaux, Gramfort, Michel, Thirion, Grisel, Blondel, Prettenhofer, Weiss, Dubourg, Vanderplas, Passos, Cournapeau, Brucher, Perrot, \& Duchesnay}]{scikit-learn}
Pedregosa, F., Varoquaux, G., Gramfort, A., {et~al.} 2011, Journal of Machine Learning Research, 12, 2825

\bibitem[{{Pereyra} \& {Magalh{\~a}es}(2004)}]{PereyraMagalhaes2004}
{Pereyra}, A., \& {Magalh{\~a}es}, A.~M. 2004, \apj, 603, 584, \dodoi{10.1086/381702}

\bibitem[{{Pineda} {et~al.}(2023){Pineda}, {Arzoumanian}, {Andre}, {Friesen}, {Zavagno}, {Clarke}, {Inoue}, {Chen}, {Lee}, {Soler}, \& {Kuffmeier}}]{Pineda2023}
{Pineda}, J.~E., {Arzoumanian}, D., {Andre}, P., {et~al.} 2023, in Astronomical Society of the Pacific Conference Series, Vol. 534, Protostars and Planets VII, ed. S.~{Inutsuka}, Y.~{Aikawa}, T.~{Muto}, K.~{Tomida}, \& M.~{Tamura}, 233, \dodoi{10.48550/arXiv.2205.03935}

\bibitem[{{Pinilla} {et~al.}(2018){Pinilla}, {Benisty}, {de Boer}, {Manara}, {Bouvier}, {Dominik}, {Ginski}, {Loomis}, \& {Sicilia Aguilar}}]{Pinilla2018}
{Pinilla}, P., {Benisty}, M., {de Boer}, J., {et~al.} 2018, \apj, 868, 85, \dodoi{10.3847/1538-4357/aae824}

\bibitem[{{Pittman} {et~al.}(2025{\natexlab{a}}){Pittman}, {Espaillat}, {Robinson}, {Thanathibodee}, {Lopez}, {Calvet}, {Zhu}, {Walter}, {Wendeborn}, {Manara}, {Campbell-White}, {Claes}, {Fang}, {Frasca}, {Gameiro}, {Gangi}, {Hern{\'a}ndez}, {K{\'o}sp{\'a}l}, {Mauc{\'o}}, {Muzerolle}, {Siwak}, {Tychoniec}, \& {Venuti}}]{Pittman2025}
{Pittman}, C.~V., {Espaillat}, C.~C., {Robinson}, C.~E., {et~al.} 2025{\natexlab{a}}, \apj, 992, 134, \dodoi{10.3847/1538-4357/adef35}

\bibitem[{{Pittman} {et~al.}(2025{\natexlab{b}}){Pittman}, {Espaillat}, {Zhu}, {Thanathibodee}, {Robinson}, {Calvet}, \& {K{\'o}sp{\'a}l}}]{Pittman2025b}
{Pittman}, C.~V., {Espaillat}, C.~C., {Zhu}, Z., {et~al.} 2025{\natexlab{b}}, arXiv e-prints, arXiv:2509.03767, \dodoi{10.48550/arXiv.2509.03767}

\bibitem[{{Planck Collaboration} {et~al.}(2016{\natexlab{a}}){Planck Collaboration}, {Adam}, {Ade}, {Aghanim}, {Alves}, {Arnaud}, {Arzoumanian}, {Ashdown}, {Aumont}, {Baccigalupi}, {Banday}, {Barreiro}, {Bartolo}, {Battaner}, {Benabed}, {Benoit-L{\'e}vy}, {Bernard}, {Bersanelli}, {Bielewicz}, {Bonaldi}, {Bonavera}, {Bond}, {Borrill}, {Bouchet}, {Boulanger}, {Bracco}, {Burigana}, {Butler}, {Calabrese}, {Cardoso}, {Catalano}, {Chamballu}, {Chiang}, {Christensen}, {Colombi}, {Colombo}, {Combet}, {Couchot}, {Crill}, {Curto}, {Cuttaia}, {Danese}, {Davies}, {Davis}, {de Bernardis}, {de Rosa}, {de Zotti}, {Delabrouille}, {Dickinson}, {Diego}, {Dole}, {Donzelli}, {Dor{\'e}}, {Douspis}, {Ducout}, {Dupac}, {Efstathiou}, {Elsner}, {En{\ss}lin}, {Eriksen}, {Falgarone}, {Ferri{\`e}re}, {Finelli}, {Forni}, {Frailis}, {Fraisse}, {Franceschi}, {Frejsel}, {Galeotta}, {Galli}, {Ganga}, {Ghosh}, {Giard}, {Gjerl{\o}w}, {Gonz{\'a}lez-Nuevo}, {G{\'o}rski}, {Gregorio}, {Gruppuso}, {Guillet}, {Hansen}, {Hanson}, {Harrison},
  {Henrot-Versill{\'e}}, {Hern{\'a}ndez-Monteagudo}, {Herranz}, {Hildebrandt}, {Hivon}, {Hobson}, {Holmes}, {Hovest}, {Huffenberger}, {Hurier}, {Jaffe}, {Jaffe}, {Jones}, {Juvela}, {Keih{\"a}nen}, {Keskitalo}, {Kisner}, {Kneissl}, {Knoche}, {Kunz}, {Kurki-Suonio}, {Lagache}, {Lamarre}, {Lasenby}, {Lattanzi}, {Lawrence}, {Leonardi}, {Levrier}, {Liguori}, {Lilje}, {Linden-V{\o}rnle}, {L{\'o}pez-Caniego}, {Lubin}, {Mac{\'\i}as-P{\'e}rez}, {Maffei}, {Maino}, {Mandolesi}, {Maris}, {Marshall}, {Martin}, {Mart{\'\i}nez-Gonz{\'a}lez}, {Masi}, {Matarrese}, {Mazzotta}, {Melchiorri}, {Mendes}, {Mennella}, {Migliaccio}, {Miville-Desch{\^e}nes}, {Moneti}, {Montier}, {Morgante}, {Mortlock}, {Munshi}, {Murphy}, {Naselsky}, {Natoli}, {N{\o}rgaard-Nielsen}, {Noviello}, {Novikov}, {Novikov}, {Oppermann}, {Oxborrow}, {Pagano}, {Pajot}, {Paoletti}, {Pasian}, {Perdereau}, {Perotto}, {Perrotta}, {Pettorino}, {Piacentini}, {Piat}, {Plaszczynski}, {Pointecouteau}, {Polenta}, {Ponthieu}, {Popa}, {Pratt}, {Prunet}, {Puget}, {Rachen},
  {Reach}, {Reinecke}, {Remazeilles}, {Renault}, {Ristorcelli}, {Rocha}, {Roudier}, {Rubi{\~n}o-Mart{\'\i}n}, {Rusholme}, {Sandri}, {Santos}, {Savini}, {Scott}, {Soler}, {Spencer}, {Stolyarov}, {Sudiwala}, {Sunyaev}, {Sutton}, {Suur-Uski}, {Sygnet}, {Tauber}, {Terenzi}, {Toffolatti}, {Tomasi}, {Tristram}, {Tucci}, {Umana}, {Valenziano}, {Valiviita}, {Van Tent}, {Vielva}, {Villa}, {Wade}, {Wandelt}, \& {Wehus}}]{Planck2016Diffuse}
{Planck Collaboration}, {Adam}, R., {Ade}, P.~A.~R., {et~al.} 2016{\natexlab{a}}, \aap, 586, A135, \dodoi{10.1051/0004-6361/201425044}

\bibitem[{{Planck Collaboration} {et~al.}(2016{\natexlab{b}}){Planck Collaboration}, {Ade}, {Aghanim}, {Alves}, {Arnaud}, {Arzoumanian}, {Ashdown}, {Aumont}, {Baccigalupi}, {Banday}, {Barreiro}, {Bartolo}, {Battaner}, {Benabed}, {Beno{\^\i}t}, {Benoit-L{\'e}vy}, {Bernard}, {Bersanelli}, {Bielewicz}, {Bock}, {Bonavera}, {Bond}, {Borrill}, {Bouchet}, {Boulanger}, {Bracco}, {Burigana}, {Calabrese}, {Cardoso}, {Catalano}, {Chiang}, {Christensen}, {Colombo}, {Combet}, {Couchot}, {Crill}, {Curto}, {Cuttaia}, {Danese}, {Davies}, {Davis}, {de Bernardis}, {de Rosa}, {de Zotti}, {Delabrouille}, {Dickinson}, {Diego}, {Dole}, {Donzelli}, {Dor{\'e}}, {Douspis}, {Ducout}, {Dupac}, {Efstathiou}, {Elsner}, {En{\ss}lin}, {Eriksen}, {Falceta-Gon{\c{c}}alves}, {Falgarone}, {Ferri{\`e}re}, {Finelli}, {Forni}, {Frailis}, {Fraisse}, {Franceschi}, {Frejsel}, {Galeotta}, {Galli}, {Ganga}, {Ghosh}, {Giard}, {Gjerl{\o}w}, {Gonz{\'a}lez-Nuevo}, {G{\'o}rski}, {Gregorio}, {Gruppuso}, {Gudmundsson}, {Guillet}, {Harrison}, {Helou},
  {Hennebelle}, {Henrot-Versill{\'e}}, {Hern{\'a}ndez-Monteagudo}, {Herranz}, {Hildebrandt}, {Hivon}, {Holmes}, {Hornstrup}, {Huffenberger}, {Hurier}, {Jaffe}, {Jaffe}, {Jones}, {Juvela}, {Keih{\"a}nen}, {Keskitalo}, {Kisner}, {Knoche}, {Kunz}, {Kurki-Suonio}, {Lagache}, {Lamarre}, {Lasenby}, {Lattanzi}, {Lawrence}, {Leonardi}, {Levrier}, {Liguori}, {Lilje}, {Linden-V{\o}rnle}, {L{\'o}pez-Caniego}, {Lubin}, {Mac{\'\i}as-P{\'e}rez}, {Maino}, {Mandolesi}, {Mangilli}, {Maris}, {Martin}, {Mart{\'\i}nez-Gonz{\'a}lez}, {Masi}, {Matarrese}, {Melchiorri}, {Mendes}, {Mennella}, {Migliaccio}, {Miville-Desch{\^e}nes}, {Moneti}, {Montier}, {Morgante}, {Mortlock}, {Munshi}, {Murphy}, {Naselsky}, {Nati}, {Netterfield}, {Noviello}, {Novikov}, {Novikov}, {Oppermann}, {Oxborrow}, {Pagano}, {Pajot}, {Paladini}, {Paoletti}, {Pasian}, {Perotto}, {Pettorino}, {Piacentini}, {Piat}, {Pierpaoli}, {Pietrobon}, {Plaszczynski}, {Pointecouteau}, {Polenta}, {Ponthieu}, {Pratt}, {Prunet}, {Puget}, {Rachen}, {Reinecke}, {Remazeilles},
  {Renault}, {Renzi}, {Ristorcelli}, {Rocha}, {Rossetti}, {Roudier}, {Rubi{\~n}o-Mart{\'\i}n}, {Rusholme}, {Sandri}, {Santos}, {Savelainen}, {Savini}, {Scott}, {Soler}, {Stolyarov}, {Sudiwala}, {Sutton}, {Suur-Uski}, {Sygnet}, {Tauber}, {Terenzi}, {Toffolatti}, {Tomasi}, {Tristram}, {Tucci}, {Umana}, {Valenziano}, {Valiviita}, {Van Tent}, {Vielva}, {Villa}, {Wade}, {Wandelt}, {Wehus}, {Ysard}, {Yvon}, \& {Zonca}}]{Planck2016a}
{Planck Collaboration}, {Ade}, P.~A.~R., {Aghanim}, N., {et~al.} 2016{\natexlab{b}}, \aap, 586, A138, \dodoi{10.1051/0004-6361/201525896}

\bibitem[{{Preibisch} {et~al.}(2002){Preibisch}, {Brown}, {Bridges}, {Guenther}, \& {Zinnecker}}]{Preibisch2002}
{Preibisch}, T., {Brown}, A. G.~A., {Bridges}, T., {Guenther}, E., \& {Zinnecker}, H. 2002, \aj, 124, 404, \dodoi{10.1086/341174}

\bibitem[{{Rizzo} {et~al.}(1998){Rizzo}, {Morras}, \& {Arnal}}]{Rizzo1998}
{Rizzo}, J.~R., {Morras}, R., \& {Arnal}, E.~M. 1998, \mnras, 300, 497, \dodoi{10.1046/j.1365-8711.1998.01916.x}

\bibitem[{{Robinson} \& {Espaillat}(2019)}]{re19}
{Robinson}, C.~E., \& {Espaillat}, C.~C. 2019, \apj, 874, 129, \dodoi{10.3847/1538-4357/ab0d8d}

\bibitem[{{Rygl} {et~al.}(2013){Rygl}, {Benedettini}, {Schisano}, {Elia}, {Molinari}, {Pezzuto}, {Andr{\'e}}, {Bernard}, {White}, {Polychroni}, {Bontemps}, {Cox}, {Di Francesco}, {Facchini}, {Fallscheer}, {di Giorgio}, {Hennemann}, {Hill}, {K{\"o}nyves}, {Minier}, {Motte}, {Nguyen-Luong}, {Peretto}, {Pestalozzi}, {Sadavoy}, {Schneider}, {Spinoglio}, {Testi}, \& {Ward-Thompson}}]{Rygl2013}
{Rygl}, K.~L.~J., {Benedettini}, M., {Schisano}, E., {et~al.} 2013, \aap, 549, L1, \dodoi{10.1051/0004-6361/201219511}

\bibitem[{{Soler} {et~al.}(2016){Soler}, {Alves}, {Boulanger}, {Bracco}, {Falgarone}, {Franco}, {Guillet}, {Hennebelle}, {Levrier}, {Martin}, \& {Miville-Desch{\^e}nes}}]{Soler2016}
{Soler}, J.~D., {Alves}, F., {Boulanger}, F., {et~al.} 2016, \aap, 596, A93, \dodoi{10.1051/0004-6361/201628996}

\bibitem[{{Tachihara} {et~al.}(1996){Tachihara}, {Dobashi}, {Mizuno}, {Ogawa}, \& {Fukui}}]{Tachihara1996}
{Tachihara}, K., {Dobashi}, K., {Mizuno}, A., {Ogawa}, H., \& {Fukui}, Y. 1996, \pasj, 48, 489, \dodoi{10.1093/pasj/48.3.489}

\bibitem[{{Tachihara} {et~al.}(2001){Tachihara}, {Toyoda}, {Onishi}, {Mizuno}, {Fukui}, \& {Neuh{\"a}user}}]{Tachihara2001}
{Tachihara}, K., {Toyoda}, S., {Onishi}, T., {et~al.} 2001, \pasj, 53, 1081, \dodoi{10.1093/pasj/53.6.1081}

\bibitem[{{Tazzari} {et~al.}(2017){Tazzari}, {Testi}, {Natta}, {Ansdell}, {Carpenter}, {Guidi}, {Hogerheijde}, {Manara}, {Miotello}, {van der Marel}, {van Dishoeck}, \& {Williams}}]{Tazzari2017}
{Tazzari}, M., {Testi}, L., {Natta}, A., {et~al.} 2017, \aap, 606, A88, \dodoi{10.1051/0004-6361/201730890}

\bibitem[{{Tothill} {et~al.}(2009){Tothill}, {L{\"o}hr}, {Parshley}, {Stark}, {Lane}, {Harnett}, {Wright}, {Walker}, {Bourke}, \& {Myers}}]{Tothill2009}
{Tothill}, N.~F.~H., {L{\"o}hr}, A., {Parshley}, S.~C., {et~al.} 2009, \apjs, 185, 98, \dodoi{10.1088/0067-0049/185/1/98}

\bibitem[{{Trapman} {et~al.}(2025){Trapman}, {Vioque}, {Kurtovic}, {Zhang}, {Rosotti}, {Pinilla}, {Carpenter}, {Cieza}, {Pascucci}, {Anania}, {Agurto-Gangas}, {Deng}, {Miley}, {P{\'e}rez}, {Sierra}, {Tabone}, {Ruiz-Rodriguez}, {Gonz{\'a}lez-Ruilova}, \& {TorresVillanueva}}]{Trapman2025}
{Trapman}, L., {Vioque}, M., {Kurtovic}, N.~T., {et~al.} 2025, \apj, 989, 10, \dodoi{10.3847/1538-4357/adc7af}

\bibitem[{{Vilas-Boas} {et~al.}(2000){Vilas-Boas}, {Myers}, \& {Fuller}}]{VilasBoas2000}
{Vilas-Boas}, J.~W.~S., {Myers}, P.~C., \& {Fuller}, G.~A. 2000, \apj, 532, 1038, \dodoi{10.1086/308586}

\bibitem[{{Vioque} {et~al.}(2023){Vioque}, {Cavieres}, {Pantaleoni Gonz{\'a}lez}, {Ribas}, {Oudmaijer}, {Mendigut{\'\i}a}, {Kilian}, {C{\'a}novas}, \& {Kuhn}}]{Vioque2023}
{Vioque}, M., {Cavieres}, M., {Pantaleoni Gonz{\'a}lez}, M., {et~al.} 2023, \aj, 166, 183, \dodoi{10.3847/1538-3881/acf75f}

\bibitem[{{Vioque} {et~al.}(2025){Vioque}, {Kurtovic}, {Trapman}, {Sierra}, {P{\'e}rez}, {Zhang}, {Curone}, {Rosotti}, {Carpenter}, {Tabone}, {Pinilla}, {Deng}, {Pascucci}, {Miley}, {Agurto-Gangas}, {Cieza}, {Anania}, {Ruiz-Rodriguez}, {Gonz{\'a}lez-Ruilova}, {TorresVillanueva}, \& {Kuznetsova}}]{Vioque2025}
{Vioque}, M., {Kurtovic}, N.~T., {Trapman}, L., {et~al.} 2025, \apj, 989, 9, \dodoi{10.3847/1538-4357/adc7b0}

\bibitem[{{Virtanen} {et~al.}(2020){Virtanen}, {Gommers}, {Oliphant}, {Haberland}, {Reddy}, {Cournapeau}, {Burovski}, {Peterson}, {Weckesser}, {Bright}, {van der Walt}, {Brett}, {Wilson}, {Millman}, {Mayorov}, {Nelson}, {Jones}, {Kern}, {Larson}, {Carey}, {Polat}, {Feng}, {Moore}, {VanderPlas}, {Laxalde}, {Perktold}, {Cimrman}, {Henriksen}, {Quintero}, {Harris}, {Archibald}, {Ribeiro}, {Pedregosa}, {van Mulbregt}, \& {SciPy 1. 0 Contributors}}]{SciPy2020}
{Virtanen}, P., {Gommers}, R., {Oliphant}, T.~E., {et~al.} 2020, Nature Medicine, 17, 261, \dodoi{10.1038/s41592-019-0686-2}

\bibitem[{{Winter} {et~al.}(2024{\natexlab{a}}){Winter}, {Benisty}, {Manara}, \& {Gupta}}]{Winter2024Lupus}
{Winter}, A.~J., {Benisty}, M., {Manara}, C.~F., \& {Gupta}, A. 2024{\natexlab{a}}, \aap, 691, A169, \dodoi{10.1051/0004-6361/202452120}

\bibitem[{{Winter} {et~al.}(2024{\natexlab{b}}){Winter}, {Benisty}, {Shuai}, {D{\^u}chene}, {Cuello}, {Anania}, {Cadiou}, \& {Joncour}}]{Winter2024Encounters}
{Winter}, A.~J., {Benisty}, M., {Shuai}, L., {et~al.} 2024{\natexlab{b}}, \aap, 691, A43, \dodoi{10.1051/0004-6361/202450842}

\bibitem[{{Winter} \& {Haworth}(2022)}]{WinterHaworth2022}
{Winter}, A.~J., \& {Haworth}, T.~J. 2022, European Physical Journal Plus, 137, 1132, \dodoi{10.1140/epjp/s13360-022-03314-1}

\bibitem[{{Yen} {et~al.}(2018){Yen}, {Koch}, {Manara}, {Miotello}, \& {Testi}}]{Yen2018}
{Yen}, H.-W., {Koch}, P.~M., {Manara}, C.~F., {Miotello}, A., \& {Testi}, L. 2018, \aap, 616, A100, \dodoi{10.1051/0004-6361/201732196}

\bibitem[{Zhang {et~al.}(2025)Zhang, Pérez, Pascucci, Pinilla, Cieza, Carpenter, Trapman, Deng, Agurto-Gangas, Sierra, Kurtovic, Ruiz-Rodriguez, Vioque, Miley, Tabone, González-Ruilova, Anania, Rosotti, TorresVillanueva, Hogerheijde, Schwarz, \& Kuznetsova}]{Zhang2025}
Zhang, K., Pérez, L.~M., Pascucci, I., {et~al.} 2025, The Astrophysical Journal, 989, 1, \dodoi{10.3847/1538-4357/addebe}

\end{thebibliography}
\bibliographystyle{aasjournal}

\appendix
\section{Supplementary Materials} \label{sec:Appendix}

\begin{figure}[h]
    \centering
    \hspace{1cm} \includegraphics[width=0.7\linewidth]{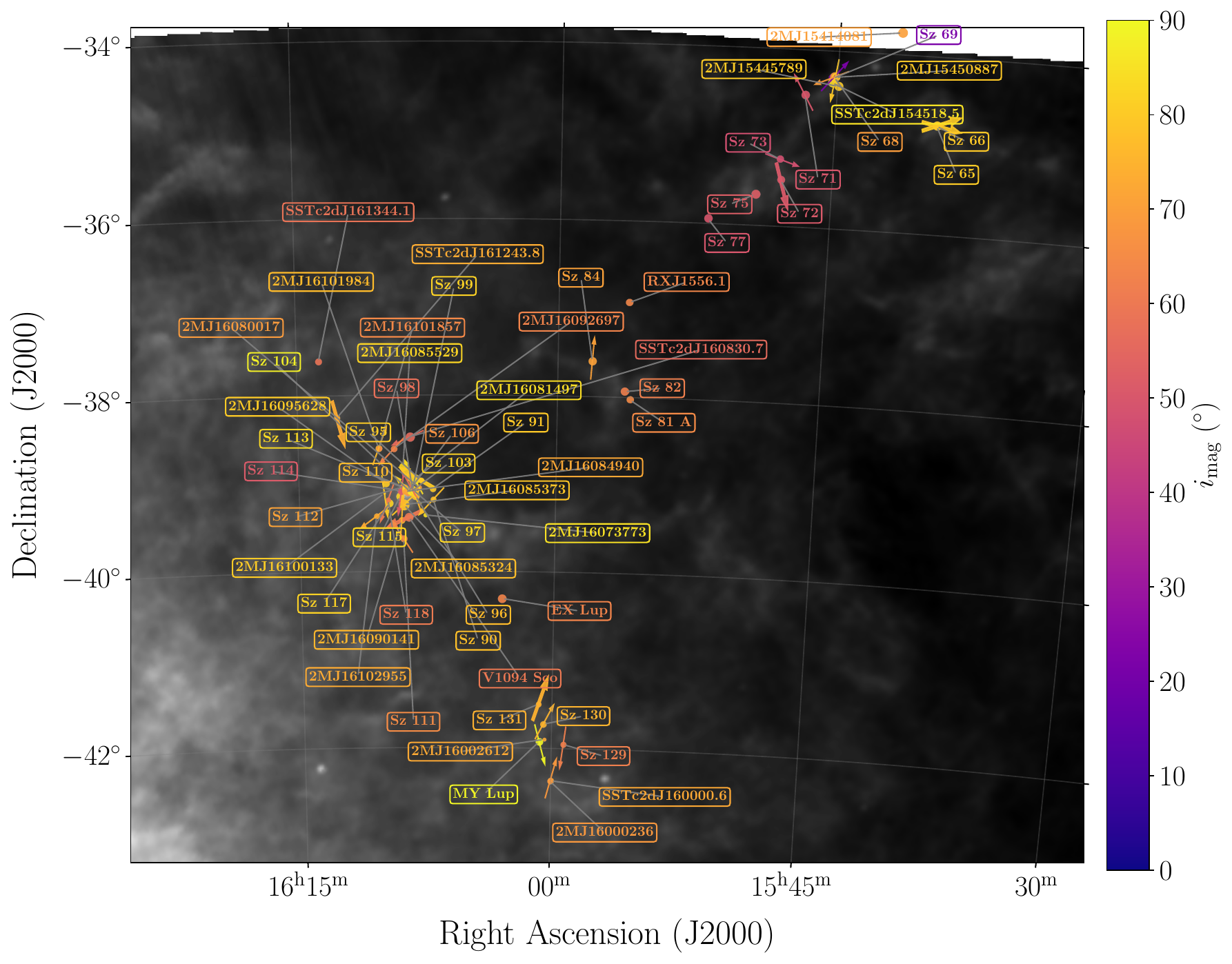}\\
    \begin{tikzpicture}
    \node[anchor=south west, inner sep=0] (image) at (0,0) {\includegraphics[width=0.37\linewidth]{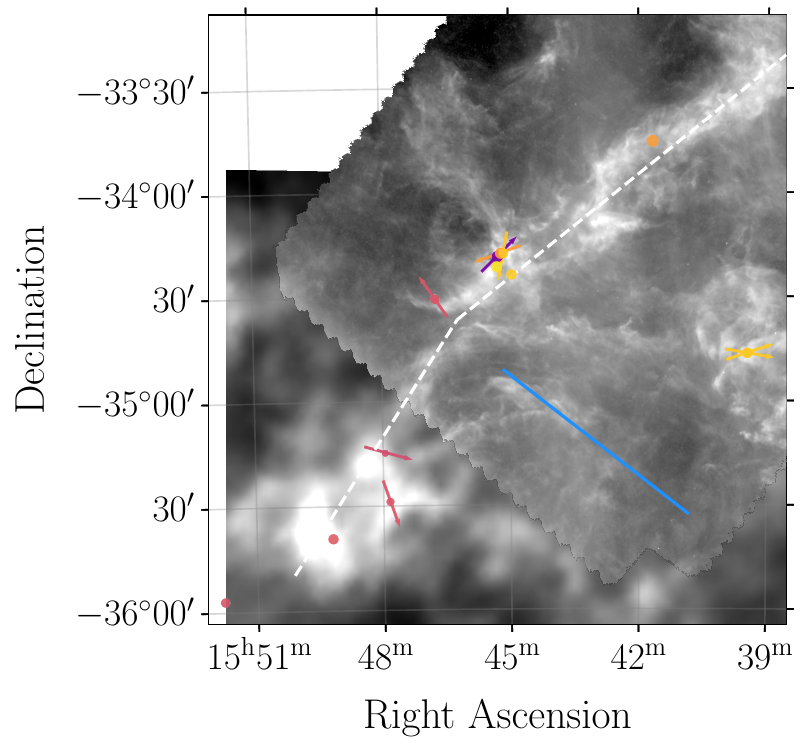}};
    \begin{scope}[x={(image.south east)},y={(image.north west)}]
            % Annotations relative to the image coordinates (0,0) to (1,1)
            \node[white, font=\large] at (0.8, 0.24) {Lupus I};
        \end{scope}
    \end{tikzpicture}
    \begin{tikzpicture}
    \node[anchor=south west, inner sep=0] (image) at (0,0) {\includegraphics[width=0.37\linewidth]{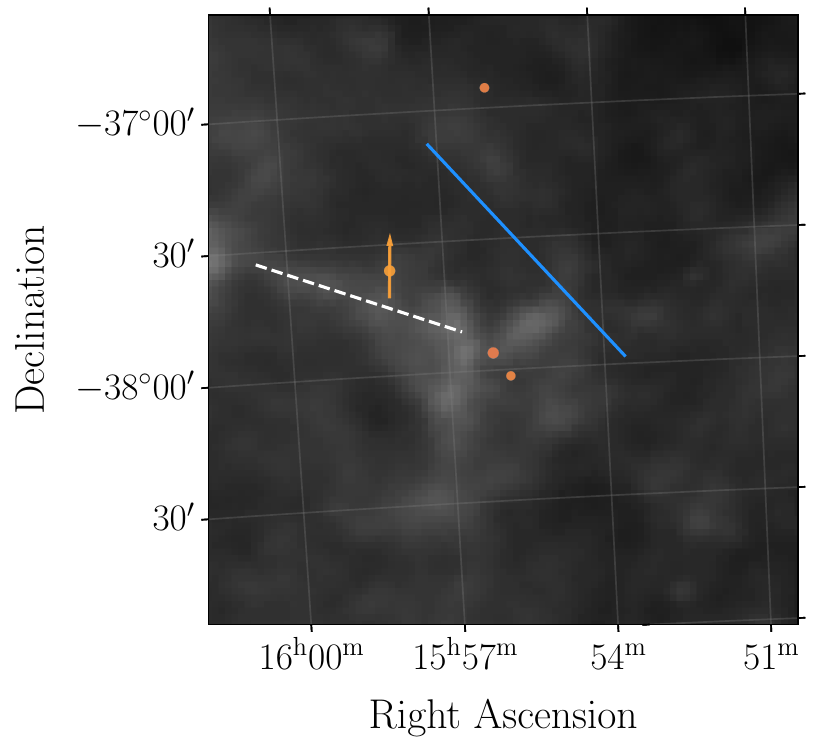}};
    \begin{scope}[x={(image.south east)},y={(image.north west)}]
            % Annotations relative to the image coordinates (0,0) to (1,1)
            \node[white, font=\large] at (0.8, 0.24) {Lupus II};
        \end{scope}
    \end{tikzpicture}
    \begin{tikzpicture}
    \node[anchor=south west, inner sep=0] (image) at (0,0) {\includegraphics[width=0.37\linewidth]{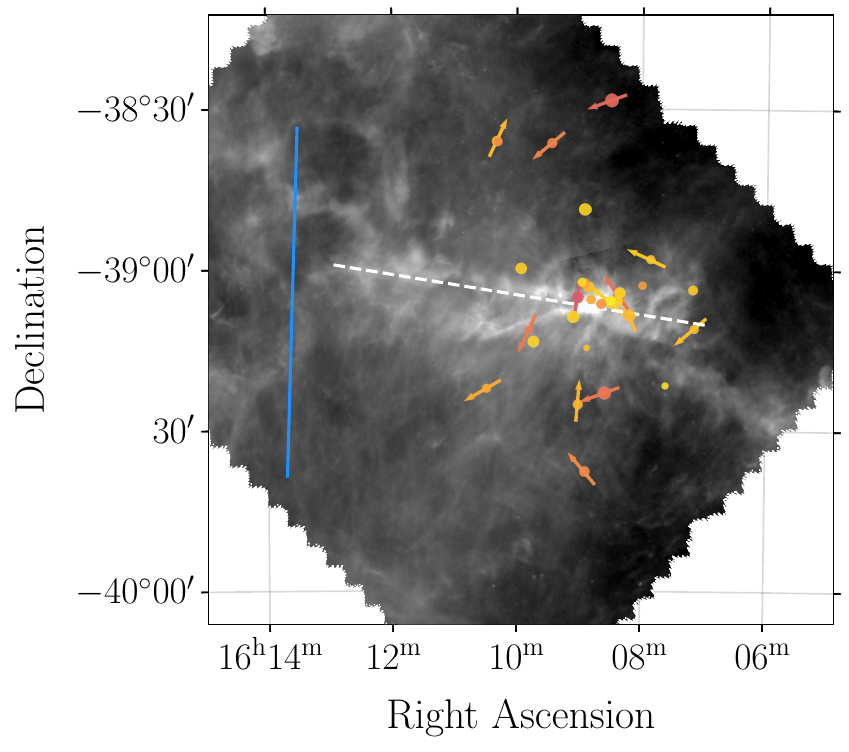}};
    \begin{scope}[x={(image.south east)},y={(image.north west)}]
            % Annotations relative to the image coordinates (0,0) to (1,1)
            \node[white, font=\large] at (0.6, 0.24) {Lupus III};
        \end{scope}
    \end{tikzpicture}
    \begin{tikzpicture}
    \node[anchor=south west, inner sep=0] (image) at (0,0) {\includegraphics[width=0.58\linewidth]{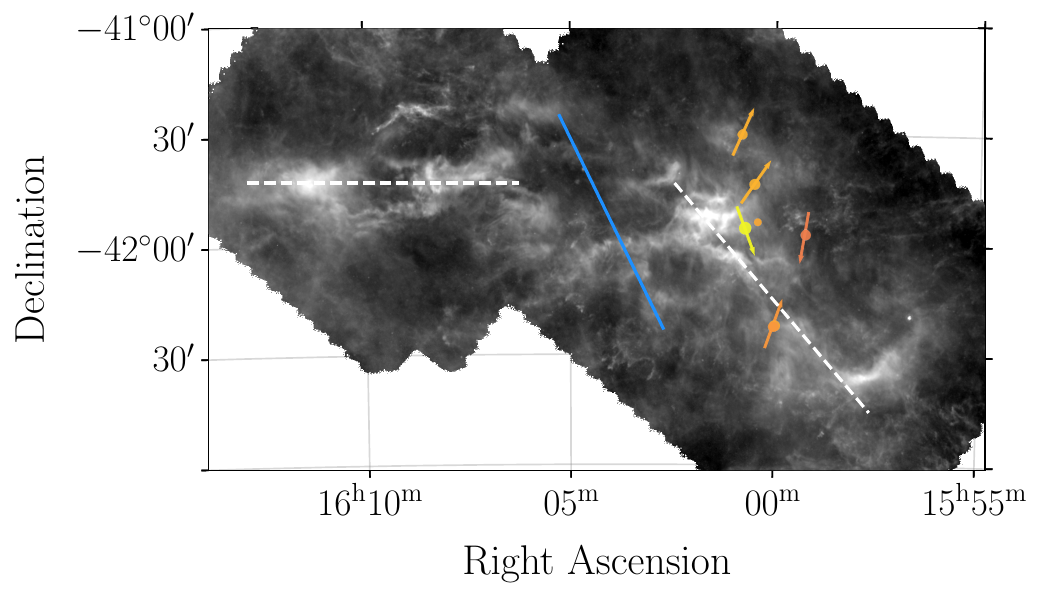}};
    \begin{scope}[x={(image.south east)},y={(image.north west)}]
            % Annotations relative to the image coordinates (0,0) to (1,1)
            \node[white, font=\large] at (0.8, 0.26) {Lupus IV};
        \end{scope}
    \end{tikzpicture}
    \caption{\textit{Top panel}: Same as Figure~\ref{fig:LupusIncl} (top left), but with CTTS names labeled. Small arrows again denote disk PAs. \textit{Bottom panels}: Zoomed regions around Lupus I (middle left), Lupus II (middle right), Lupus III (bottom left), and Lupus IV (bottom right). Dashed white lines show the approximate orientation and extent of the primary filaments in the plane of the sky. Blue lines show the mean magnetic field direction, and the line length is 3~pc in the plane of the sky to give a broad indication of the detected correlation scale. High-resolution background images show the Herschel SPIRE 250~\micron\ dust continuum maps from the Herschel Gould Belt Survey \citep{Andre2010}, and low resolution images show the IRIS 100 \micron\ map \citep{IRIS2005}. Points are colored by \imag\ according to the color bar in the top panel.
    }
    \label{fig:LupusApp}
\end{figure}
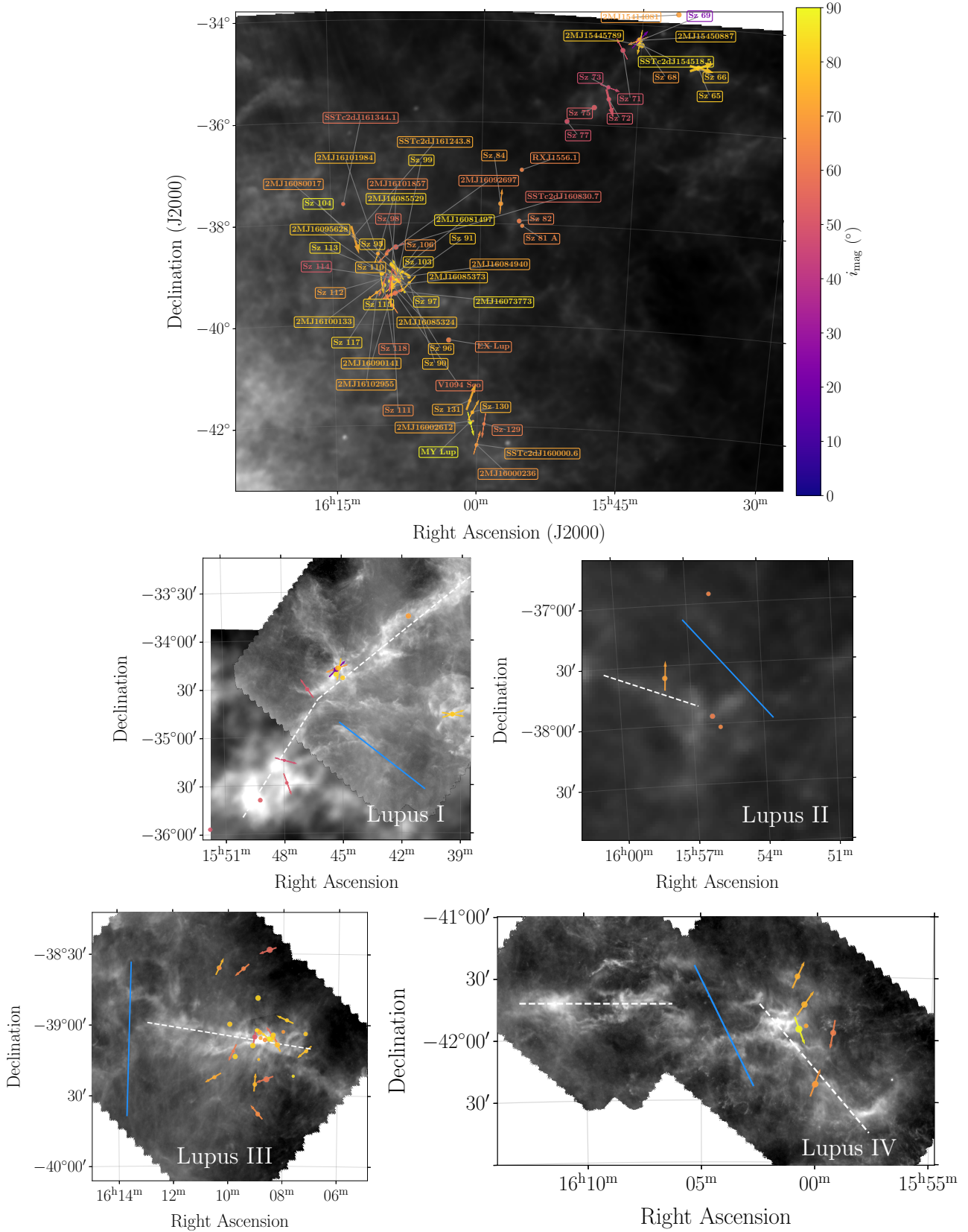

Figure~\ref{fig:LupusApp} shows the labeled version of Figure~\ref{fig:LupusIncl}, as well as zoomed images of the individual subregions with high-resolution Herschel SPIRE maps, when available.
Section~\ref{Appsec:clusteringDesign} describes our design and validation of the clustering procedure. It includes consensus matrices to visualize the clustering procedure results (Figures~\ref{fig:matrix} and \ref{fig:matrix_4D}), as well as clustering procedure tests (Figure~\ref{fig:clusterTests}).  Section~\ref{Appsec:flowresults} presents the full accretion flow model results (Table~\ref{tab:flowresults} and Figure~\ref{fig:AllFits}), along with \imag\ validation tests (Figures~\ref{fig:UniformTest} and \ref{fig:InclDep}). Finally, Section~\ref{Appsec:otherregions} discusses potential \imag\ trends in other star-forming regions (Figure~\ref{fig:TaurusOrion}).

\subsection{Clustering Algorithm Design and Testing} \label{Appsec:clusteringDesign}

Here we present the results of our tests to determine the design and reliability of our \texttt{HDBSCAN} clustering procedure. 
We design the algorithm to be optimized to categorize the Lupus CTTSs into the four appropriate Lupus subregions based on their 3D spatial locations alone.
We choose \texttt{cluster_selection_method}=\textit{excess of mass} to prioritize the most stable groups. The \texttt{min_cluster_size} hyperparameter defines the minimum number of CTTSs that must be connected for \texttt{HDBSCAN} to recognize them as a cluster. It must be greater than 1, as a group containing a single member is not meaningful, and it should not exceed the size of the smallest group that we aim to identify.
The smallest subregion is Lupus II, with 3 on-cloud CTTSs and 1 nearby off-cloud CTTS. These factors restrict \texttt{min_cluster_size} to be 2, 3, or 4. Then, the \texttt{min_samples} hyperparameter sets the strictness of the algorithm, with larger values causing more CTTSs to be marked as noise rather than being assigned to a group. \texttt{HDBSCAN}'s default choice is to set \texttt{min_samples} equal to \texttt{min_cluster_size}.

\begin{figure}
    \centering
    \begin{tikzpicture}
    \node[anchor=south west, inner sep=0] (image) at (0,0) {\includegraphics[width=\linewidth]{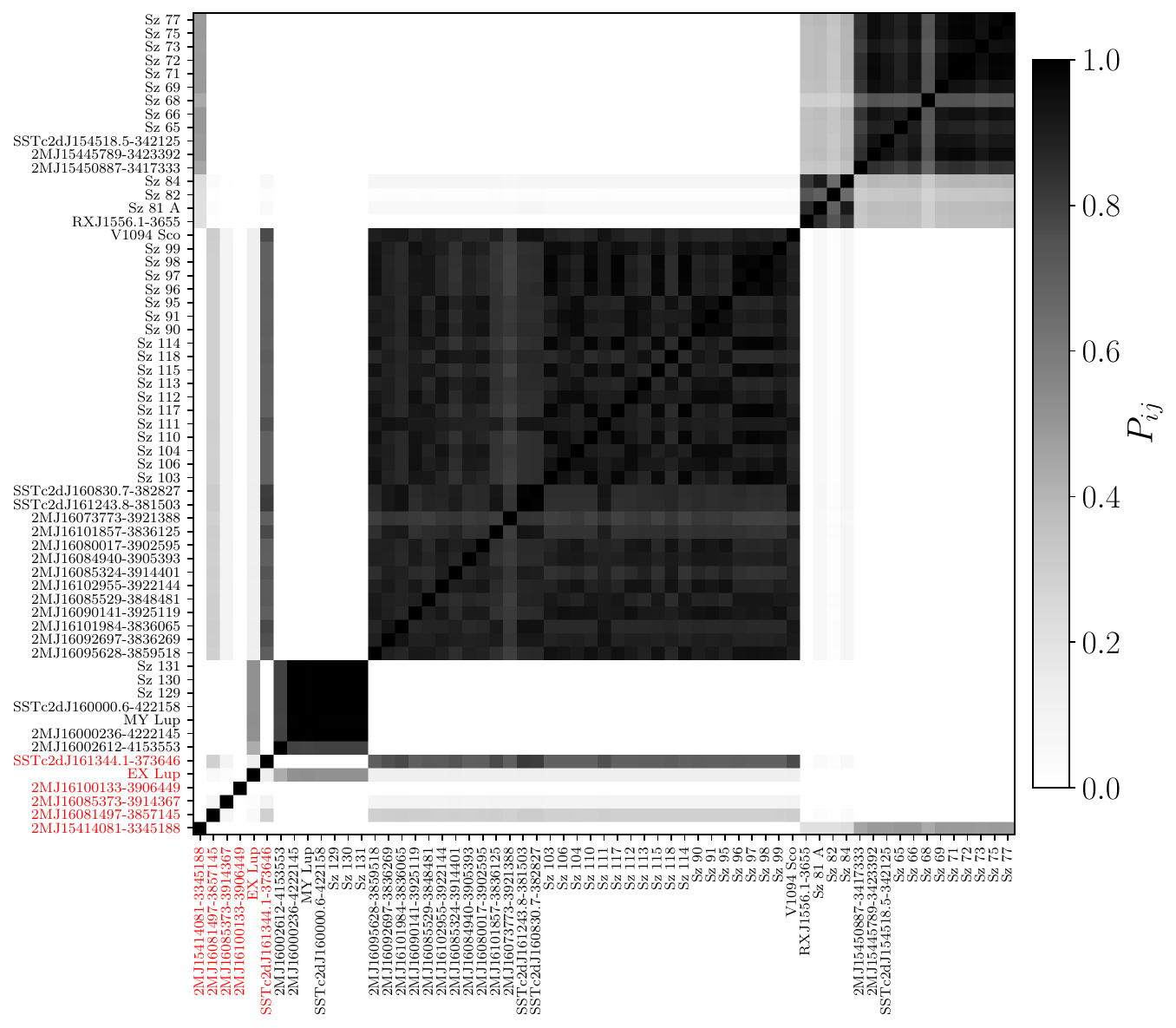}};
    \begin{scope}[x={(image.south east)},y={(image.north west)}, text opacity=0.85]
            \node[white, font=\huge] at (0.795, 0.91) {I};
            \node[white, font=\huge] at (0.703, 0.805) {II};
            \node[white, font=\huge] at (0.502, 0.568) {III};
            \node[white, font=\huge] at (0.276, 0.3105) {IV};
        \end{scope}
    \end{tikzpicture}
    \caption{Consensus matrix for the 3D groups found from 5000 MC iterations of \texttt{HDBSCAN}, using \texttt{min_cluster_size}=3, \texttt{min_samples}=3, and \texttt{cluster_selection_method}=\textit{excess of mass}. Dark squares indicate groups with high pairwise co-membership probabilities ($P_{ij}$), and Roman numerals indicate the associated Lupus subregions. Red labels indicate CTTSs that are marked as noise in more than 30\% of the MC iterations, all of which are off-cloud sources.}
    \label{fig:matrix}
\end{figure}

Only two sets of [\texttt{min_cluster_size}, \texttt{min_samples}] successfully classify the CTTSs into the four Lupus subregions: [3, 3] and [4, 3]. However, the latter results in a high probability ($\sim$0.6) of merging Lupus II sources into Lupus I, so we choose [3, 3] as the final values.\footnote{We note that the 4D group results are the same whether the chosen hyperparameters are [3, 3] or [4, 3].} Figure~\ref{fig:matrix} shows the consensus matrix \citep[that is, the pairwise co-membership probabilities, $P_{ij}$; see][]{Monti2003} resulting from 5000 MC iterations of \texttt{HDBSCAN} using \texttt{min_cluster_size}=3 and \texttt{min_samples}=3. When implementing the Monte Carlo (MC) procedure with these hyperparameters, the true subregion members all have mean pairwise co-membership probabilities ($P_{ij,{\rm mean}}$) greater than 0.5 and individual noise probabilities ($P_{\rm noise}$) below 0.3. Therefore, we set $P_{\rm min}$ to 0.5 and $P_{\rm noise,max}$ to 0.3.

\begin{figure}
    \centering
    \begin{tikzpicture}
    \node[anchor=south west, inner sep=0] (image) at (0,0) {\includegraphics[width=\linewidth]{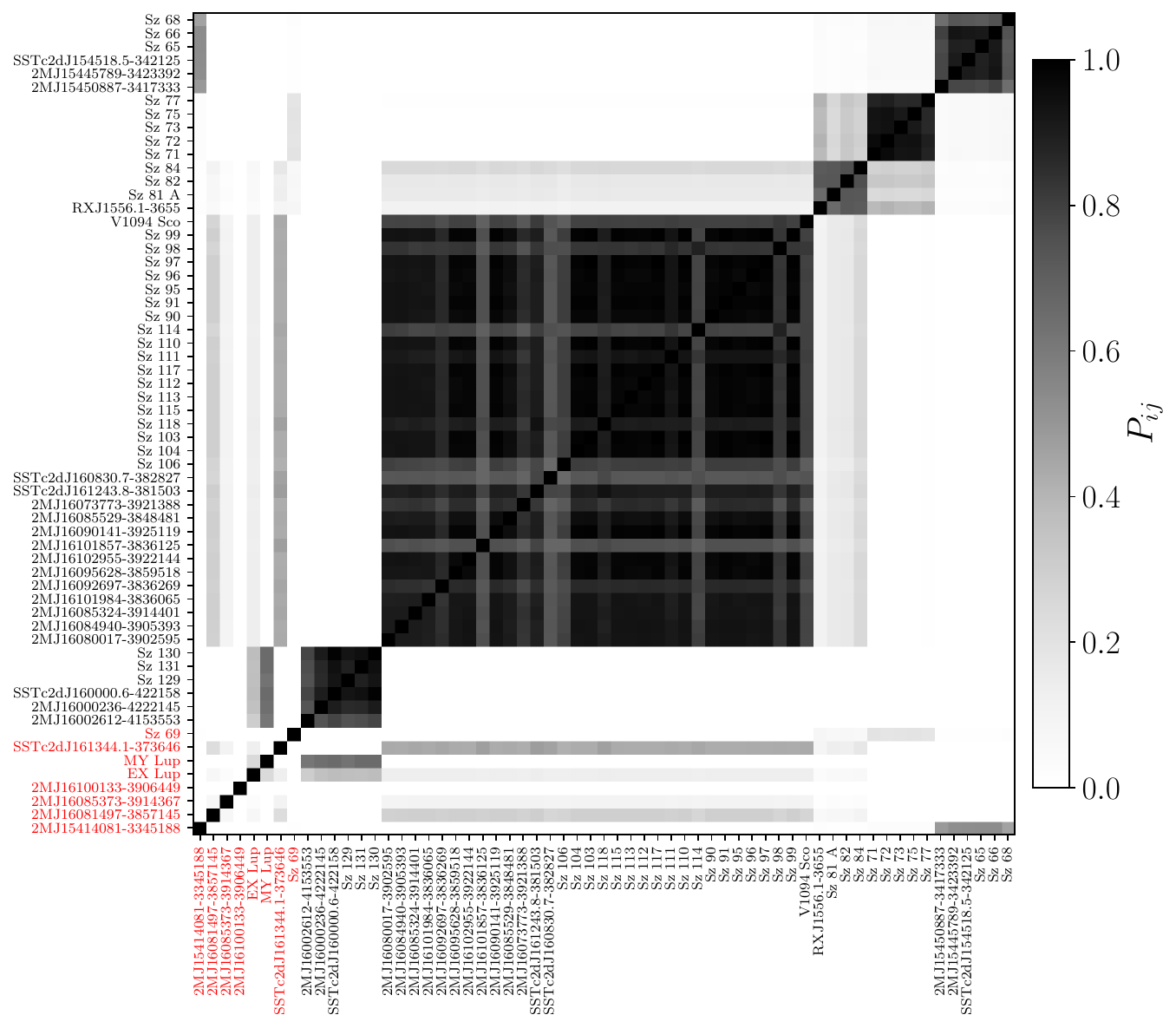}};
    \begin{scope}[x={(image.south east)},y={(image.north west)}, text opacity=0.85]
            % Annotations relative to the image coordinates (0,0) to (1,1)
            \node[white, font=\LARGE] at (0.828, 0.97) {I};
            \node[white, font=\LARGE] at (0.828, 0.93) {NW};
            \node[white, font=\LARGE] at (0.765, 0.894) {I};
            \node[white, font=\LARGE] at (0.765, 0.8625) {SE};
            \node[white, font=\huge] at (0.715, 0.815) {II};
            \node[white, font=\huge] at (0.515, 0.58) {III};
            \node[white, font=\huge] at (0.292, 0.33) {IV};
        \end{scope}
    \end{tikzpicture}
    \caption{Same as Figure~\ref{fig:matrix}, but for the 4D groups. Roman numerals indicate the assigned groups, and red labels indicate CTTSs that are marked as noise in more than 30\% of the MC iterations.}
    \label{fig:matrix_4D}
\end{figure}

\begin{figure}
    \centering
    \includegraphics[width=0.49\linewidth]{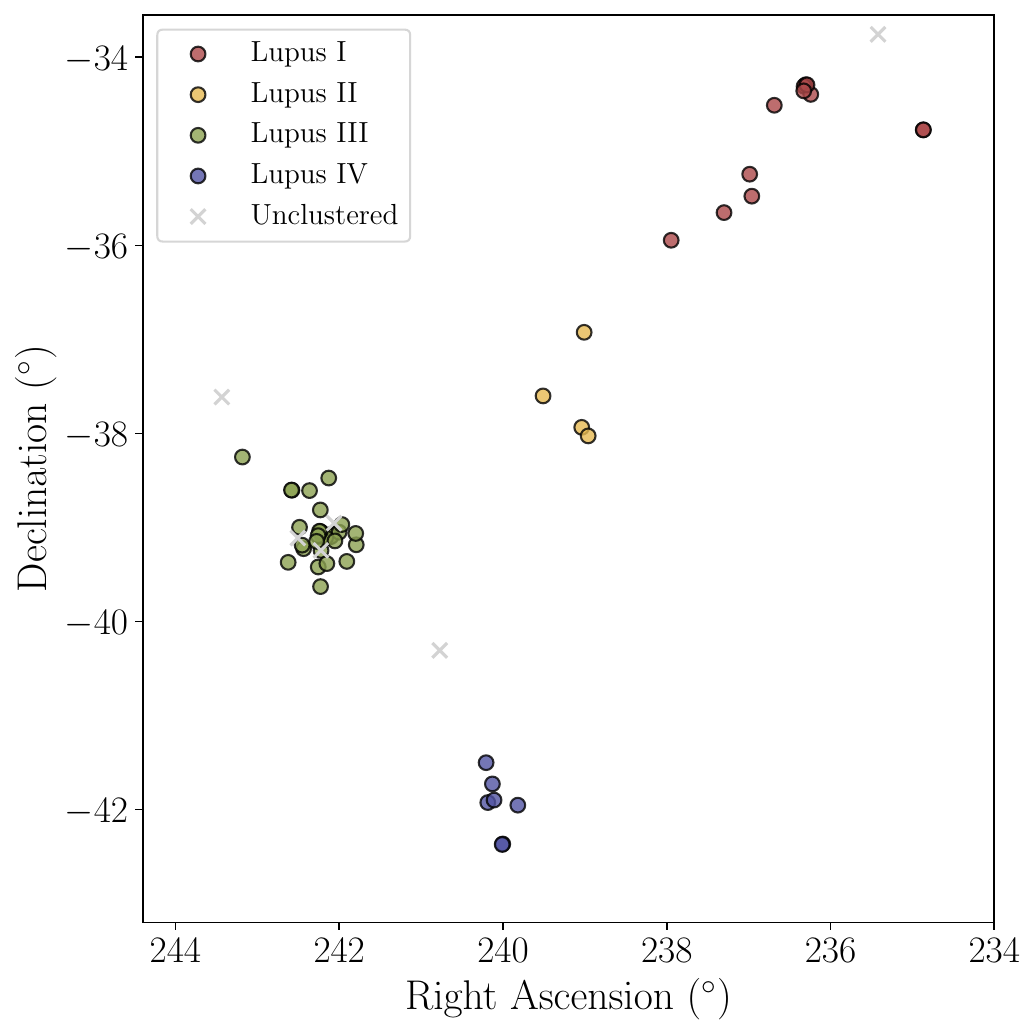}
    \includegraphics[width=0.49\linewidth]{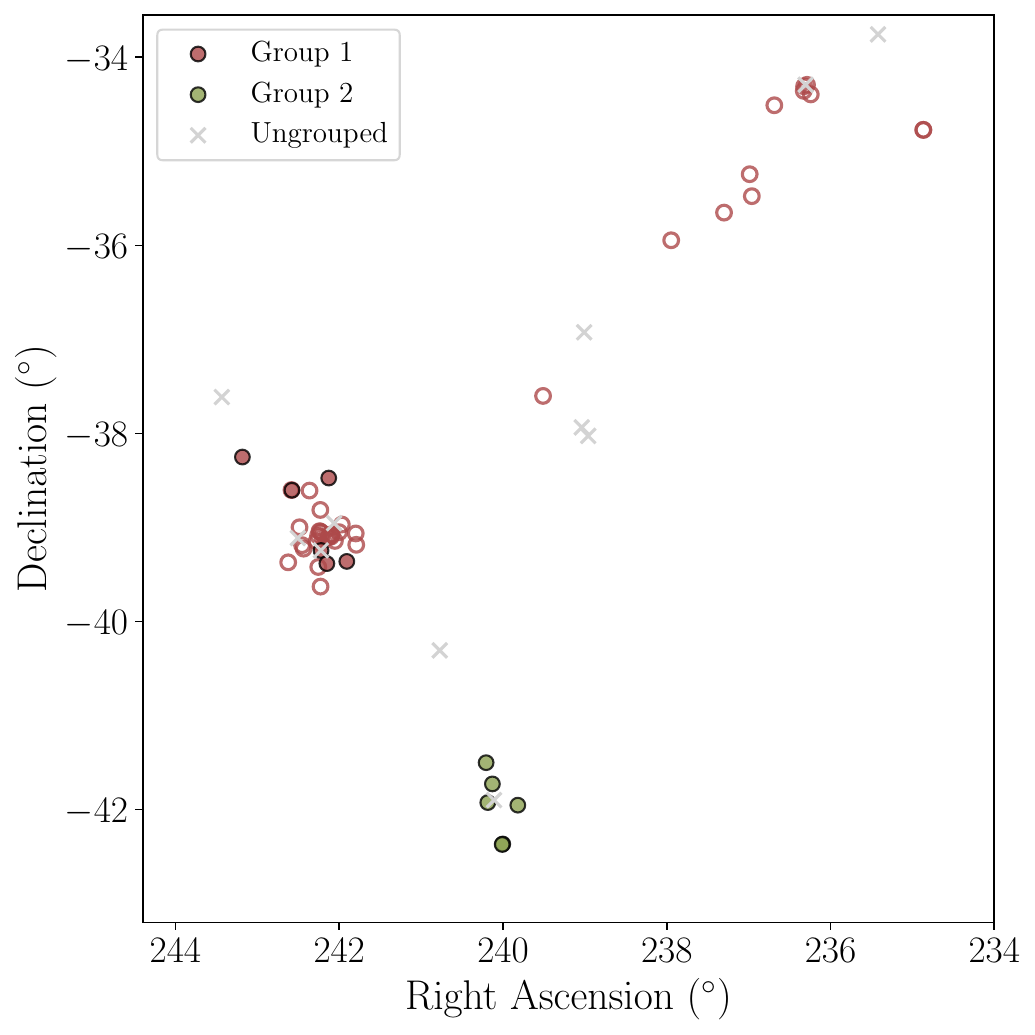}
    \includegraphics[width=\linewidth]{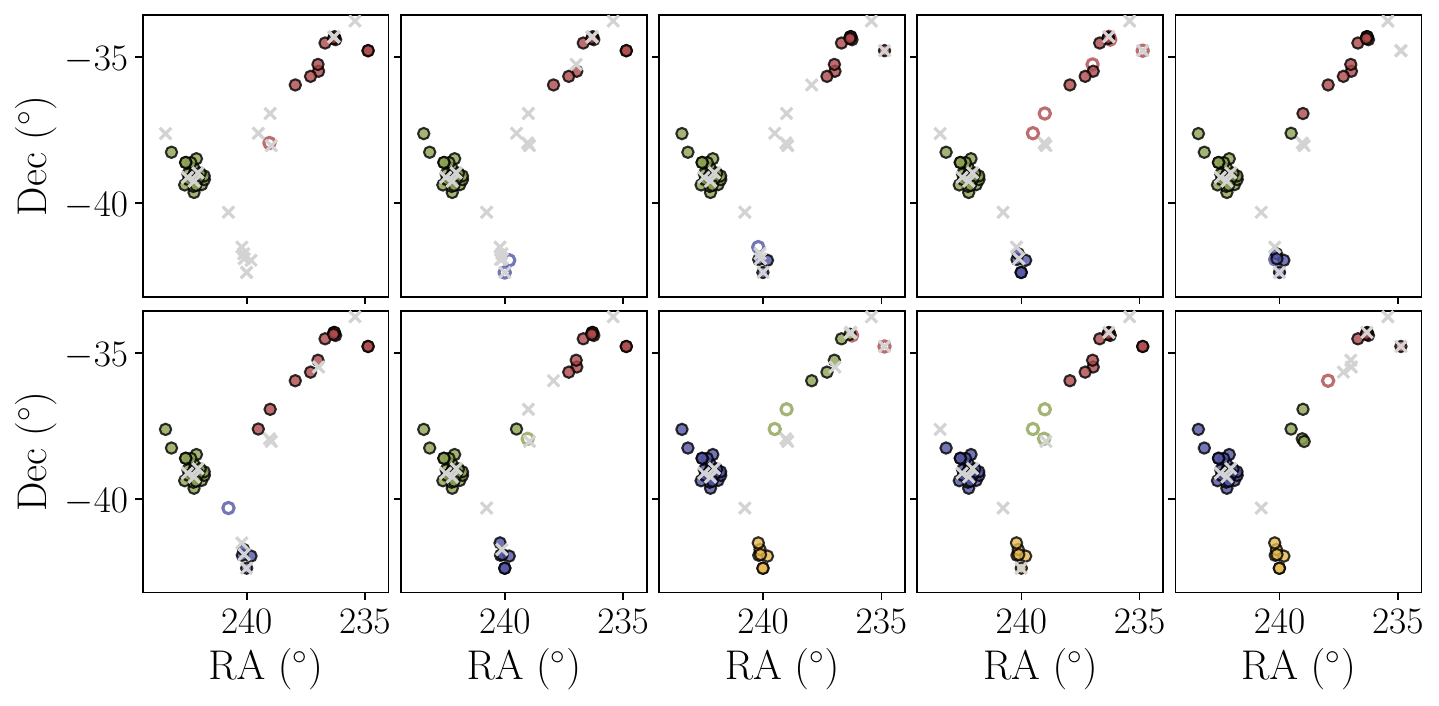}
    \caption{Clustering procedure tests, where points are colored by cluster assignment. Solid points indicate high-probability group members ($P_{ij,{\rm mean}}>P_{\rm min}$), whereas unfilled points indicate lower-probability members ($P_{ij,{\rm mean}}\leq P_{\rm min}$). Gray x markers indicate ungrouped sources marked as noise ($P_{\rm noise}>P_{\rm noise,max}$).
    Top left panel: Successful clustering of the four Lupus subregions achieved with \texttt{HDBSCAN} using \texttt{min_cluster_size}=3, \texttt{min_samples}=3, and \texttt{cluster_selection_method}=\textit{excess of mass} and coherence clustering thresholds of $P_{\rm min}$=0.5 and $P_{\rm noise,max}$=0.3, taking only the 3D coordinates into account. These parameters are used for all clustering processes. Top right panel: Groups found when assigning randomly-generated \imag\ values that are consistent with the observed \imag\ distribution to each MC sample as the fourth dimension in \texttt{HDBSCAN}. 
    Bottom panels: Groups found when perturbing randomly-assigned \imag\ values by our measured $\sigma_{i_{\rm mag}}$ values, shown for a representative set of the 100 total iterations. 
    % Plotted seeds: [194, 411, 178, 85, 200, 425, 119, 428, 123, 423]
    }
    \label{fig:clusterTests}
\end{figure}

As described in Section~\ref{sec:SampleAnalysis}, a star is considered a member of a group if it has a pairwise co-membership probability $P_{ij}$ greater than $P_{\rm min}$ with at least one other group member. Then, it is considered a high-probability member if its $P_{ij,{\rm mean}}$ is also greater than $P_{\rm min}$. The $P_{\rm noise,max}$ threshold of 0.3 means that any CTTS that is marked as noise, rather than a group member, in more than 30\% of the MC iterations is labeled ``ungrouped''. This results in six CTTSs being marked as ungrouped based on their 3D coordinates|2MASSJ15414081-3345188, SSSTc2dJ161344.1-373646, 2MASSJ16081497-3857145, 2MASSJ16085373-3914367, 2MASSJ16100133-3906449, and EX~Lup|all of which are in the off-cloud category \citep[see Table~\ref{tab:sample} notes and][]{Luhman2020}. These are indicated with gray x markers in Figure~\ref{fig:clusterTests} (top left).

The application of this optimized procedure to the intended four dimensions (3D spatial location plus \imag) is shown in Figure~\ref{fig:LupusIncl} (right), and the associated consensus matrix is shown in Figure~\ref{fig:matrix_4D}. We validate the results here using two tests. First, we test whether the inclusion of \imag\ as a fourth dimension produces spurious correlations by assigning random \imag\ values to each CTTS at each MC realization.
Specifically, in each MC step, we draw each \imag\ from a distribution that is uniform in $\cos i_{\rm mag}$ between 0 and 0.7, which is consistent with our observed Lupus \imag\ distribution (see Appendix~\ref{Appsec:InclReliability}).
The \texttt{HDBSCAN} results for this test are shown in Figure~\ref{fig:clusterTests} (top right). In this case, most of the Lupus I, II, and III CTTSs are weakly classified into one group, and most of the the Lupus IV CTTSs are in a second group.
These groups clearly do not correspond to those in Figure~\ref{fig:LupusIncl}, demonstrating that the observed correlations do not result from the addition of a fourth dimension alone.

In the next test, we initially assign random \imag\ values to each CTTS (again drawn from a uniform distribution in $\cos i_{\rm mag}$ between 0 and 0.7), and then perturb these values by the measured $\sigma_{i_{\rm mag}}$ values during each MC realization. This will show whether the observed correlations are due to correlated uncertainties. We repeat this test for 100 random initial sets of \imag, and the results for 10 representative runs are shown in Figure~\ref{fig:clusterTests} (bottom). We can clearly see that the inclusion of random \imag\ values perturbed by our $\sigma_{i_{\rm mag}}$ does not produce any consistent groups, and they do not resemble the correlations of the observed \imag\ values. In most cases, three or fewer total groups are identified, with a maximum of four groups identified (compared to the five we find for our observed sample). Only rarely are all grouped CTTSs marked as high-confidence members with $P_{ij,{\rm mean}}>P_{\rm min}$. Both tests thus demonstrate that our clustering results are not caused by a) the inclusion of \imag\ as the fourth dimension, b) the $\sigma_{i_{\rm mag}}$ values used to perturb the \imag\ values, or c) any overpowering influence of spatial locations irrespective of the measured \imag.

\subsection{Accretion flow model results and validation tests} \label{Appsec:flowresults}

\begin{figure}
    \centering
    \begin{tikzpicture}
    \centering
    \node[anchor=south west, inner sep=0] (image) at (0,0) {\includegraphics[width=0.95\textwidth]{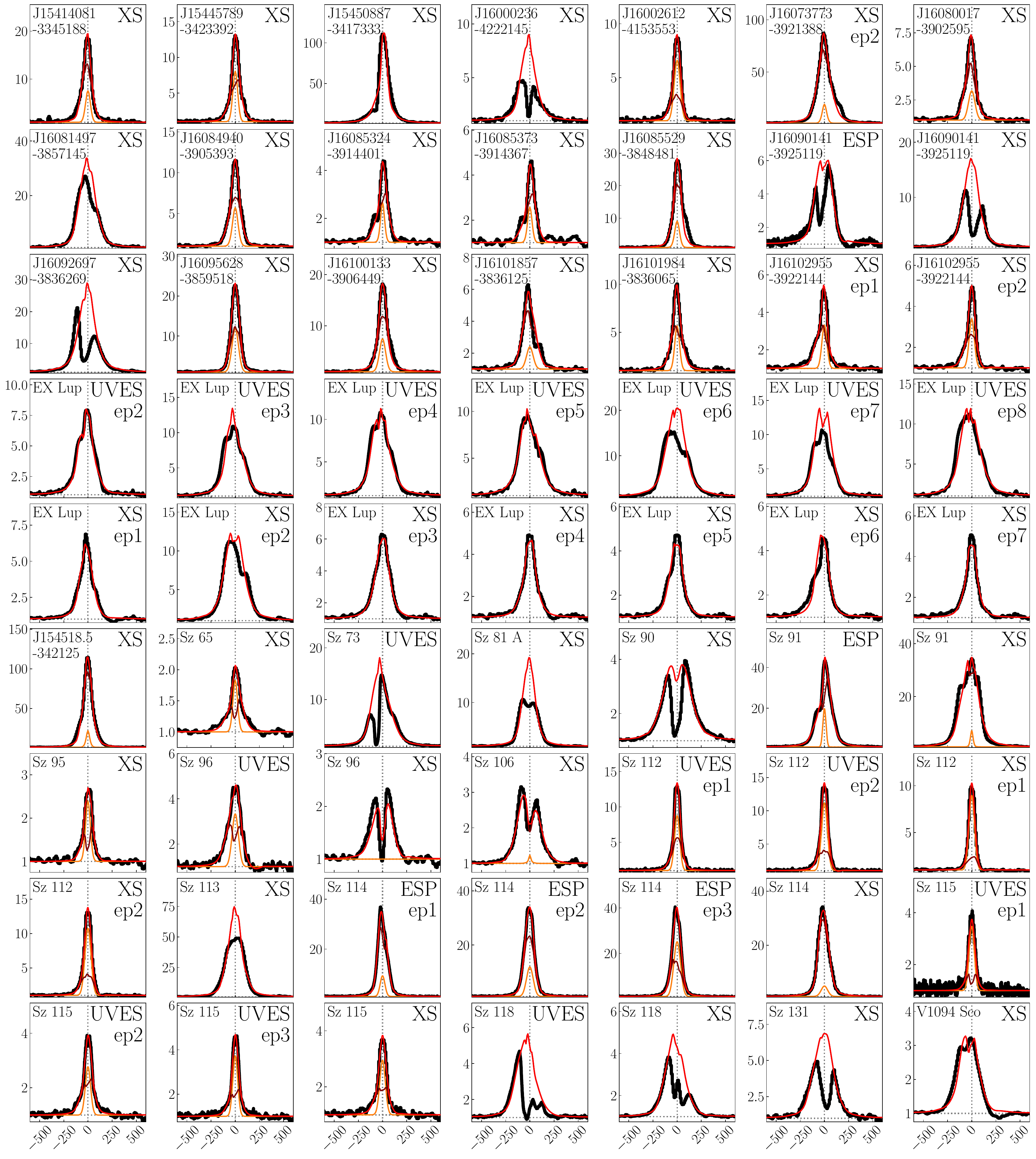}};
    \begin{scope}[x={(image.south east)},y={(image.north west)}]
            \node[black, font=\large] at (0.5, -0.01) {Velocity (km/s)};
            \node[black, rotate=90, font=\large] at (-0.01, 0.5) {Flux normalized to continuum};
        \end{scope}
    \end{tikzpicture}
    \caption{New flow model fits to VLT \halpha\ profiles from X-Shooter (XS), UVES, and ESPRESSO (ESP). Continuum-normalized \halpha\ observations are shown in black; Gaussian components representative of the chromosphere are orange, where applicable; magnetospheric contributions are maroon; and total model fits are bright red. The $x$-axes span $\pm$600~km/s. Absorption features that cannot be attributed to a magnetospheric origin are masked out in the fitting process.}
    \label{fig:AllFits}
\end{figure}

Table~\ref{tab:flowresults} presents the accretion flow model results for our Lupus sample, which come from \cite{Pittman2025} and this work as indicated in the table caption. Figure~\ref{fig:AllFits} shows the model fits to all new observations modeled here. Sections~\ref{Appsec:InclReliability}--\ref{Appsec:ChromTest} present validation tests.

\newpage
\startlongtable
\begin{deluxetable*}{>{\raggedright\arraybackslash}p{3.6cm}lllllp{1.75cm}p{1.75cm}}
\tabletypesize{\footnotesize}
\tablecaption{Weighted-mean accretion flow model results \label{tab:flowresults}}
\tablehead{
\colhead{Object} & \colhead{\ri} & \colhead{\rw} & \colhead{\mdot$_{\rm flow}$} & \colhead{\tmax} & \colhead{\imag} & \colhead{$A$} & \colhead{$\sigma$}\\
\colhead{} & \colhead{(R$_{\star}$)} & \colhead{(R$_{\star}$)} & \colhead{($10^{-9}$\msun/yr)} & \colhead{(K)} & \colhead{(\degrees)} & \colhead{(Norm. flux)} & \colhead{(km/s)}
}
\startdata
\textbf{Lupus I} \\ \hline
2MASSJ15445789-3423392 & $3.38 \pm 1.68$ & $0.90 \pm 0.67$ & $0.04 \pm 0.02$ & $11208 \pm 1062$ & $80 \pm 7$ & $6.63 \pm 0.18$ & $25.14 \pm 107.76$ \\
2MASSJ15450887-3417333 & $6.08 \pm 0.24$ & $1.61 \pm 0.19$ & $3.41 \pm 0.37$ & $7858 \pm 94$ & $80 \pm 5$ & \dots & \dots \\
SSTc2dJ154518.5-342125 & $3.75 \pm 0.25$ & $0.92 \pm 0.22$ & $0.07 \pm 0.02$ & $11865 \pm 763$ & $85 \pm 5$ & $21.14 \pm 1.76$ & $16.56 \pm 1.87$ \\
Sz 65 & $4.12 \pm 2.33$ & $0.83 \pm 0.68$ & $0.48 \pm 0.34$ & $10438 \pm 1301$ & $80 \pm 9$ & $1.1 \pm 33.5$ & $23.44 \pm 7.07$ \\
Sz 66\tablenotemark{a} & $1.78 \pm 0.11$ & $0.23 \pm 0.03$ & $1.12 \pm 0.24$ & $8220 \pm 178$ & $79 \pm 5$ & \dots & \dots \\
Sz 68\tablenotemark{a} & $2.82 \pm 0.25$ & $0.83 \pm 0.19$ & $34.78 \pm 4.51$ & $8263 \pm 156$ & $69 \pm 5$ & \dots & \dots \\
Sz 69\tablenotemark{a} & $4.05 \pm 0.15$ & $0.98 \pm 0.67$ & $0.60 \pm 0.14$ & $9919 \pm 979$ & $22 \pm 15$ & \dots & \dots \\
Sz 71\tablenotemark{a} & $1.64 \pm 0.10$ & $1.10 \pm 0.18$ & $1.74 \pm 0.31$ & $9189 \pm 239$ & $50 \pm 5$ & $8.84 \pm 0.04$ & $24.69 \pm 0.11$ \\
Sz 72\tablenotemark{a} & $3.50 \pm 0.00$ & $0.18 \pm 0.00$ & $2.66 \pm 0.13$ & $8509 \pm 25$ & $51 \pm 5$ & \dots & \dots \\
Sz 73 & $2.34 \pm 0.09$ & $1.42 \pm 0.49$ & $3.20 \pm 0.90$ & $10237 \pm 631$ & $48 \pm 5$ & \dots & \dots \\
Sz 75\tablenotemark{a} & $2.83 \pm 0.14$ & $3.16 \pm 0.00$ & $40.37 \pm 2.73$ & $8407 \pm 30$ & $52 \pm 5$ & \dots & \dots \\
\hline \textbf{Lupus II} \\ \hline
Sz 81 A & $2.08 \pm 0.25$ & $1.38 \pm 0.56$ & $0.88 \pm 0.57$ & $11306 \pm 1135$ & $65 \pm 10$ & \dots & \dots \\
Sz 82\tablenotemark{a} & $2.86 \pm 0.12$ & $0.94 \pm 0.09$ & $11.24 \pm 1.71$ & $7987 \pm 113$ & $62 \pm 5$ & $7.17 \pm 0.02$ & $25.85 \pm 0.07$ \\
Sz 84\tablenotemark{a} & $1.90 \pm 0.12$ & $1.44 \pm 0.18$ & $0.63 \pm 0.15$ & $10535 \pm 329$ & $70 \pm 5$ & \dots & \dots \\
\hline \textbf{Lupus III} \\ \hline
2MASSJ16073773-3921388 & $3.53 \pm 0.27$ & $2.03 \pm 0.39$ & $0.27 \pm 0.06$ & $11485 \pm 861$ & $85 \pm 5$ & $18.02 \pm 1.57$ & $21.92 \pm 48.32$ \\
2MASSJ16080017-3902595 & $1.67 \pm 0.46$ & $1.34 \pm 0.55$ & $0.12 \pm 0.06$ & $11999 \pm 890$ & $69 \pm 11$ & $2.2 \pm 2330.8$ & $28.2 \pm 8702.3$ \\
2MASSJ16084940-3905393 & $2.51 \pm 1.39$ & $1.02 \pm 0.68$ & $0.30 \pm 0.22$ & $11446 \pm 1253$ & $72 \pm 10$ & $5.39 \pm 0.12$ & $30.23 \pm 0.99$ \\
2MASSJ16085324-3914401 & $3.48 \pm 2.31$ & $0.64 \pm 0.51$ & $0.16 \pm 0.10$ & $10925 \pm 1258$ & $77 \pm 10$ & $1.91 \pm 0.11$ & $21.94 \pm 1.63$ \\
2MASSJ16085529-3848481 & $3.12 \pm 1.33$ & $0.79 \pm 0.60$ & $0.05 \pm 0.03$ & $11504 \pm 964$ & $82 \pm 5$ & $9.44 \pm 0.56$ & $24.41 \pm 4.51$ \\
2MASSJ16090141-3925119 & $2.39 \pm 0.14$ & $1.48 \pm 0.38$ & $0.31 \pm 0.04$ & $10971 \pm 507$ & $76 \pm 5$ & \dots & \dots \\
2MASSJ16092697-3836269 & $2.36 \pm 0.35$ & $1.55 \pm 0.69$ & $7.03 \pm 6.24$ & $8598 \pm 957$ & $63 \pm 14$ & \dots & \dots \\
2MASSJ16095628-3859518 & $3.34 \pm 1.55$ & $1.14 \pm 0.70$ & $0.04 \pm 0.02$ & $11623 \pm 955$ & $80 \pm 7$ & $10.47 \pm 5.16$ & $32.45 \pm 1.75$ \\
2MASSJ16101857-3836125 & $2.14 \pm 0.86$ & $1.23 \pm 0.58$ & $0.07 \pm 0.03$ & $12108 \pm 816$ & $65 \pm 19$ & $0.9 \pm 1619.6$ & $22.7 \pm 7788.3$ \\
2MASSJ16101984-3836065 & $2.02 \pm 0.65$ & $1.29 \pm 0.57$ & $0.10 \pm 0.05$ & $11998 \pm 856$ & $77 \pm 10$ & $3.5 \pm 1100.1$ & $24.8 \pm 5376.0$ \\
2MASSJ16102955-3922144 & $1.61 \pm 0.33$ & $1.27 \pm 0.41$ & $0.20 \pm 0.08$ & $11981 \pm 734$ & $73 \pm 7$ & $2.46 \pm 0.06$ & $26.16 \pm 0.85$ \\
SSTc2dJ160830.7-382827\tablenotemark{a} & $1.58 \pm 0.09$ & $0.07 \pm 0.01$ & $0.21 \pm 0.04$ & $10631 \pm 513$ & $56 \pm 5$ & \dots & \dots \\
Sz 90 & $2.42 \pm 0.94$ & $0.97 \pm 0.71$ & $2.07 \pm 1.22$ & $10690 \pm 1171$ & $78 \pm 12$ & \dots & \dots \\
Sz 91 & $6.74 \pm 0.15$ & $0.88 \pm 0.11$ & $0.88 \pm 0.13$ & $9333 \pm 154$ & $80 \pm 5$ & $18.81 \pm 0.17$ & $16.45 \pm 0.17$ \\
Sz 95 & $5.10 \pm 2.05$ & $1.10 \pm 0.74$ & $0.56 \pm 0.42$ & $9391 \pm 1191$ & $80 \pm 10$ & $1.55 \pm 0.04$ & $26.52 \pm 0.90$ \\
Sz 96 & $2.63 \pm 0.54$ & $1.34 \pm 0.47$ & $0.81 \pm 0.28$ & $10242 \pm 584$ & $77 \pm 5$ & $2.28 \pm 0.02$ & $32.50 \pm 0.39$ \\
Sz 97\tablenotemark{a} & $2.63 \pm 0.14$ & $2.58 \pm 0.20$ & $1.59 \pm 0.36$ & $11360 \pm 408$ & $83 \pm 5$ & $1.58 \pm 0.11$ & $25.58 \pm 2.89$ \\
Sz 98\tablenotemark{a} & $2.94 \pm 1.28$ & $1.50 \pm 0.38$ & $2.45 \pm 2.09$ & $9143 \pm 613$ & $58 \pm 8$ & \dots & \dots \\
Sz 99\tablenotemark{a} & $2.94 \pm 0.01$ & $2.14 \pm 0.12$ & $0.63 \pm 0.03$ & $10208 \pm 153$ & $82 \pm 5$ & $4.85 \pm 0.05$ & $29.01 \pm 0.36$ \\
Sz 103\tablenotemark{a} & $2.05 \pm 0.07$ & $0.57 \pm 0.05$ & $0.67 \pm 0.10$ & $10611 \pm 115$ & $82 \pm 5$ & \dots & \dots \\
Sz 104\tablenotemark{a} & $1.92 \pm 0.01$ & $1.03 \pm 0.10$ & $0.47 \pm 0.01$ & $10151 \pm 176$ & $88 \pm 5$ & $7.32 \pm 0.03$ & $29.60 \pm 0.16$ \\
Sz 106 & $4.22 \pm 1.97$ & $0.97 \pm 0.73$ & $0.13 \pm 0.08$ & $10901 \pm 1139$ & $67 \pm 22$ & $0.3 \pm 15710.7$ & $11.9 \pm 77310.1$ \\
Sz 110\tablenotemark{a} & $1.67 \pm 0.00$ & $1.82 \pm 0.26$ & $3.42 \pm 1.09$ & $9375 \pm 221$ & $77 \pm 5$ & $6.25 \pm 0.05$ & $29.03 \pm 0.25$ \\
Sz 111\tablenotemark{a} & $3.33 \pm 0.00$ & $1.77 \pm 0.00$ & $1.39 \pm 0.00$ & $10250 \pm 3$ & $65 \pm 5$ & $0.01 \pm 0.00$ & $0.01 \pm 0.00$ \\
Sz 112 & $1.90 \pm 0.32$ & $1.13 \pm 0.30$ & $0.23 \pm 0.06$ & $11144 \pm 550$ & $71 \pm 5$ & $8.98 \pm 0.03$ & $29.17 \pm 0.11$ \\
Sz 113 & $4.50 \pm 0.60$ & $1.44 \pm 0.65$ & $0.34 \pm 0.23$ & $10791 \pm 1104$ & $84 \pm 5$ & \dots & \dots \\
Sz 114 & $2.39 \pm 0.08$ & $0.56 \pm 0.05$ & $0.10 \pm 0.01$ & $12208 \pm 198$ & $51 \pm 5$ & $13.37 \pm 0.04$ & $33.21 \pm 0.10$ \\
Sz 115 & $4.05 \pm 0.85$ & $1.13 \pm 0.32$ & $0.51 \pm 0.20$ & $9283 \pm 595$ & $82 \pm 5$ & $2.16 \pm 0.01$ & $25.88 \pm 0.17$ \\
Sz 117\tablenotemark{a} & $4.09 \pm 0.54$ & $1.45 \pm 0.31$ & $1.39 \pm 0.25$ & $8941 \pm 237$ & $82 \pm 5$ & \dots & \dots \\
Sz 118 & $2.07 \pm 0.30$ & $1.86 \pm 0.27$ & $1.33 \pm 0.42$ & $12416 \pm 587$ & $61 \pm 5$ & \dots & \dots \\
V1094 Sco & $1.89 \pm 1.33$ & $1.18 \pm 0.68$ & $14.21 \pm 12.26$ & $8767 \pm 834$ & $60 \pm 12$ & \dots & \dots \\
\hline \textbf{Lupus IV} \\ \hline
2MASSJ16000236-4222145 & $1.87 \pm 0.36$ & $1.46 \pm 0.60$ & $0.62 \pm 0.35$ & $11639 \pm 956$ & $68 \pm 10$ & \dots & \dots \\
2MASSJ16002612-4153553 & $1.94 \pm 0.91$ & $1.23 \pm 0.62$ & $0.27 \pm 0.21$ & $10906 \pm 1254$ & $73 \pm 9$ & $5.32 \pm 0.10$ & $30.30 \pm 0.73$ \\
MY Lup\tablenotemark{a} & $7.17 \pm 0.48$ & $0.98 \pm 0.19$ & $0.71 \pm 0.14$ & $10120 \pm 319$ & $89 \pm 5$ & \dots & \dots \\
SSTc2dJ160000.6-422158\tablenotemark{a} & $4.45 \pm 0.56$ & $1.13 \pm 0.30$ & $0.31 \pm 0.09$ & $9080 \pm 415$ & $72 \pm 5$ & $1.64 \pm 0.01$ & $23.38 \pm 0.24$ \\
Sz 129\tablenotemark{a} & $3.06 \pm 0.25$ & $0.96 \pm 0.17$ & $5.24 \pm 0.81$ & $8129 \pm 169$ & $62 \pm 5$ & $5.89 \pm 0.01$ & $33.96 \pm 0.08$ \\
Sz 130\tablenotemark{a} & $3.16 \pm 0.20$ & $0.69 \pm 0.21$ & $1.94 \pm 0.30$ & $8167 \pm 272$ & $75 \pm 5$ & $13.75 \pm 0.04$ & $28.80 \pm 0.10$ \\
Sz 131 & $1.97 \pm 0.52$ & $1.10 \pm 0.67$ & $1.10 \pm 0.74$ & $10982 \pm 1278$ & $74 \pm 10$ & \dots & \dots \\
\hline \textbf{Off-Cloud} \\ \hline
2MASSJ15414081-3345188 & $2.34 \pm 0.46$ & $1.28 \pm 0.50$ & $0.03 \pm 0.01$ & $12374 \pm 640$ & $70 \pm 8$ & $6.67 \pm 0.35$ & $24.67 \pm 2.12$ \\
2MASSJ16081497-3857145 & $2.89 \pm 0.21$ & $1.83 \pm 0.55$ & $0.16 \pm 0.03$ & $10824 \pm 824$ & $84 \pm 5$ & \dots & \dots \\
2MASSJ16085373-3914367 & $3.46 \pm 2.35$ & $0.81 \pm 0.60$ & $0.02 \pm 0.02$ & $10580 \pm 1373$ & $79 \pm 9$ & $1.97 \pm 0.14$ & $22.70 \pm 2.03$ \\
2MASSJ16100133-3906449 & $2.10 \pm 0.88$ & $0.83 \pm 0.56$ & $0.42 \pm 0.25$ & $11453 \pm 1251$ & $80 \pm 7$ & $6.7 \pm 241.9$ & $30.3 \pm 1171.2$ \\
EX Lup & $2.30 \pm 0.00$ & $1.31 \pm 0.07$ & $6.79 \pm 0.89$ & $9275 \pm 108$ & $63 \pm 5$ & \dots & \dots \\
RXJ 1556.1-3655\tablenotemark{a} & $3.78 \pm 0.33$ & $0.47 \pm 0.56$ & $9.88 \pm 4.70$ & $7779 \pm 443$ & $64 \pm 6$ & \dots & \dots \\
SSTc2dJ161243.8-381503\tablenotemark{a} & $5.61 \pm 0.42$ & $1.43 \pm 0.22$ & $1.41 \pm 0.32$ & $9228 \pm 217$ & $73 \pm 5$ & \dots & \dots \\
SSTc2dJ161344.1-373646\tablenotemark{a} & $2.34 \pm 0.04$ & $1.29 \pm 0.09$ & $0.33 \pm 0.03$ & $11560 \pm 273$ & $58 \pm 5$ & $4.67 \pm 0.10$ & $18.79 \pm 0.42$ \\
Sz 77\tablenotemark{a} & $2.44 \pm 0.74$ & $0.72 \pm 0.44$ & $2.59 \pm 1.63$ & $8939 \pm 733$ & $50 \pm 9$ & $2.70 \pm 0.02$ & $19.86 \pm 0.17$ \\
 \enddata
 \tablenotetext{}{Accretion flow model results calculated as the weighted mean of the available epochs. Parameter uncertainties are taken to be the standard deviation of the top 1000 models, where the contribution of each model is scaled by exp(-\chisq/2). The \imag\ uncertainties are set to be at least 5\degrees, which is the typical \imag\ grid spacing. When perturbing \imag\ by its uncertainty, we use a truncated normal to prevent non-physical inclinations. Columns are the inner magnetospheric truncation radius \ri, the width of the flow at the disk \rw, the accretion rate \mdot, the maximum temperature of the flow \tmax, the magnetospheric inclination \imag\, and the amplitude ($A$) and width ($\sigma$) of the Gaussian used to represent the chromosphere, when applicable. $a$: Results from \cite{Pittman2025}. A table with results for all individual epochs is available in the online journal in machine-readable format.
 }
\end{deluxetable*}

\subsubsection{The measured \imag\ distribution} \label{Appsec:InclReliability}

\begin{figure}
    \centering
    \includegraphics[width=0.39\linewidth]{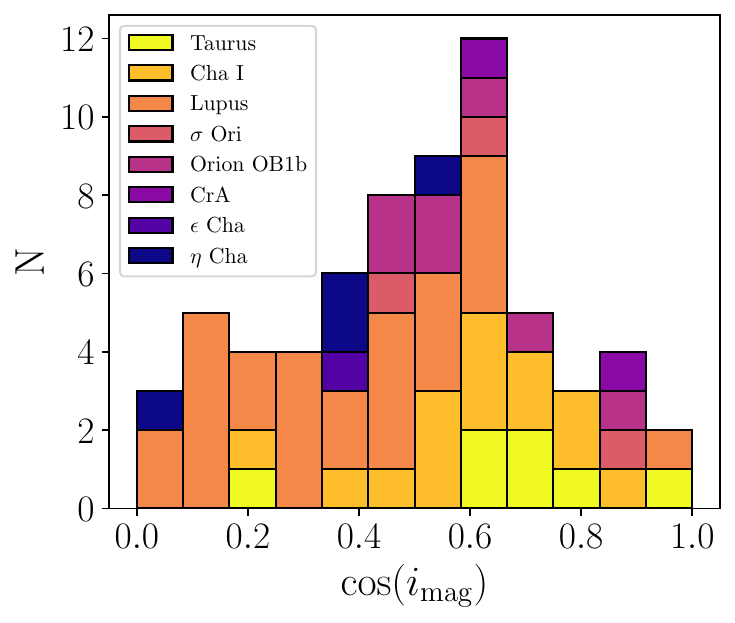}
    \includegraphics[width=0.6\linewidth]{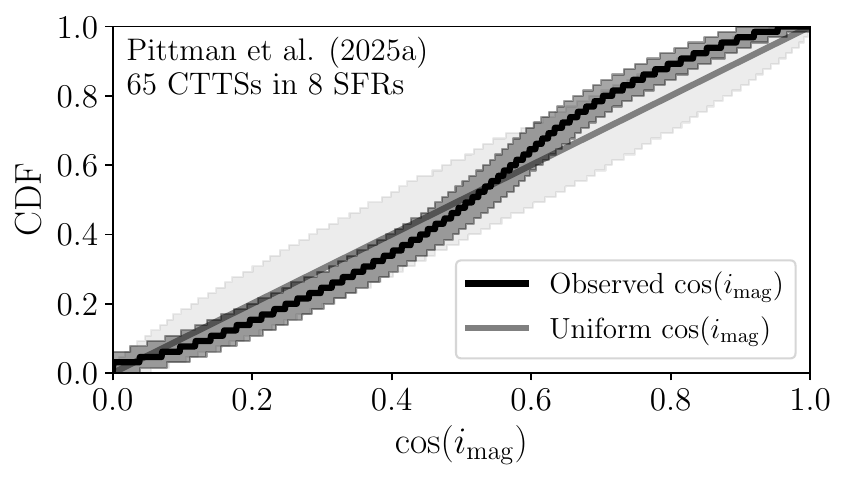}
    \includegraphics[width=0.39\linewidth]{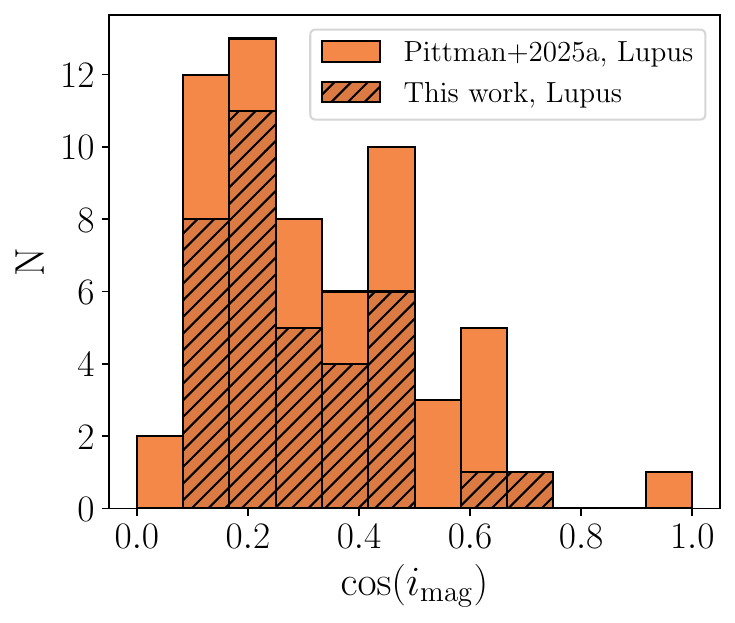}
    \includegraphics[width=0.6\linewidth]{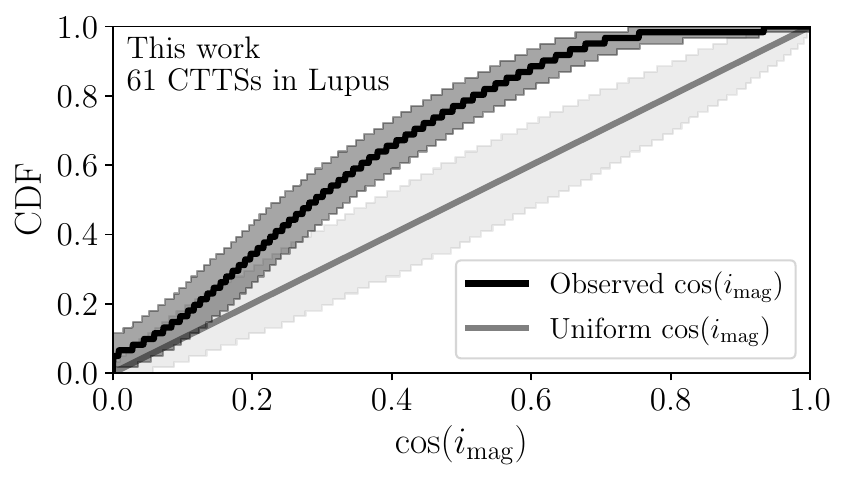}
    \caption{Inclination uniformity test. \textit{Top}: \cosi\ histogram (left) and CDF (right) for the eight star forming regions (SFRs) in \cite{Pittman2025}. The CDF figure shows the median and 95\% confidence interval of the observed \cosi\ values, as well as those of a uniform distribution. The observed \cosi\ values are consistent with uniformity, within uncertainties. \textit{Bottom}: Same as above, but for the Lupus sample in this work. These \cosi\ values are inconsistent with uniformity.}
    \label{fig:UniformTest}
\end{figure}

A sample of random inclinations should be uniform in \cosi\ between 0 and 1, with a median \cosi\ of 0.5 (corresponding to 60\degrees). Our 61 \imag\ measurements for Lupus, which are dominated by highly-inclined CTTSs in the Lupus~III region, have a median \cosi\ of 0.3 (see Figure~\ref{fig:UniformTest}, bottom left). 
We wish to confirm that this bias towards higher \imag\ is not a product of a bias in the modeling procedure, so we examine \imag\ values measured from the flow model for other samples of CTTSs. 
\cite{Pittman2025} applied the same modeling procedure to 65 CTTSs in Taurus, Chamaeleon I, $\epsilon$~Cha, $\eta$~Cha, Lupus (25 of which are included in this work), Orion OB1b, $\sigma$~Ori, and Corona Australis (CrA). The distribution of their \cosi\ results are shown in Figure~\ref{fig:UniformTest} (top left), and this has a median \cosi\ of 0.5, in agreement with a uniform distribution.

To characterize the uniformity further, we perform the following Monte Carlo cumulative distribution function (CDF) analysis. For each of the \cosi\ samples \citep[this work and that of][]{Pittman2025}, we first create 2000 CDFs drawn from a uniform distribution of the same sample size as the \cosi\ measurements. We show the median of the 2000 CDFs as the thick gray lines in Figure~\ref{fig:UniformTest} (right column), which is a straight line from 0 to 1. We then take the 2.5th and 97.5th percentiles of the 2000 CDFs to calculate the 95\% confidence interval (shown as the shaded gray region). This spans the range of CDFs that are likely to be consistent with \cosi\ being uniform between 0 to 1. We then create 2000 CDFs drawn from the observed \cosi\ distributions perturbed by their standard deviations ($\sigma_{i_{\rm mag}}$), and then calculate their 95\% confidence interval (shown as the shaded black region). We see that the \citep[][]{Pittman2025} sample from 8 SFRs is consistent with uniformity within the 95\% confidence intervals (Figure~\ref{fig:UniformTest}, top right). Conversely, the Lupus sample in this work is inconsistent with uniformity within the 95\% confidence intervals (Figure~\ref{fig:UniformTest}, bottom right). 
Instead, the Lupus sample is approximately uniform in $\cos i_{\rm mag}$ in the narrower band between 0 and 0.7, within uncertainties.
This comparison shows that the non-uniformity of \imag\ is a feature of the Lupus sample in particular, rather than of the modeling procedure itself. We discuss the geometric cause of this non-uniformity in Section~\ref{subsec:context}.

\subsubsection{Effect of inclination on the \halpha\ profile} \label{Appsec:InclEffect}

\begin{figure}
    \centering
    \includegraphics[trim=0 60pt 57pt 5pt, clip, height=0.15\textheight]{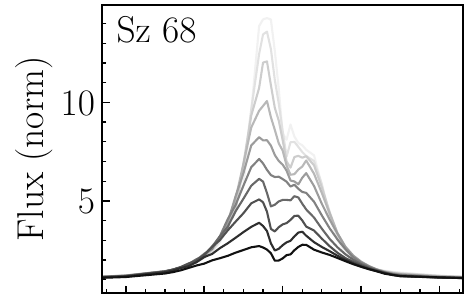}
    \includegraphics[trim=30pt 60pt 57pt 5pt, clip, height=0.15\textheight]{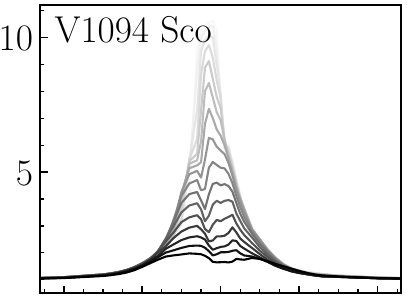}
    \includegraphics[trim=30pt 60pt 0 5pt, clip, height=0.15\textheight]{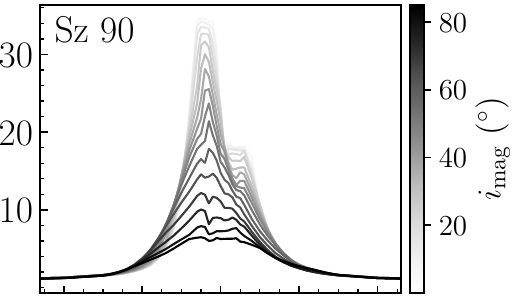}
    \includegraphics[trim=2pt 0pt 56pt 8pt, clip, height=0.212\textheight]{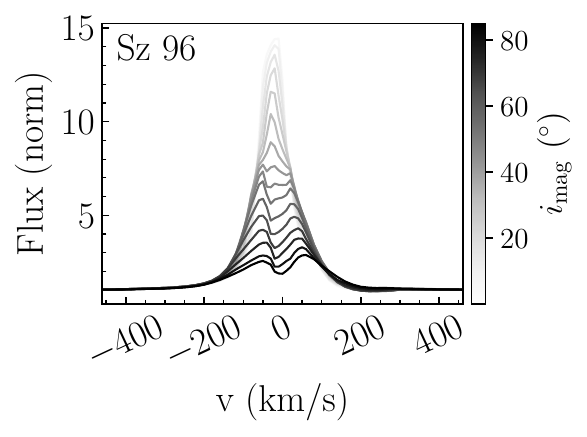}
    \includegraphics[trim=30pt 0pt 57pt 4pt, clip, height=0.212\textheight]{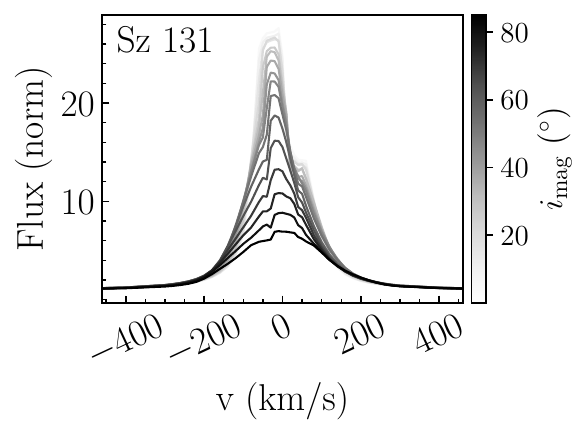}
    \includegraphics[trim=30pt 0pt 0 4pt, clip, height=0.212\textheight]{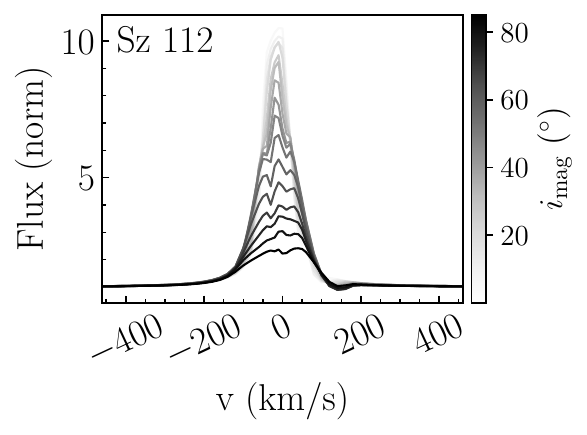}
    \caption{\halpha\ profile dependence on \imag, with all other accretion flow model parameters fixed.}
    \label{fig:InclDep}
\end{figure}

Figure~\ref{fig:InclDep} shows the effect of \imag\ on the observed \halpha\ profile for 6 CTTSs in the sample. For this test, the values of \ri, \rw, \mdot, and \tmax\ are fixed to values near those in Table~\ref{tab:flowresults}. The exact \halpha\ profile morphology associated with a given \imag\ value depends on both the stellar parameters and the other flow model parameters. In Sz~68, for example, the highly inclined models have brighter peaks on the blue wing, whereas in Sz~131, the brighter peaks are in the red wing. In V1094~Sco, the lower-inclination models show self-absorption on the blue side, whereas in Sz~90 they are on the red side. Therefore, there is no single \halpha\ profile element that would cause different CTTSs to cluster towards having similar \imag\ values.

\subsubsection{Degeneracy Tests} \label{sec:validation}

We checked whether the measured inclinations are correlated with any other system parameters. We found a moderate correlation between \imag\ and $\log_{10}$\mstar\ (with a Pearson correlation coefficient of $p=-0.4$), where \mstar\ comes from \cite{Pittman2025} and \cite{alcala14,alcala17}. This can be attributed to the different mass distributions of the samples in each subregion, which also have different inclination distributions. In our sample, the median \mstar\ is 0.30~\msun\ in Lupus I, 0.37~\msun\ in Lupus II, 0.16~\msun\ in Lupus III, 0.21~\msun\ in Lupus IV, and 0.32~\msun\ in the off-cloud CTTS.
We found no correlation between \imag\ and the stellar-mass-normalized \mdot\ (\mdot/${\rm M}_\star^2$).

In \cite{Pittman2025}, we noted that our corner plot analysis did not indicate any significant degeneracy between \imag\ and other flow model parameters, either on an individual-object scale or on the conglomerate results from the full sample. Therefore, it is unlikely that the measured \imag\ values result from degeneracy with the other model parameters.

\subsubsection{Significance of the chromosphere} \label{Appsec:ChromTest}

We also tested whether the inclusion of a Gaussian component representative of the chromosphere in the lower accretors influenced their measured inclinations.
There are eleven CTTSs whose best-fit inclinations change by more than 10\degrees\ when the Gaussian component is not included in the fitting procedure. Inspection of the fits shows that eight of these CTTSs clearly require a chromospheric contribution, as the magnetospheric model alone is unable to account for both the narrow central emission and the broad wings. The magnetospheric model alone can explain the \halpha\ profiles of the remaining three CTTSs; however, they all have spectral types of M3 or later and accretion rates below $5\times10^{-10}$~\msunyr. In this parameter space, the chromosphere should be a significant contribution to the \halpha\ profile. Therefore, we conclude that these eleven inclinations that depend on the inclusion of the chromospheric component remain trustworthy.

\subsection{Other star-forming regions} \label{Appsec:otherregions}

\begin{figure}[h!]
    \centering
    \begin{tikzpicture}
    \centering
    \node[anchor=south west, inner sep=0] (image) at (0,0) {\includegraphics[width=0.6\linewidth]{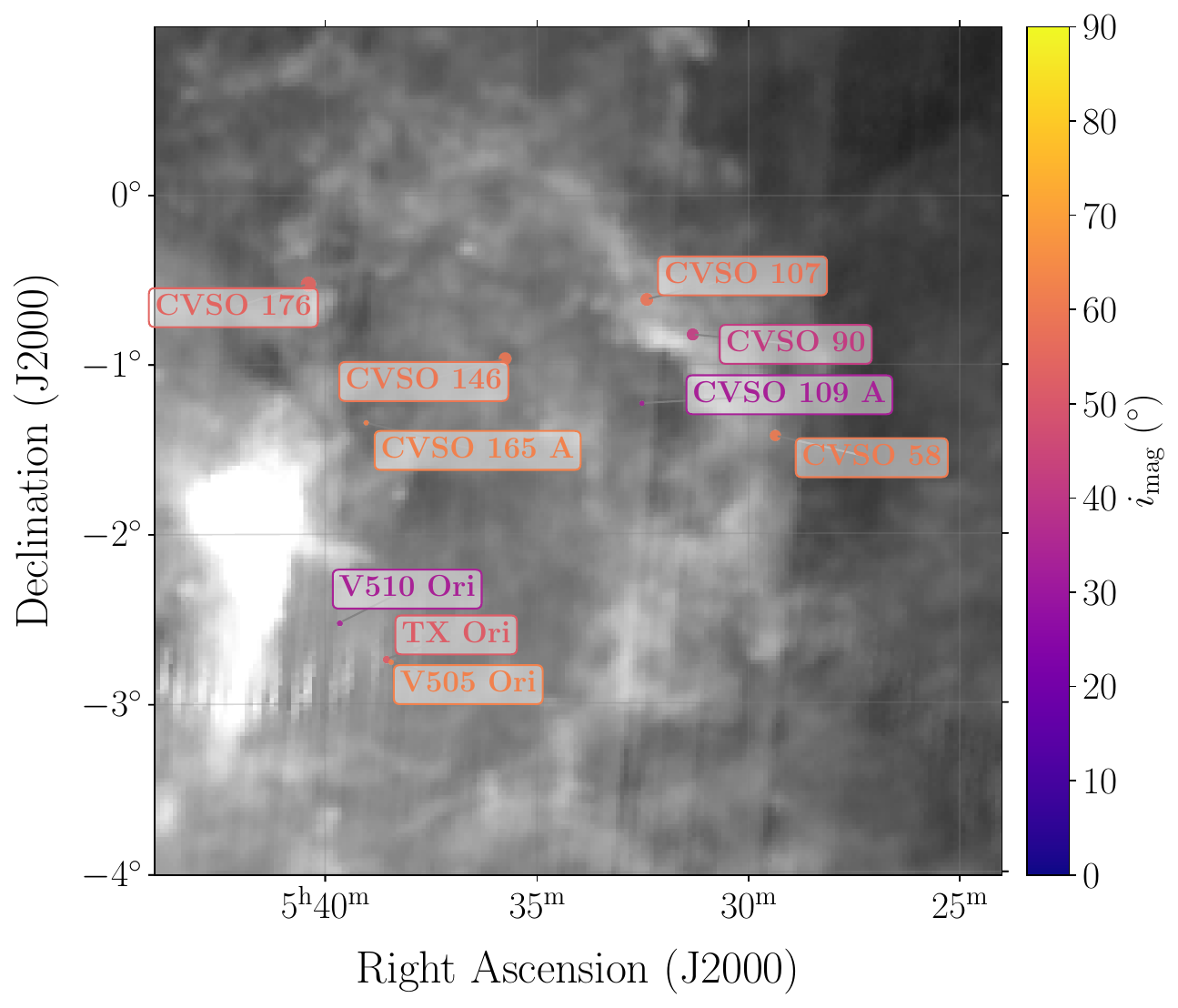}};
    \begin{scope}[x={(image.south east)},y={(image.north west)}]
            % Annotations relative to image coordinates (0,0) to (1,1)
            \node[white, font=\large] at (0.41, 0.69) {Orion OB1b};
            \node[white, font=\large] at (0.275, 0.355) {$\sigma$ Ori};
        \end{scope}
    \end{tikzpicture}
    \begin{tikzpicture}
    \centering
    \node[anchor=south west, inner sep=0] (image) at (0,0) {\includegraphics[width=0.39\linewidth]{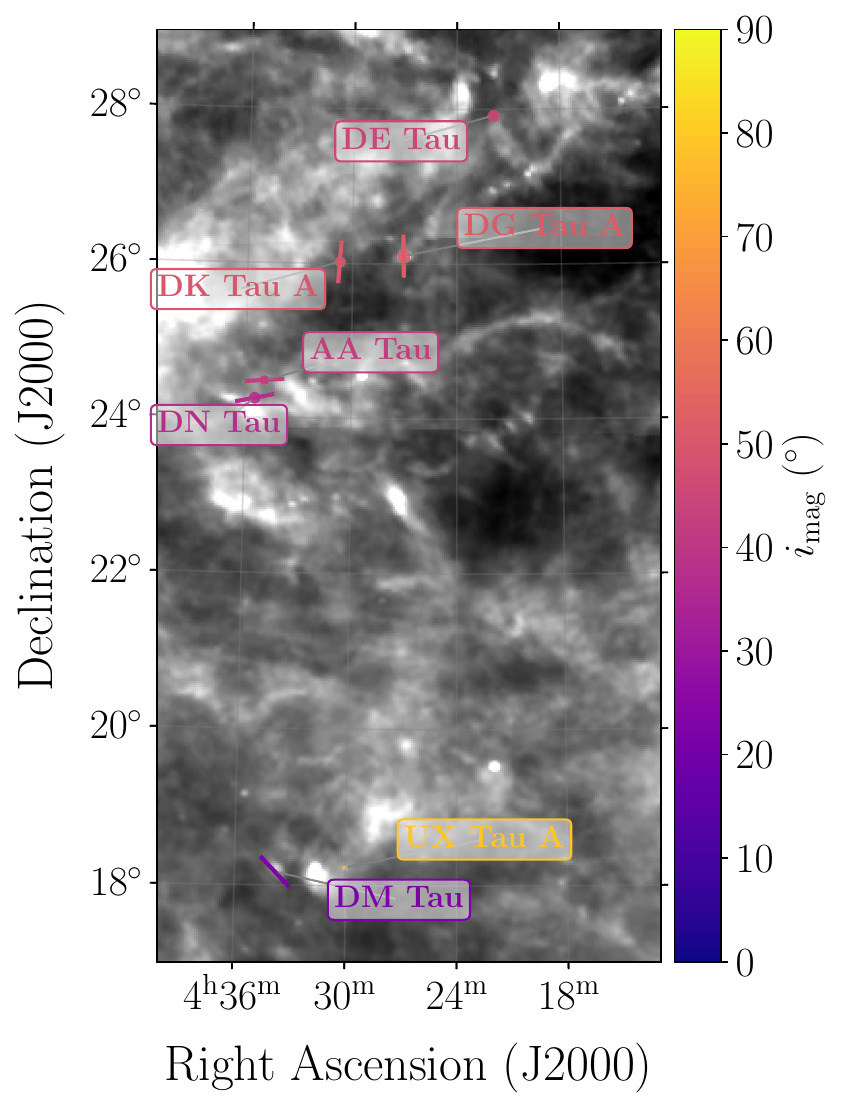}};
    \begin{scope}[x={(image.south east)},y={(image.north west)}]
            % Annotations relative to image coordinates (0,0) to (1,1)
            \node[white, font=\large] at (0.6, 0.54) {Taurus};
        \end{scope}
    \end{tikzpicture}
    \begin{tikzpicture}
    \centering
    \node[anchor=south west, inner sep=0] (image) at (0,0) {\includegraphics[width=0.62\linewidth]{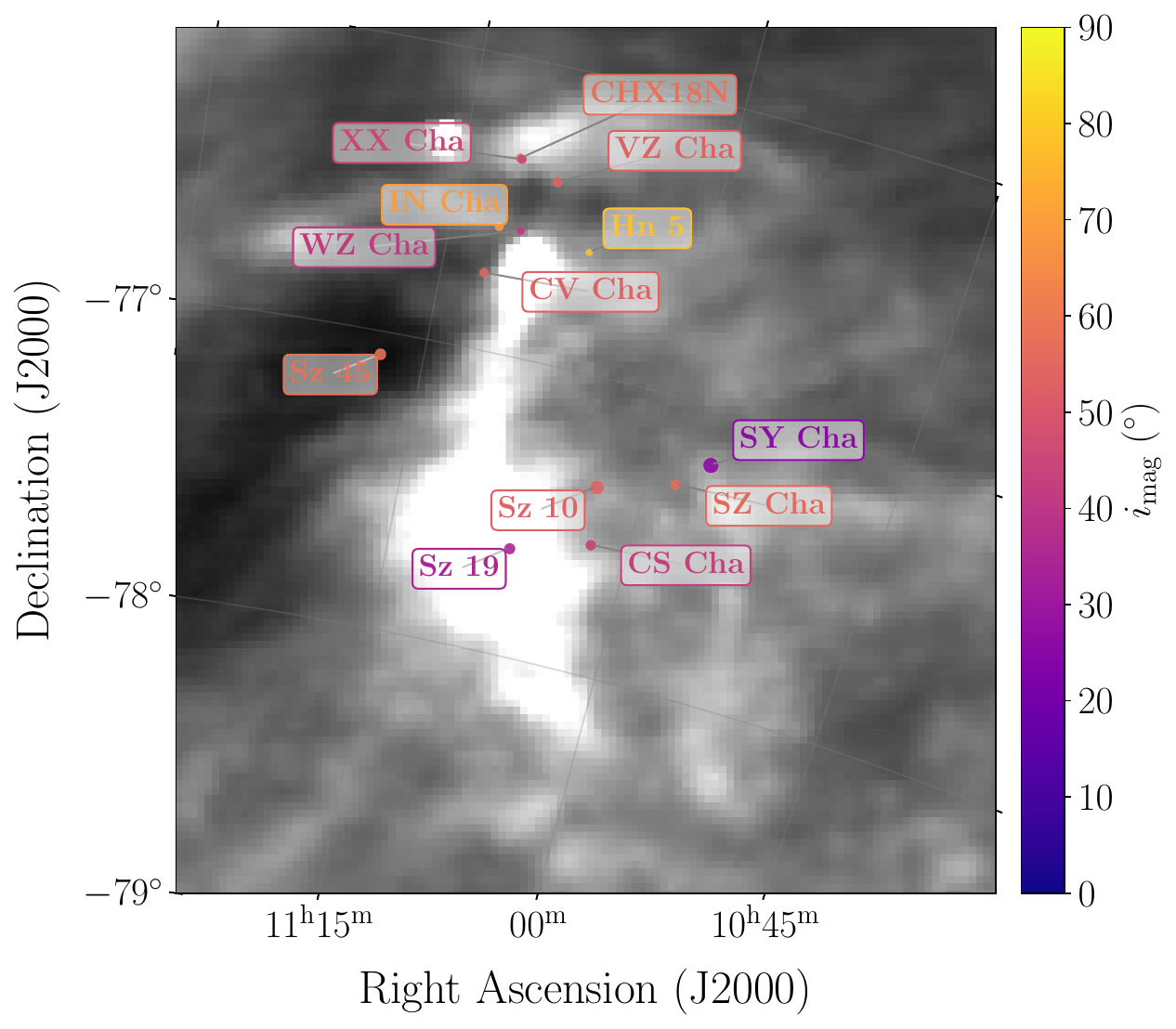}};
    \begin{scope}[x={(image.south east)},y={(image.north west)}]
            % Annotations relative to image coordinates (0,0) to (1,1)
            \node[white, font=\large] at (0.5, 0.2) {Chamaeleon I};
        \end{scope}
    \end{tikzpicture}
    \caption{Maps of the Orion (top left), Taurus (top right), and Chamaeleon I (bottom) targets from \cite{Pittman2025} with IRIS 100 \micron\ maps \citep{IRIS2005} shown in grayscale. The color bar shows the magnetospheric inclination found from the accretion flow model. Point sizes indicate \gaia\ such that more distant sources appear smaller. Position angles (PAs) compiled by \cite{Aizawa2020} are shown as colored bars. They are determined from $0^\circ\leq {\rm PA} < 180^\circ$ rather than $0^\circ\leq {\rm PA} < 360^\circ$. The \imag\ and PA values in Orion OB1b and Taurus may be spatially correlated, though the samples are too small to confirm.
    }
    \label{fig:TaurusOrion}
\end{figure}

\cite{Aizawa2020} did not find any significant nonuniformity in PAs in the Taurus Molecular Cloud, the Upper Scorpius OB Association, the $\rho$ Ophiuchi Cloud Complex, or the Orion Nebula Cluster. Still, our results do not indicate random distributions of inclinations within individual star-forming regions (though the sample sizes are too small to be conclusive).
Five out of seven nominal magnetospheric inclinations in Orion OB1b are within the 8\degrees\ range between 54--62\degrees\ (with the remaining two being 33\degrees\ and 42\degrees; see Figure~\ref{fig:TaurusOrion}, top left). This is notable given that the median stellar separation between these stars with reliable \gaia\ is 20~pc. It could indicate the influence of nearby expanding shells from the massive stars in the region, but more observations and analysis are required to expand the sample to test for correlations in Orion OB1b.

The 5 northernmost Taurus systems (AA Tau, DE Tau, DG Tau~A, DK Tau~A, and DN Tau) have nominal inclinations in the 14\degrees\ range between 38--52\degrees, 
and the PAs of the pairs AA/DN Tau and DK/DG Tau~A are aligned within 7\degrees\ (Figure~\ref{fig:TaurusOrion}, top right). In 
Chamaeleon I, conversely, \imag\ spans the range between 26--78\degrees\ somewhat uniformly (Figure~\ref{fig:TaurusOrion}, bottom). In the future, we will extend our analysis to include all $\sim$50 CTTSs in the Chamaeleon I region presented in \citet[][and references therein]{manara23PPVII}, as well as 37 CTTSs in Taurus from the GHOsT project \citep{Alcala2021,Gangi2022}, to examine whether statistically-significant \imag\ correlations appear when more CTTSs are included.

\end{document}